\documentclass[pra, reprint, twocolumn, superscriptaddress, amsmath, amssymb, aps]{revtex4-2}

\usepackage[english]{babel}
\usepackage{graphicx}
\graphicspath{{./Images/},{./ImagesAppendix/},{./images/},{./imagesAppendix/}}

\usepackage{amsmath, amssymb, amsfonts}
\usepackage{physics}
\usepackage{bm}
\usepackage{braket}
\usepackage{array}
\usepackage{multirow}
\usepackage{float}
\usepackage{commath}
\usepackage{verbatim}
\usepackage{color}
\usepackage{enumitem}
\usepackage{dsfont}
\usepackage{mathtools}
\usepackage[colorlinks=true, citecolor=darkBlue, linkcolor=darkBlue, urlcolor=blue]{hyperref}
\usepackage{amssymb}
\usepackage{tabularx}

\definecolor{darkBlue}{rgb}{0.08, 0.13, 0.4}

\usepackage{tikz}
\usetikzlibrary{positioning, arrows.meta}
\usepackage{pgfplots}

\definecolor{THc}{rgb}{0.9,0.3,0.2}

\newcommand{\canc}[1]{}

\begin{document}

\title{Quantum Coherence and Anomalous Work Extraction in Qubit Gate Dynamics}
\author{F. Perciavalle}
\affiliation{Dipartimento di Fisica, Universit\`a della Calabria, 87036 Arcavacata di Rende (CS), Italy}
\affiliation{INFN--Gruppo collegato di Cosenza}

\author{N. Lo Gullo}
\affiliation{Dipartimento di Fisica, Universit\`a della Calabria, 87036 Arcavacata di Rende (CS), Italy}
\affiliation{INFN--Gruppo collegato di Cosenza}

\author{F. Plastina}
\affiliation{Dipartimento di Fisica, Universit\`a della Calabria, 87036 Arcavacata di Rende (CS), Italy}
\affiliation{INFN--Gruppo collegato di Cosenza}

\date{\today}

\begin{abstract}
We develop a framework based on the Kirkwood-Dirac quasiprobability distribution to quantify the contribution of coherence to work extraction during generic, cyclic quantum evolutions. In particular, we focus on ``anomalous processes'', counterintuitive scenarios in which, due to the negativity of the quasiprobability distribution, work can be extracted even when individual processes are associated with energy gain. Applying this framework to qubits undergoing sequences of single- and two-qubit gate operations, we identify specific conditions under which such anomalous work exchanges occur. Furthermore, we analyze the quasiprobabilistic structure of deep quantum circuits and establish a compositional relation linking the work statistics of full circuits to those of their constituent gates. Our work highlights the role of coherence in the thermodynamics of quantum computation and provides a foundation for systematically studying potential thermodynamic relevance of specific quantum circuits.

\end{abstract}

\maketitle

\section{Introduction}
Quantum coherence lies at the heart of many quantum technologies and fundamental quantum phenomena~\cite{streltsov2017colloquium, xi2015quantum, streltsov2015measuring, li2016quantum, wu2021experimental}. Recent advances have revealed that coherence plays a crucial role~\cite{anders2016scirep, francica2019pre, santos2019njp,  francica2020quantum, shi2022entanglement, shi2020quantum, gour2022PRXQ, francica2024work, rodrigues2024, onishchenko2024natcom} in quantum thermodynamics~\cite{vinjanampathy2016quantum,goold2016role,alicki2018introduction,campbell2025roadmap,korzekwa2016extraction,talkner2016aspects, allahverdyan2014nonequilibrium, campisi2011colloquium, campaioli2024colloquium}. In particular, the presence of initial coherence during work extraction processes challenges the classical notions of thermodynamic constraints, allowing for anomalous processes that defy classical intuition. Quantum thermodynamic features are often explored through work statistics, which capture the impact of energy fluctuations in quantum processes~\cite{tasaki2000jarzynski, plastina2014irreversible,silva2008statistics, talkner2016aspects}. However, standard approaches to work statistics for quantum systems often rely on projective energy measurements, particularly within the widely adopted two-point measurement (TPM) scheme~\cite{gherardini2024quasiprobabilities, diaz2020quantum, batalhao2014experimental, mazzola2013meausring, dorner2013extracting, oftelie2025measurement, solfanelli2025universal}. Although projective measurements are  commonly implemented, they inherently destroy quantum coherence of the initial state in the energy eigenbasis, making the protocol invasive and leading to violations of the first law~\cite{imparato2024Q, acin2017nogo}, or of the no-signalling-in-time condition~\cite{gherardini2024quasiprobabilities, leggett1985quantum}. To overcome this limitation, quasiprobability distributions, in particular the Kirkwood-Dirac quasiprobabilities (KDQs)~\cite{kirkwood1933quantum,dirac1945on,gherardini2024quasiprobabilities,  arvidsson2024properties}, have emerged as powerful alternatives and have attracted a lot of interest, both theoretically~\cite{lostaglio2023kirkwood, pezzutto2025nonpositive, santini2023work, pei2023exploring, chakrabarty2025probing, donati2024energetics, thio2025kirkwood, donelli2025impact} and experimentally~\cite{li2025experimental,hernandez2024projective, Hernandez2024interferometry}. This framework allows for a consistent description of quantum work statistics that preserve and reveal the effects of coherence. Notably, KDQs can take negative or complex values, intrinsically signaling nonclassical features such as quantum interference and anomalous energy transitions, which can either enhance or hinder the extraction of work.

In this paper, we investigate the quasiprobability structure of work distribution in quantum circuits, focusing on how the coherence of the input state enables anomalous thermodynamic behavior. Using a systematic framework, we decompose the KDQs of deep circuits into those of their constituent gates, showing that the overall unitary transformation can exhibit negative KDQs, and thus anomalous work extraction, even if individual gates do not. We also identify conditions under which global KDQs simplify to combinations of single-gate KDQs. Through examples with universal single- and two-qubit gates, we show how coherence-induced anomalies enable work extraction beyond classical limits, offering insight into thermodynamic advantages of specific quantum circuits.

The remainder of the paper is organized as follows: in Sec.~\ref{sec:qthermo_cyclic} we provide an overview on the thermodynamics of cyclic transformations in the presence of initial coherences, by introducing the KDQs. We show that energy varying processes can lead to anomalous contributions to work extraction. In sec.~\ref{sec:single_qubit}, we explore the nonclassical thermodynamic features of single-qubit transformations; we first study the most general unitary transformation and then elementary quantum gates as the Hadamard and the $\pi/8$ gate.
In Sec.~\ref{sec:decomposition}, we provide a framework for relating deep quantum circuit KDQs with those of their constituent gates, and to decompose them into contributions from these individual components. Then, in Sec.~\ref{sec:two_qbits}, we study the KDQs of two-qubit gates, showing how they simplify when the input state is factorized, and identifying the gates for which KDQs retain a simplified form even in the presence of initial entanglement. In Sec.~\ref{sec:example}, we analyze the thermodynamic properties of a representative two-qubit circuit by applying the decomposition of KDQs and the results on two-qubit gates, both derived in the preceding sections. Finally, we give some concluding remarks in Sec.~\ref{secconclu}.

\section{Quantum thermodynamics of cyclic transformations}\label{sec:qthermo_cyclic}
We consider a cyclic transformation described by the unitary evolution $\hat{\mathcal{U}}$ over the time interval $t \in [0, \tau]$. The initial/final Hamiltonian is $\hat{\mathcal{H}}=\sum_k E_k \hat{\Pi}_k$, with eigenvalues $\{E_k\}$ and eigenprojectors $\{\hat{\Pi}_k =\ket{E_k}\bra{E_k}\}$. The system is initially prepared in the input state $\hat{\rho}$ and, after the unitary evolution, its energy change can be described by the extractable work 
\begin{equation}
\mathcal{W_{\mathcal{U}}}=\operatorname{Tr}\Bigl[\hat{\mathcal{H}}\Bigl(\hat{\rho} - \hat{\mathcal{U}}\hat{\rho}\hat{\mathcal{U}}^{\dagger}\Bigr)\Bigr].    
\end{equation}

One of the most widely used approaches to measure the work performed on/extracted from a quantum system is the TPM scheme. 
This protocol involves performing two projective measurements in the eigenbasis of the system's Hamiltonian, first at time $t=0$, and, later, at the time $t=\tau$, at the end of the transformation. The outcomes of these measurements form pairs $\left(E_i,E_f\right)$, which are distributed according to a joint probability distribution $p_{if}^{\mathcal{U}}(\hat{\rho})$, which depends on the initial state $\hat{\rho}$ and the unitary dynamics $\hat{\mathcal{U}}$. However, marginalizing the latter distribution over all the possible outcomes of the first measurement does not always lead to the right probability of getting outcome $E_f$ from the second measurement, $p_f^{\mathcal{U}}(\hat{\rho})$; so that the procedure turns out to be invasive~\cite{gherardini2024quasiprobabilities}. One of the major sources of this invasiveness is the incompatibility between the initial state and the eigenbasis of the initial Hamiltonian: $\bigl[\hat{\rho},\hat{\Pi}_i\bigr]\neq 0$; in this case, the first projective measurement irreversibly destroys any coherences present in $\hat{\rho}$ with respect to that basis, thereby altering the subsequent system evolution. 

To investigate work statistics within noninvasive thermodynamic frameworks, the TPM joint probability distribution is replaced by quasiprobability distributions. Among these, the Kirkwood-Dirac quasiprobabilities (KDQs)~\cite{kirkwood1933quantum,dirac1945on,perarnau2017nogo, pei2023exploring, gherardini2024quasiprobabilities, lostaglio2023kirkwood, hernandez2024projective, thio2025kirkwood, santini2023work, donelli2025impact} are widely adopted and are defined as
\begin{equation}
    q_{if}^{\mathcal{U}}(\hat{\rho})=\operatorname{Tr}\left[\hat{\mathcal{U}}^{\dagger}\hat{\Pi}_f \hat{\mathcal{U}} \hat{\Pi}_i \hat{\rho} \right];
\end{equation}
KDQs are linear in $\hat{\rho}$ and 
their marginal sums over initial and final indices are always properly normalized. However, differently from canonical probability distributions, they can become negative and nonreal. 

The work extracted from a cyclic transformation can be expressed in terms of the KDQs as follows
\begin{equation}
    \mathcal{W}_{\mathcal{U}}[\hat{\rho}]=\sum_{i\neq f}q_{if}^{\mathcal{U}}(\hat{\rho})(E_i - E_f)=\sum_{i\neq f}\mathrm{Re}q_{if}^{\mathcal{U}}(\hat{\rho})(E_i - E_f);
\end{equation}
the real part of the KDQ is the so called Margenau-Hill quasiprobability (MHQ)~\cite{margenau1961correlation,gherardini2024quasiprobabilities, pei2023exploring, yi2025observing,yoshimura2025quasiprobability, bizzarri2025quasiprobability}, which can be written as $\mathrm{Re}q_{if}^{\mathcal{U}}(\hat{\rho})=\frac{1}{2}\operatorname{Tr}\left[\Bigl\{\hat{\mathcal{U}}^{\dagger}\hat{\Pi}_f \hat{\mathcal{U}}, \hat{\Pi}_i\Bigr\}\hat{\rho}\right]$. While the real part directly contributes to the extracted work, the imaginary part does not influence this first moment, but plays a role in higher-order moments (see appendix~\ref{app:real_im}) and, crucially, will come into play as well when evaluating KDQ's of unitary processes expressed in terms of gates.

\begin{figure}[t!]
\centering
\includegraphics[width=0.9\linewidth]{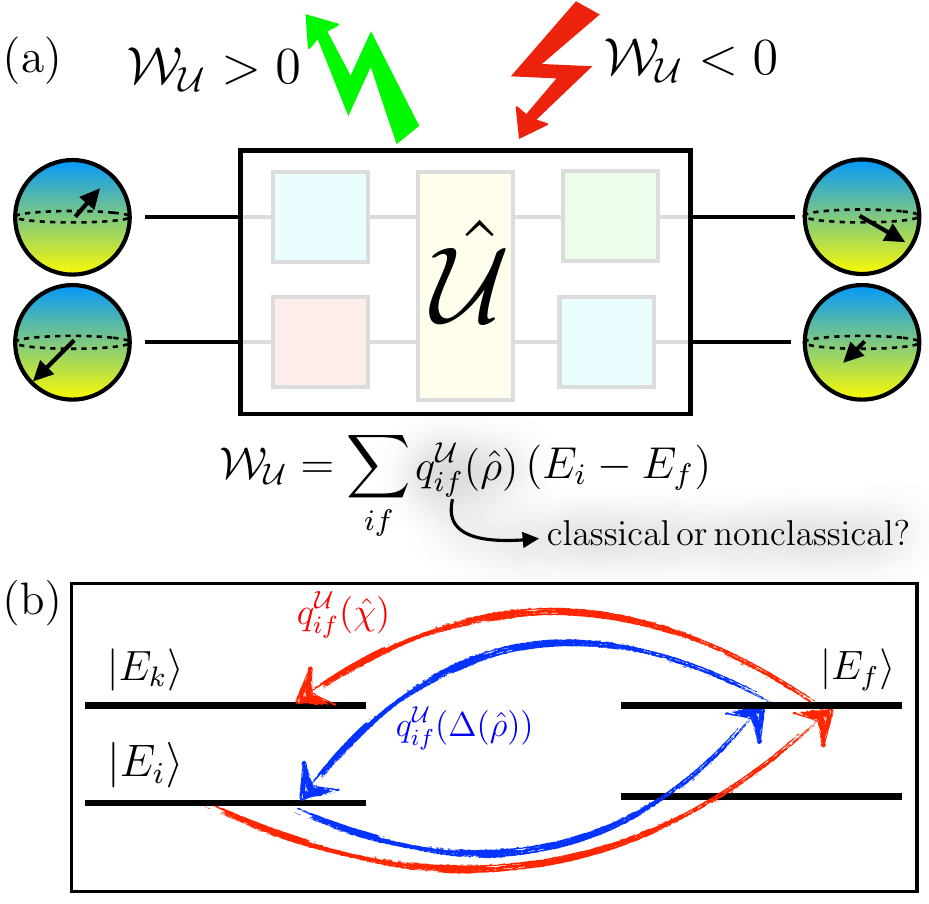}
\caption{Panel (a): sketch of the protocol for a two-qubit system, whose state is transformed by a unitary $\hat{\mathcal{U}}$ decomposed into elementary gates constituting a quantum circuit. We explore the role of the coherences and of the quantum circuit in the extractable work and in its nonclassical statistics. Panel (b): Pictorial representation of population and coherent parts of the KDQs in a single-qubit system. The blue arrows refer to the population part, the red arrows to the coherent part that potentially lead to nonclassicality.}
\label{fig:sketch}
\end{figure}

In our scheme, the initial/final Hamiltonian is the noninteracting $\hat{\mathcal{H}}=E\sum_{j=1}^L \hat{Z}_j$, where $\hat{Z}_j$ is the Pauli matrix acting on the computational basis such that $\hat{Z}_j\ket{\uparrow_j}=\ket{\uparrow_j}$ and $\hat{Z}_j\ket{\downarrow_j}=-\ket{\downarrow_j}$ ($\hat{X}_j$ and $\hat{Y}_j$ are the remaining ones), $E$ fixes the single qubit energy, $L$ is the number of qubits. Our goal is to investigate how the nonclassicality of the work distribution is influenced by the properties of the transformation, that we will treat as a quantum circuit decomposed in elementary gates. The protocol is sketched in Fig.~\ref{fig:sketch}(a), with the example of a two-qubit system.

The input quantum state can be always written as 
\begin{equation}
\hat{\rho}=\sum_{ik}\lambda_{ik}\ket{E_i}\bra{E_k}=\Delta(\hat{\rho})+\hat{\chi},    
\end{equation}
where $\Delta(\hat{\rho})=\sum_i \lambda_{ii}\ket{E_i}\bra{E_i}$ is the dephased version of the state $\hat{\rho}$ in the Hamiltonian eigenbasis and $\hat{\chi}=\sum_{i\neq k}\lambda_{ik}\ket{E_i}\bra{E_k}$ contains the associated coherences. $\hat{\chi}$ can be interpreted as the coherence injected into the mixed state $\Delta(\hat{\rho})$. Since the KDQs are linear in $\hat{\rho}$, they can be easily split into population and coherent contributions 
\begin{equation}
q_{if}^\mathcal{U}(\hat{\rho})=q_{if}^\mathcal{U}(\Delta(\hat{\rho})) + q_{if}^\mathcal{U}(\hat{\chi}),     
\end{equation}
where $q_{if}^\mathcal{U}(\Delta(\hat{\rho}))=\bigl| k_{if}^{\mathcal{U}}\bigr|^2\lambda_{ii}$ and $q_{if}^\mathcal{U}(\hat{\chi})=\sum_k{}^{'}\left(k_{kf}^{\mathcal{U}}\right)^{*}k_{if}^{\mathcal{U}}\lambda_{ik}$, see appendix~\ref{app:popcoh} for the detailed derivation. The primed sum $\sum_k{}^{'}$ indicates summation over all the $k\neq i$, while the quantity $k_{if}^{\mathcal{U}}=\braket{E_f|\hat{\mathcal{U}}|E_i}$ represents the transition amplitude between two different energy eigenstates. The population part of the KDQ is given by the transition probability of the process $\ket{E_i}\rightarrow\ket{E_f}$ weighted by the population $\lambda_{ii}$ and always contributes positively to the KDQ. On the other hand, the coherent part of the KDQ arises from the product of the transition amplitudes of the process $\ket{E_i}\rightarrow\ket{E_f}$ and of the time-reverse process $\ket{E_f}\rightarrow\ket{E_k}$ (with $k\neq i$). The mixing of potentially negative and complex transition amplitudes, each weighted by different coherences, can result in negative and complex coherent parts of the KDQs, which may ultimately lead to negative total KDQs. The two contributions to KDQ are pictorially represented in Fig. \ref{fig:sketch}(b) for a two-level system.

\begin{figure*}[t!]
\centering
\includegraphics[width=\linewidth]{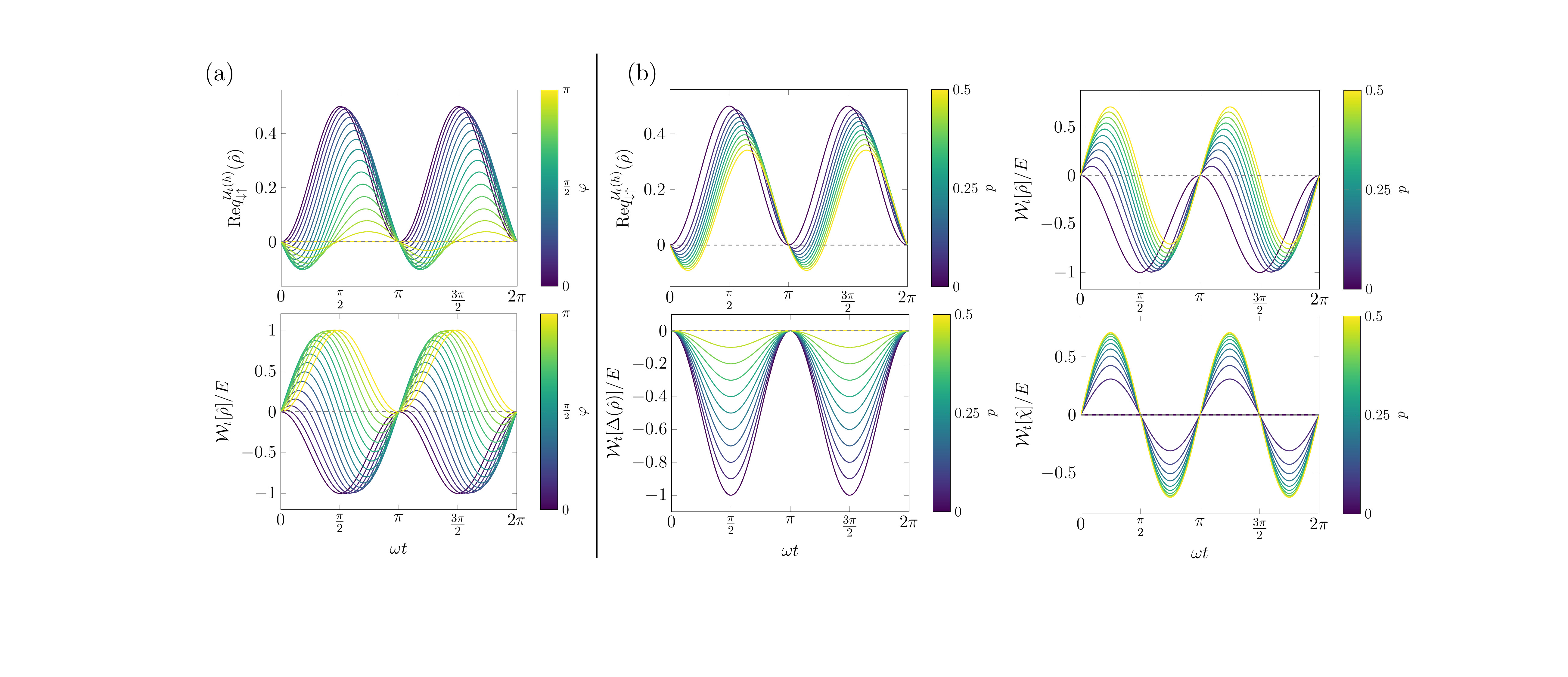}
\caption{Real part of $q_{\downarrow\uparrow}$ and work extracted from a pure state $\ket{\psi(p,\varphi)}=\sqrt{p}\ket{\uparrow}+e^{i\varphi}\sqrt{1-p}\ket{\downarrow}$ that undergoes a transformation described by $\hat{\mathcal{H}}_h=h\left(\hat{X}+\hat{Z}\right)$, with $h>0$. Panel (a): the population is fixed to $p=\frac{1}{2}$ and different values of the phase $\varphi$ are considered. Panel (b): the phase is fixed to $\varphi=\pi/2$ and different values of the population $p\in[0,0.5]$ are considered.}
\label{fig:hadamard_features}
\end{figure*}
The negativity of the real part of KDQs, i.e. negativity of MHQs, signals the presence of ``anomalous processes" that contribute in a counterintuitive way to the work extraction. For instance, processes $\ket{E_i}\rightarrow\ket{E_f}$ associated to larger final eigenergies $E_f > E_i$ with negative KDQ anomalously contribute positively to work extraction, making them advantageous in contrast to the classical scenario. On the other hand, processes corresponding to smaller final energies $E_f < E_i$ with negative KDQ negatively impact work extraction and are thus detrimental.

Moreover, the injection of coherences in thermal states allows for the possibility of positive work extraction during a cyclic transformation. In particular, in a cyclic transformation with a thermal input state, the work satisfies the Jarzynski equality (JE)~\cite{jarzynski1997nonequilibrium,jarzynski1997equilibrium} which implies $\mathcal{W}_{\mathcal{U}}\leq 0$. This means that no work can be extracted. However, the introduction of coherence can potentially lead to a violation of this bound. In the specific case of a qubit, any classical state represented in energy the basis without population inversion (i.e. with population imbalanced toward the low energy state) is thermal with positive temperature and thus satisfies $\mathcal{W}_{\mathcal{U}}\leq 0$. Thus, the injection of coherence is essential to achieve positive work extraction. In addition, in the absence of population inversion, the nonpositivity of $q_{if}^\mathcal{U}(\hat{\chi})$ is a necessary but not sufficient condition for this to occur, as discussed in detail in appendix~\ref{app:popcoh}.

\section{KDQs of single-qubit transformations}\label{sec:single_qubit}
We now consider single-qubit operations; we first focus on the most general unitary transformation and then on the specific Hadamard and $\pi/8$ gates. Let $\hat{\rho}$ be a input state, with populations $\lambda_{\uparrow\uparrow}=p$ and $\lambda_{\downarrow\downarrow}=1-p$, and coherences $\lambda_{\uparrow\downarrow}=\gamma=|\gamma|e^{i\varphi}$ and $\lambda_{\downarrow\uparrow}=\gamma^*=|\gamma|e^{-i\varphi}$, where $p \in [0,1]$ and $|\gamma|\leq\sqrt{p(1-p)}$ to guarantee positivity of the state. The most general single-qubit unitary transformation can be expressed as $\hat{R}(\theta,\boldsymbol{n})=e^{-i (\theta/2)\boldsymbol{n}\cdot \hat{\boldsymbol{\sigma}}}$ up to a global phase factor, with $\hat{\boldsymbol{\sigma}}=\left(\hat{X},\hat{Y},\hat{Z}\right)$. The extractable work through $\hat{R}(\theta,\boldsymbol{n})$ is
\begin{equation}
\mathcal{W}_R[\hat{\rho}]=2E\left[(n_x^2+n_y^2)\sin^2(\theta/2)-2\mathrm{Re}q_{\downarrow\uparrow}^{R}(\hat{\rho})\right],
\label{eq:work_R}
\end{equation}
see appendix~\ref{app:single_qb} for the detailed derivation. The MHQ $\mathrm{Re}q_{\downarrow\uparrow}^R(\hat{\rho})$ is crucial and its negativity signals the presence of anomalous processes $\ket{\downarrow}\rightarrow\ket{\uparrow}$ and enhances the work extraction. 

The extractable work can be split in population and coherent parts: $\mathcal{W}_R[\hat{\rho}] = \mathcal{W}_R[\Delta(\hat{\rho})]+\mathcal{W}_R[\hat{\chi}]$. Since $\Delta(\hat{\rho})$ of a qubit is always thermal, the population term $\mathcal{W}_R[\Delta(\hat{\rho})]$ is always negative. Thus, the possibility of having positive extractable work relies entirely on the coherent part of the work (see appendix~\ref{app:single_qb})
\begin{equation}
    \mathcal{W}_R[\hat{\chi}] = -4E\:\mathrm{Re}q_{\downarrow\uparrow}^R(\hat{\chi})
\end{equation}
that is, it depends on the coherent part of the MHQ associated to process $\ket{\downarrow} \rightarrow \ket{\uparrow}$. To recap, negative $\mathrm{Re}q_{\downarrow\uparrow}^R(\hat{\chi})$ is essential for achieving positive $\mathcal{W}_R[\hat{\chi}]$, which can counteract the detrimental negative $\mathcal{W}_R[\Delta(\hat{\rho})]$ and ultimately exceed it, leading to a overall positive extracted work.

Let us now move to elementary single-qubit gates. We recall that any unitary transformation $\hat{\mathcal{U}}$ can be expressed as a sequence of elementary quantum gates chosen from a universal gate set. In particular, we consider the approximately universal set of single-qubit quantum gates given by $\{ \hat{H},\hat{T}\}$, see \cite{nielsen2012quantum}, with $\hat{H}=\frac{1}{\sqrt{2}}\left(\hat{X}+\hat{Z}\right)$ being the Hadamard gate and $\hat{T}=\ket{\uparrow}\bra{\uparrow}+e^{i\pi/4}\ket{\downarrow}\bra{\downarrow}$ being the $\pi/8$ gate.

We first observe that the gate $\hat{T}$ has the structure $\hat{P}_{\phi}=\ket{\uparrow}\bra{\uparrow}+ e^{i\phi}\ket{\downarrow}\bra{\downarrow}$, so that it commutes with both the projectors $\hat{\Pi}_{\uparrow}$ and $\hat{\Pi}_{\downarrow}$; therefore, its associated KDQs are trivial:
\begin{equation}
q_{\downarrow\uparrow}^{P_{\phi}}(\hat{\rho})=q_{\uparrow\downarrow}^{P_{\phi}}(\hat{\rho})=0, 
    \quad
    q_{\downarrow\downarrow}^{P_{\phi}}(\hat{\rho})=1-p, \quad 
    q_{\uparrow\uparrow}^{P_{\phi}}(\hat{\rho})=p.
\end{equation}
On the other hand, the KDQs associated to the Hadamard gate are (see appendix~\ref{app:explicit_Hadamard-like})
\begin{align}
&q^H_{\downarrow\downarrow}(\hat{\rho})=\frac{(1-p)-\gamma^*}{2},
\quad
q^H_{\downarrow\uparrow}(\hat{\rho})=\frac{(1-p)+\gamma^*}{2},\nonumber\\
&q^H_{\uparrow\downarrow}(\hat{\rho})=\dfrac{p-\gamma}{2},
\quad
q^H_{\uparrow\uparrow}(\hat{\rho})=\dfrac{p+\gamma}{2}.
\quad
\end{align}
Thus, using Eq.~\eqref{eq:work_R}, we can compute the work extractable through an Hadamard gate, reading
\begin{align}
    \mathcal{W}_H [\hat{\rho}]=\underbrace{E(2p-1)}_{\mathcal{W}_H[\Delta(\hat{\rho})]} + \underbrace{(-2|\gamma|E\cos\varphi)}_{\mathcal{W}_H[\hat{\chi}]}. 
\end{align}
We observe that $\mathrm{Re}q_{\downarrow\uparrow}^H(\hat{\chi})<0$ for $\cos\varphi <0$. Among these values of the phases, only those with $|\cos\varphi|>\frac{1-2p}{2|\gamma|}$ produce positive extractable work. On the other hand, the condition $\mathrm{Re}q_{\downarrow\uparrow}^H(\hat{\rho})<0$ can be obtained for $|\gamma||\cos\varphi | >1-p$ (with $\cos\varphi < 0$). At the same time, $|\gamma|\leq \sqrt{p(1-p)}$, which implies $\sqrt{p(1-p)}|\cos\varphi|>1-p$ that cannot be satisfied in the absence of population inversion ($p\leq \frac{1}{2}$). Thus, $\mathrm{Re}q_{\downarrow\uparrow}^H(\hat{\rho})\geq 0$ and so the processes $\ket{\downarrow}\rightarrow\ket{\uparrow}$ do not anomalously contribute to the work extraction. The coherence can however suppress the effectiveness of the process making the extractable work positive. An example case is given by the choice $(p,|\gamma|,\varphi)=(\frac{1}{2},\frac{1}{2},\pi)$ for which the extractable work is positive, $\mathcal{W}_H[\hat{\rho}] = E$, and the nonzero KDQs are $q_{\downarrow\downarrow}^H(\hat{\rho})=q_{\uparrow\downarrow}^H(\hat{\rho})=\frac{1}{2}$. The injection of coherences do not lead to negative values of the MHQs of interest, yet they are responsible for violating the thermal bound $\mathcal{W}\leq 0$ by suppressing the detrimental processes $\ket{\downarrow}\rightarrow\ket{\uparrow}$. 


\subsection{KDQs and work extraction in the Hadamard-like time evolution}
We now consider the example of a quantum time evolution that, at a specific time, realizes the Hadamard gate. It is instructive to explore the thermodynamic behavior during the system’s dynamics, as it offers insight into various nonclassical features that may emerge. The Hadamard gate $\hat{H}=\frac{1}{\sqrt{2}}\left(\hat{X}+\hat{Z}\right)$ can be realized by the evolution of a quantum system under the Hamiltonian $\hat{\mathcal{H}}_h=h\left(\hat{X}+\hat{Z}\right)$ (with $h>0$) until time $t_H =\frac{\pi}{2\sqrt{2}h}$. Indeed, at time $t_H$, the latter unitary becomes 
\begin{equation}
\hat{\mathcal{U}}_{h}(t_H)\coloneqq\hat{\mathcal{U}}_H=-i\hat{H}, 
\end{equation}
and reproduces the Hadamard gate up to a global phase that does not affect the quasiprobabilities and, in general, the time evolution of the system.

At $t\neq t_H$, it is possible that the MHQ $\mathrm{Re} q_{\downarrow\uparrow}(\hat{\rho})$ becomes negative, meaning that anomalous processes come into play. To show that, we first consider as input the pure state with $p=\frac{1}{2}$, i.e. an equatorial state on the Bloch sphere. It is the relative phase between $\ket{\uparrow}$ and $\ket{\downarrow}$ that controls the work extraction. In Fig.~\ref{fig:hadamard_features}(a) we report the dynamics of $\mathrm{Re}q_{\downarrow\uparrow}^{\mathcal{U}_h(t)}(\hat{\rho})$ and extractable work for different values of the relative phase $\varphi$ with fixed $p=\frac{1}{2}$. We observe the presence of anomalous work extraction processes which is marked by the presence of negative $\mathrm{Re}q_{\downarrow\uparrow}^{\mathcal{U}_h(t)}(\hat{\rho})$. All the KDQs $\mathrm{Re}q_{\downarrow\uparrow}^{\mathcal{U}_h(t)}(\hat{\rho})$ become positive at $\omega t_H=\frac{\pi}{2}$, when the Hadamard gate is realized; however, anomalous processes with negative MHQ are found in the \textit{pre-Hadamard} time window $\omega t \in (0,\pi/2)$, for any value of the phase except $\varphi=0$ and $\varphi=\pi$. Thus, a complex relative factor between the superposed states is required to get anomalous positive contribution to the extractable work; the larger nonclassical work extraction contribution is obtained in the neighborood of $\varphi=\pi/2$. In the \textit{post-Hadamard} time window, all of the $\mathrm{Re}q_{\downarrow\uparrow}^{\mathcal{U}_h(t)}(\hat{\rho})$ remain positive, excluding the presence of anomalous processes contributing positively to work extraction. However, it is still possible to have negative $\mathrm{Re}q_{\uparrow\downarrow}^{\mathcal{U}_h(t)}(\hat{\rho})=-\mathrm{Re}q_{\downarrow\uparrow}^{\mathcal{U}_h(t)}(\hat{\rho}) +\frac{1}{2}\sin^2(\omega t)$, which can lead to the presence of detrimental anomalous processes that negatively contribute to the extractable work. 

After having analyzed the KDQs, the question that arises is: what is the effect of these anomalous processes on the total extracted work? We first observe that, for $p=\frac{1}{2}$, the total contribution to the work is only given by the coherent part of the state, see Eq.~\eqref{eq:W(t)}. Coherences are necessary to get nonzero extractable work. In the pre-Hadamard time window, the phases in the neighborood of $\varphi=\pi/2$ are advantageous for extracting work, with the latter growing basically linearly; while, in the post-Hadamard time window, phases in the neighborood of $\varphi=\pi$ are advantageous with a decay that is smooth and nonlinear. The anomalous processes ($\mathrm{Re}q_{\downarrow\uparrow}^{\mathcal{U}_h(t)}(\hat{\rho})>0$) present in the pre-Hadamard time window enhance the work extraction at short times (giving rise to a linear growth). 

\begin{figure}[!t]
\centering
\includegraphics[width=\linewidth]{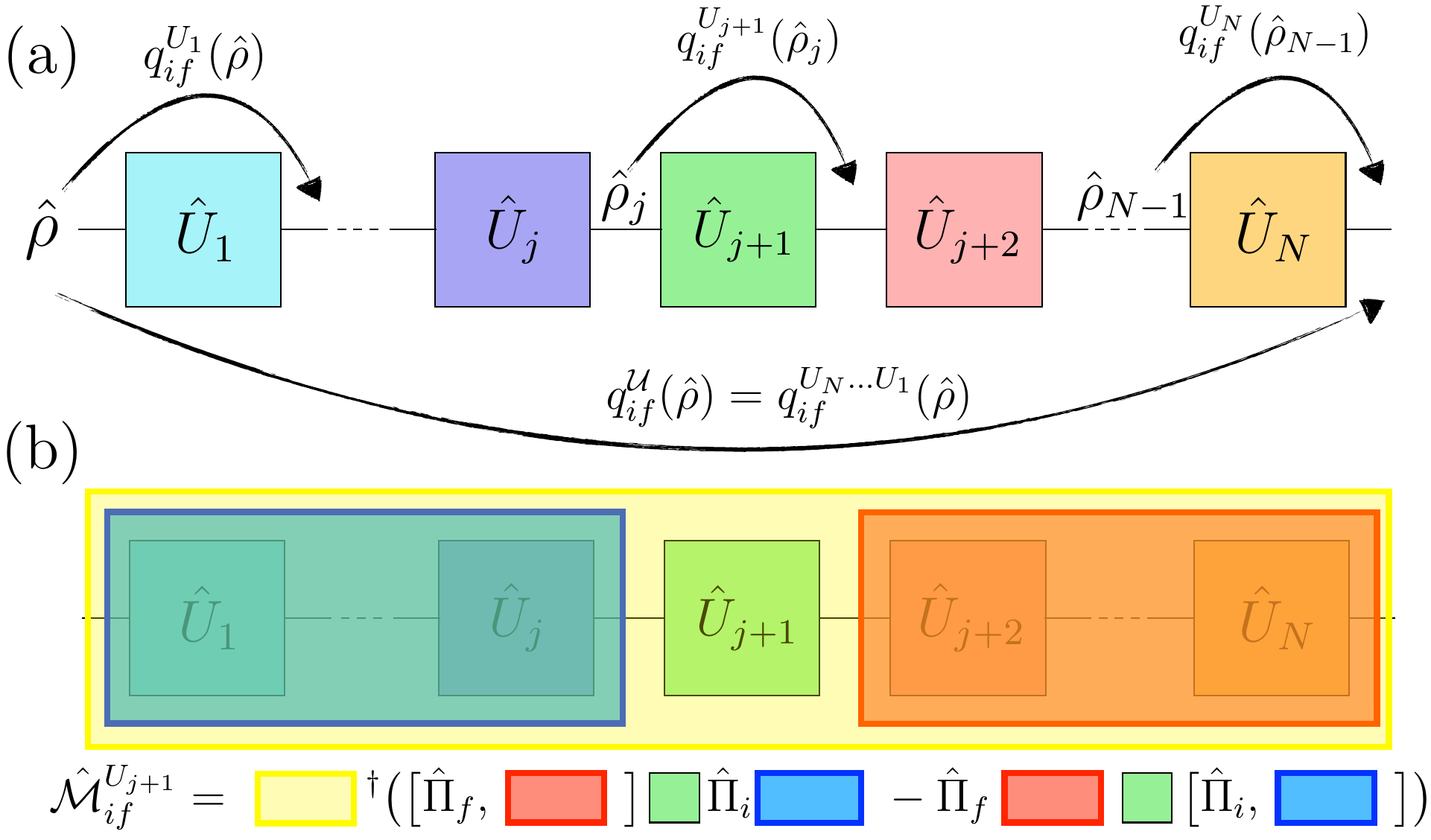}
\caption{Panel (a): Sketch of the KDQs of a unitary transformation decomposed in multiple gates, with a pictorial representation of the full KDQ $q_{if}^{U_N \ldots U_1}(\hat{\rho})$ and the three possible KDQs associated to the constituent gates: (i) $q_{if}^{U_{j+1}}(\hat{\rho}_j)$ with $\hat{U}_{j+1}$ not in the border of $\mathcal{U}$, (ii) $q_{if}^{U_1}(\hat{\rho})$, (iii) $q_{if}^{U_{N}}(\hat{\rho}_{N-1})$. Panel (b): pictorial representation of the structure of the operator $\hat{\mathcal{M}}_{if}^{U_{j+1}}$ whose derivation is given in Eq.~\eqref{eq:M_operator_main}. The yellow block represents the operator $\hat{U}_N \ldots \hat{U}_1$, the blue block represents the operator $\hat{U}_j \ldots \hat{U}_1$, The red block represents the operator $\hat{U}_{N} \ldots \hat{U}_{j+2}$ and the green block is the constituent gate $\hat{U}_{j+1}$.}
\label{fig:nonclassical_sketch}
\end{figure}
In Fig.~\ref{fig:hadamard_features}(b) we perform the same type of analysis by fixing the phase to $\varphi=\pi/2$ and varying the population in the interval $p\in[0,0.5]$. In this case, it is possible that $\mathcal{W}_t[\Delta(\hat{\rho})]< 0$, so that positive coherent part of the work is not sufficient to get total positive extractable work. We observe that the MHQ $\mathrm{Re}q^{\mathcal{U}_h(t)}_{\downarrow\uparrow}(\hat{\rho})$ is negative at short times, when the total extractable work is positive and grows. 

We also compare $\mathcal{W}_t[\hat{\chi}]$ with $\mathcal{W}_t[\Delta(\hat{\rho})]=\mathcal{W}_t[\hat{\rho}]-\mathcal{W}_t[\hat{\chi}]$ showing that, in the pre-Hadamard time window, $\mathcal{W}_t[\hat{\chi}]$ dominates on the negative $\mathcal{W}_t[\Delta(\hat{\rho})]$, causing the extractable work to be positive. 
Finally, we observe that a state with population $p=\frac{1}{2}$ is the one that reaches the larger negative values of $\mathrm{Re}q^{\mathcal{U}_h(t)}_{\downarrow\uparrow}(\hat{\rho})$ and larger values of positive extractable work. On the other hand, by sending $p\rightarrow 0$ we approach the $T=0$ thermal case, where the system is in the ground state $\ket{\downarrow}$, and the work is dominated by its large and negative population contribution. 

\section{Decomposition of KDQs of deep circuits}\label{sec:decomposition}
Anomalous processes can impact the work distribution of complex unitary transformations. In the context of single-qubit transformations, while the individual gates from the approximately universal set $\{ \hat{H},\hat{T}\}$ exhibit positive MHQs, certain combinations of these gates may not preserve this property. Consequently, for a quantum circuit represented by a unitary transformation $\hat{\mathcal{U}}$, the relationship between its quasiprobabilities and those of the individual gates used in its decomposition is expected to be nontrivial. Uncovering such relationship could potentially simplify the characterization of quasiprobabilities in complex or deep circuits by revealing underlying patterns that govern their structure.

In the following, we consider a generic quantum circuit of depth $N$:  $\hat{\mathcal{U}}=\hat{U}_N\ldots\hat{U}_1$, with the constituent gates $\hat{U}_j$ belonging to an approximately universal set. We are interested in finding a relation between the KDQs associated to the full deep quantum circuit $\hat{\mathcal{U}}$ and those of the constituent gates. 
The latter can be located in three different positions: inside the circuit [case (i)] or at the borders, meaning that it is the first gate $\hat{U}_1$ [case (ii)] or the last one, $\hat{U}_N$ [case (iii)]. The three cases are pictorially depicted in Fig.~\ref{fig:nonclassical_sketch}(a). In each of the three cases, the difference between the full KDQ and those associated to the constituent gates $\hat{U}_{j+1}$ is
\begin{widetext}
    \begin{align}
&q_{if}^{\mathcal{U}}(\hat{\rho}) - q_{if}^{U_{j+1}}(\hat{\rho}_j) = \operatorname{Tr} \bigl[ \hat{\mathcal{M}}_{if}^{U_{j+1}} \hat{\rho} \bigr],
\label{eq:discrepancy_KDQ}\\
&\hat{\mathcal{M}}_{if}^{U_{j+1}}=
    \begin{cases}
        \bigl(\hat{U}_N \ldots\hat{U}_1\bigr)^{\dagger} \bigl[ \hat{\Pi}_f, \hat{U}_N \ldots\hat{U}_2 \bigr]\hat{U}_1\hat{\Pi}_i  \; \; \;\; \textrm{if} \;\; j=0,\\
        \bigl(\hat{U}_{N}\ldots\hat{U}_1\bigr)^{\dagger} \bigl(\bigl[ \hat{\Pi}_f,\hat{U}_{N}\ldots\hat{U}_{j+2} \bigr]\hat{U}_{j+1}\hat{\Pi}_i \hat{U}_{j}\ldots\hat{U}_1 - 
    \hat{\Pi}_f\hat{U}_{N}\ldots\hat{U}_{j+1}\bigl[\hat{\Pi}_i, \hat{U}_{j}\ldots\hat{U}_1 \bigr]
    \bigr) \; \; \;\; \textrm{if} \;\; j=1,\ldots,N-2,\\
    -\bigl(\hat{U}_{N}\ldots\hat{U}_1\bigr)^{\dagger}\hat{\Pi}_f \hat{U}_N \bigl[ \hat{\Pi}_i, \hat{U}_{N-1}\ldots\hat{U}_1\bigr] \; \; \;\; \textrm{if} \;\; j=N-1,
    \end{cases}
    \label{eq:M_operator_main}
\end{align}
\end{widetext}
with $j=0,\ldots,N-1$, $\hat{\rho}_j \coloneqq (\hat{U}_j \ldots \hat{U}_1)\hat{\rho}(\hat{U}_j \ldots \hat{U}_1)^{\dagger}$ and $\hat{\rho}_{j=0}\coloneqq\hat{\rho}$. The operator $\hat{\mathcal{M}}_{if}^{U_{j+1}}$ depends on the incompatibility between the initial Hamiltonian projectors and the gates that precede $\hat{U}_{j+1}$ together with the incompatibility between the final Hamiltonian projectors and the gates that follow $\hat{U}_{j+1}$. Its structure is pictorially depicted in Fig.~\ref{fig:nonclassical_sketch}(b) and its detailed derivation is reported in appendix~\ref{app:decomposition}. If the reduced circuit preceding and following the constituent gate in question commute with the initial and final projectors, $\hat{\mathcal{M}}_{if}^{U_{j+1}}$ is zero and thus the KDQ reduces to that of the constituent gate. Otherwise, the KDQ of the full circuit gets modified.

From Eq.~\eqref{eq:discrepancy_KDQ}, it follows that 
\begin{equation}
    q_{if}^{\mathcal{U}}(\hat{\rho})=\frac{1}{N}\sum_{j=0}^{N-1} q_{if}^{U_{j+1}}(\hat{\rho}_j)+\frac{1}{N}\mathcal{Q}_{if}^{\mathcal{U}}(\hat{\rho}),
    \label{eq:full_correction}
\end{equation}
with $\mathcal{Q}_{if}^{\mathcal{U}}(\hat{\rho})\coloneqq\frac{1}{N}\sum_{j=0}^{N-1} \operatorname{Tr}\Bigl[\hat{\mathcal{M}}_{if}^{U_{j+1}} \hat{\rho} \Bigr]$. The full KDQ is the weighted sum of the KDQs associated to all the constituent gates that decompose $\hat{\mathcal{U}}$ corrected by $\mathcal{Q}_{if}^{\mathcal{U}}(\hat{\rho})$ that depends on the incompatibility between Hamiltonian projectors and the various parts of the circuit.

\subsection{Decomposition of KDQs of single-qubit circuits and minimality of the HTH circuit}
\begin{figure}[!t]
\centering
\includegraphics[width=\linewidth]{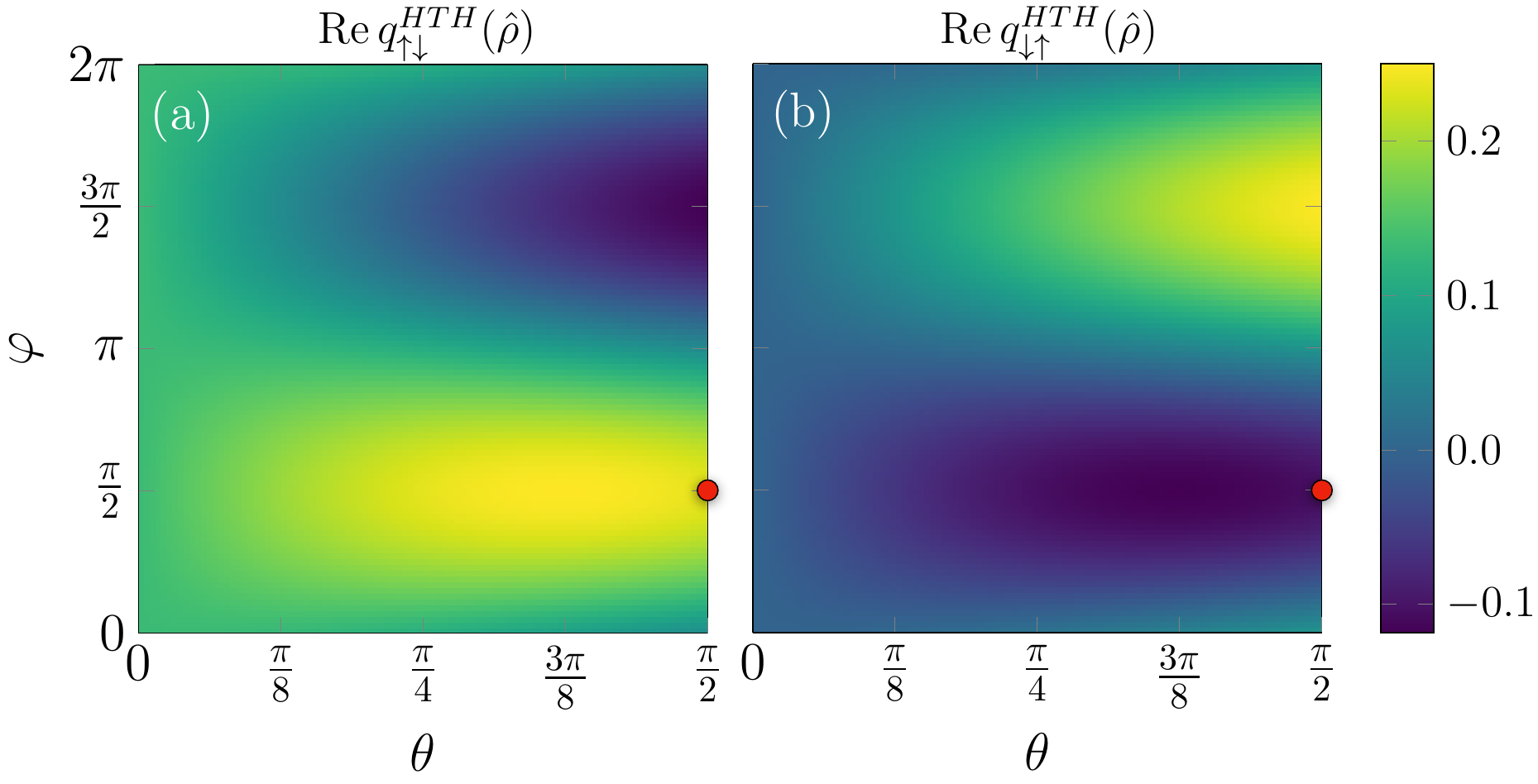}
\caption{MHQs associated to the $\hat{H}\hat{T}\hat{H}$ circuit that contribute to the work extraction for an input state $\hat{\rho}(\theta,\varphi)=\ket{\psi(\theta,\varphi)}\bra{\psi(\theta,\varphi)}$ with $\ket{\psi(\theta,\varphi)}=\cos(\theta/2)\ket{\downarrow}+e^{i\varphi}\sin(\theta/2)\ket{\uparrow}$. The two panels report the MHQs associated to the two different processes $\ket{\uparrow}\rightarrow\ket{\downarrow}$ (panel (a)) and $\ket{\downarrow}\rightarrow\ket{\uparrow}$ (panel (b)). The red dot indicates the values of $(\theta,\varphi)$ for which an in-depth analysis is performed in the main text.}
\label{fig:HTH}
\end{figure}
In this section, we provide an operational example of decomposition of KDQs, by showing that, given the approximately universal set $\{ \hat{H},\hat{T}\}$, the circuit $\hat{H}\hat{T}\hat{H}$ is the minimal circuit in which the full KDQs do not coincide with those of the constituent gates. 

Since the gate $\hat{T}$ commutes with the initial/final Hamiltonian projectors, it simplifies the structure of the KDQs associated with the full circuit. 
Let us first consider circuits with depth $N=2$ written as $\hat{\mathcal{U}}=\hat{V}\hat{U}$ for which Eqs.~\eqref{eq:-V} and~\eqref{eq:-U} hold. The KDQs of the full circuit correspond to those of its constituent gates if at least one of the commutators $[\hat{\Pi}_i,\hat{U}]$ or $[\hat{\Pi}_f,\hat{V}]$ is zero. The possible circuits are $\hat{H}\hat{H}=\hat{\mathds{1}}$, $\hat{H}\hat{T}$, $\hat{T}\hat{H}$ and $\hat{T}\hat{T}$ for which the KDQs are either trivial ($\hat{H}\hat{H}$) or they correspond to that of the Hadamard gate. 

Thus, all of the single-qubit circuits with depth $N=2$ made up of gates belonging to the aforementioned set, have full KDQs that correspond to those of the Hadamard gate.

We then move to circuits with depth $N=3$, described by $\hat{\mathcal{U}}=\hat{M}\hat{V}\hat{U}$ for which Eqs. \eqref{eq:MVU-V},~\eqref{eq:MVU-U} and~\eqref{eq:MVU-M} hold. In this case, it is sufficient that 
\begin{align}
&[\hat{\Pi}_f,\hat{M
}\hat{V}]=0 \,\, \textrm{or} \,\,
[\hat{\Pi}_f,\hat{M}]=[\hat{\Pi}_i,\hat{U}]=0 \,\, \textrm{or} \,\,
[\hat{\Pi}_i,\hat{V}\hat{U}]=0
\label{eq:commutation_decomposition}
\end{align}
to guarantee that the total KDQs can be trivially written as the KDQs of one of the constituent gates. 

The possible circuits with depth $N=3$ made up of the gates $\hat{H}$ and $\hat{T}\}$ are $\hat{H}\hat{H}\hat{T}$, $\hat{T}\hat{H}\hat{H}$, $\hat{H}\hat{T}\hat{H}$, $\hat{T}\hat{T}\hat{H}$, $\hat{H}\hat{T}\hat{T}$, $\hat{T}\hat{H}\hat{T}$, $\hat{H}\hat{H}\hat{H}$ and $\hat{T}\hat{T}\hat{T}$. For any of these circuits we evaluate the commutators in Eq.~\eqref{eq:commutation_decomposition} that are widely simplified by using the identities $[\hat{\Pi}_k,\hat{T}]=[\hat{\Pi}_k,\hat{T}\hat{T}]=0$, that hold for any $k$. The results are reported in table~\ref{tab:commutation-table}.

\renewcommand{\arraystretch}{1.35}  
\begin{table}[ht]
\centering
\begin{tabular}{ |c|c|c|c| } 
\hline
$\hat{M}\hat{V}\hat{U}$ & $[\hat{\Pi}_f,\hat{M}\hat{V}]$ & $[\hat{\Pi}_f,\hat{M}]; \,[\hat{\Pi}_i,\hat{U}]$ & $[\hat{\Pi}_i,\hat{V}\hat{U}]$\\
\hline
$\hat{H}\hat{H}\hat{T}$ & $=0$ & $\neq 0; \, =0$  & $\neq 0$ \\ 
$\hat{T}\hat{H}\hat{H}$ & $\neq 0$ & $=0; \, \neq 0$  & $=0$ \\ 
$\hat{H}\hat{T}\hat{H}$ & $\neq 0$ & $\neq 0; \, \neq 0$  & $\neq 0$ \\ 
$\hat{T}\hat{T}\hat{H}$ & $=0$ & $=0; \, \neq 0$  & $\neq 0$ \\ 
$\hat{H}\hat{T}\hat{T}$ & $\neq 0$ & $\neq 0; \, =0$  & $=0$ \\ 
$\hat{T}\hat{H}\hat{T}$ & $\neq 0$ & $=0; \, =0$  & $\neq 0$ \\ 
$\hat{H}\hat{H}\hat{H}$ & $=0$ & $\neq 0; \, \neq 0$  & $=0$ \\ 
$\hat{T}\hat{T}\hat{T}$ & $=0$ & $=0; \, =0$  & $=0$ \\ 
\hline
\end{tabular}
\caption{Commutation relations of Eq.~\eqref{eq:commutation_decomposition} for different circuits with depth $N=3$.}
\label{tab:commutation-table}
\end{table}
Out of these eight possible transformations, only $\hat{H}\hat{T}\hat{H}$ does not satisfy any of the three commutation conditions of Eq.~\eqref{eq:commutation_decomposition} and so its full KDQs cannot be trivially reduced to that of one of the constituent gates. For this reason, we identify it as the minimal circuit for which the full KDQs do not simply reduce to those of the constituent gates. A possible consequence of the latter observation is that the work distribution of the full circuit $\hat{H}\hat{T}\hat{H}$ can be potentially characterized by anomalous processes with $\mathrm{Re}q_{\downarrow\uparrow}^{HTH}<0$ or $\mathrm{Re}q_{\uparrow\downarrow}^{HTH}<0$, even though such anomalies are absent in the constituent gates.

To illustrate this explicitly, we consider a system initialized in the pure input state $\ket{\psi(\theta,\varphi)}=\cos(\theta/2)\ket{\downarrow} + e^{i\varphi}\sin(\theta/2)\ket{\uparrow}$, where $\theta$ is the mixing angle and $\varphi$ the relative phase. The mixing angle and the population are related through the relation $\cos(\theta/2)=\sqrt{1-p}$ and $\sin(\theta/2)=\sqrt{p}$, $\theta \leq \pi/2$ guarantees no population inversion. While the state evolves under the transformation described by the circuit $\hat{\mathcal{U}} = \hat{H} \hat{T} \hat{H}$, we are interested in analyzing the behavior of the MHQs that contribute to the work, shown in Fig.~\ref{fig:HTH}. 

We find that, for both MHQs of interest, there are significant regions in the parameter space where the work distribution is influenced by nonclassical anomalous processes. Such nonclassicality is advantageous for work extraction when the MHQ associated with the transition $\ket{\uparrow} \rightarrow \ket{\downarrow}$ becomes negative, while that associated with the reverse process $\ket{\downarrow} \rightarrow \ket{\uparrow}$ remain positive. For this reason, we analyze in detail the decomposition of the KDQs at the coordinates $(\theta = \pi/2, \varphi = \pi/2)$ (indicated by a red dot in the figure), where $\mathrm{Re}\, q_{\uparrow\downarrow}^{HTH}(\hat{\rho})$ is largely positive and $\mathrm{Re}\, q_{\downarrow\uparrow}^{HTH}(\hat{\rho})$ is largely negative. Moreover, as will become evident shortly, these specific values of $(\theta, \varphi)$ exemplify a scenario in which the full circuit exhibits anomalous processes, whereas the contribution to work extraction from the single constituent gates does not.

The input state is $\hat{\rho} = \ket{\psi}\bra{\psi}$, where $\ket{\psi} = \ket{\psi(\pi/2,\pi/2)} = \frac{1}{\sqrt{2}}(\ket{\uparrow} + i\ket{\downarrow})$. The first gate $\hat{H}$ transforms $\hat{\rho}\rightarrow\hat{\rho}_H=\hat{H}\hat{\rho}\hat{H}^{\dagger}$, with $\hat{\rho}_H=\ket{\psi_H}\bra{\psi_H}$ and $\ket{\psi_H}=\frac{1}{\sqrt{2}}(\ket{\uparrow}-i\ket{\downarrow})$; the KDQ associated to this gate and to the process $\ket{\downarrow}\rightarrow\ket{\uparrow}$ is $\mathrm{Re}q_{\downarrow\uparrow}^H(\hat{\rho})=\frac{1}{4}$. At the same time, we have $\mathrm{Re}q_{\uparrow\downarrow}^H(\hat{\rho})=\frac{1}{4}$, and so the extractable work is identically zero. Subsequently, the gate $\hat{T}$ transforms $\hat{\rho}_H\rightarrow\hat{\rho}_{TH}=\hat{T}\hat{H}\hat{\rho}\bigl(\hat{T}\hat{H}\bigr)^{\dagger}$, with $\hat{\rho}_{TH}=\ket{\psi_{TH}}\bra{\psi_{TH}}$ and $\ket{\psi_{TH}}=\frac{1}{\sqrt{2}}(\ket{\uparrow}+ e^{-i\pi/4}\ket{\downarrow})$. Since the transformation is given by the $\pi/8$ gate, we have $q_{\downarrow\uparrow}^T(\hat{\rho}_{H})=q_{\uparrow\downarrow}^T(\hat{\rho}_{H})=0$. Finally, the application of the second $\hat{H}$ gate transforms the state according to $\hat{\rho}_{TH}\rightarrow\hat{\rho}_{HTH}=\hat{H}\hat{T}\hat{H}\hat{\rho}\bigl(\hat{H}\hat{T}\hat{H}\bigr)^{\dagger}$, with $\hat{\rho}_{HTH}=\ket{\psi_{HTH}}\bra{\psi_{HTH}}$ and $\ket{\psi_{HTH}}=\frac{1}{2}\bigl[(1+e^{-i\pi/4})\ket{\uparrow}+ (1-e^{-i\pi/4})\ket{\downarrow}\bigr]$. The two KDQs of interest are $\mathrm{Re}q_{\downarrow\uparrow}^H(\hat{\rho}_{TH})=\frac{1}{4}\left(\frac{2+\sqrt{2}}{2}\right)$ and $\mathrm{Re}q_{\uparrow\downarrow}^H(\hat{\rho}_{TH})=\frac{1}{4}\left(\frac{2-\sqrt{2}}{2}\right)$, both of which are positive, indicating the absence of anomalous processes.

Let us now consider the full transformation, represented by 
\begin{equation}
    \hat{H}\hat{T}\hat{H}=\frac{1}{2}
    \begin{bmatrix}
        1 + e^{i\pi/4} & 1 - e^{i\pi/4} \\
        1 - e^{i\pi/4} & 1 + e^{i\pi/4}
    \end{bmatrix}.
\end{equation}
The straightforward computation of the KDQs leads to
\begin{align}
&\mathrm{Re}q_{\downarrow\uparrow}^{HTH}(\hat{\rho})=\frac{1-\sqrt{2}}{4}<0, \\& \mathrm{Re}q_{\uparrow\downarrow}^{HTH}(\hat{\rho})=\frac{1}{4}>0,
\end{align}
which means that, differently from the constituent gates, the extractable work receives anomalous positive contributions from the process $\ket{\downarrow}\rightarrow\ket{\uparrow}$.

\section{KDQs of two-qubit gates}\label{sec:two_qbits}
As mentioned above, any unitary transformation $\hat{\mathcal{U}}$ can be expressed as a sequence of elementary quantum gates chosen from a universal gate set. We have also mentioned that the set $\{\hat{H},\hat{T}\}$ is an approximately universal set of single-qubit quantum gates~\cite{nielsen2012quantum}; furthermore, the set $\{ \hat{H},\hat{T},\hat{U}_{\rm CNOT}\}$ is approximately universal for any unitary transformation acting on $L$-qubit states~\cite{nielsen2012quantum, dawson2005solovay}. The CNOT gate is the two-qubit gate that flips (leaves unchanged) the second qubit depending on whether the first qubit is in state $\ket{\downarrow}$ ($\ket{\uparrow}$).

In a two-qubit system the initial/final Hamiltonian is $\hat{\mathcal{H}}=E\bigl(\hat{Z}\otimes\hat{\mathds{1}} + \hat{\mathds{1}}\otimes \hat{Z} \bigr)$ and so the projectors are $\hat{\Pi}_{\uparrow\uparrow}=\ket{\uparrow\uparrow}\bra{\uparrow\uparrow}$, $\hat{\Pi}_{\uparrow\downarrow}=\ket{\uparrow\downarrow}\bra{\uparrow\downarrow}$, $\hat{\Pi}_{\downarrow\uparrow}=\ket{\downarrow\uparrow}\bra{\downarrow\uparrow}$ and $\hat{\Pi}_{\downarrow\downarrow}=\ket{\downarrow\downarrow}\bra{\downarrow\downarrow}$. To simplify matters, we rename the two-qubit eigenbasis as $\{\ket{\downarrow\downarrow},\ket{\downarrow\uparrow},\ket{\uparrow\downarrow},\ket{\uparrow\uparrow} \}=\{\ket{E_0},\ket{E_1}, \ket{E_2}, \ket{E_3} \}$; the associated eigenergies are $\{E_0 = -2E, E_1 = 0, E_2 = 0, E_3=2E\}$. Thus, the extractable work through a generic two-qubit unitary cyclic transformation $\hat{\mathcal{U}}$, is 
\begin{align}
    \mathcal{W}_\mathcal{U}[\hat{\rho}]=&\sum_{i\neq f}q^\mathcal{U}_{if} (E_i - E_f)=\nonumber\\&2E \bigl[(q_{10}^\mathcal{U}-q_{01}^\mathcal{U}) + (q_{20}^\mathcal{U}-q_{02}^\mathcal{U}) + 2(q_{30}^\mathcal{U}-q_{03}^\mathcal{U}) +\nonumber\\& (q_{31}^\mathcal{U}-q_{13}^\mathcal{U}) + (q_{32}^\mathcal{U}-q_{23}^\mathcal{U})\bigr].
        \label{eq:work_2qbt}
\end{align}
For brevity, we have omitted the explicit dependence of $q_{if}^{\mathcal{U}}$ on $\hat{\rho}$, that is now a two-qubit state. We observe that 10 out of the 16 total KDQs contribute to the work extraction.

Our interest is to explore the features of the KDQs for two-qubit gates that can decompose a generic two-qubit circuit $\hat{\mathcal{U}}$. The latter can be decomposed in gates belonging to the set $\{\hat{H},\hat{T},\hat{U}_{\rm CNOT}\}$, thus each constituent gate is either CNOT or the tensor product of local gates $\hat{U}\otimes \hat{V}$, in which $\hat{U}$ and $\hat{V}$ are built as sequences of gates belonging to the approximately universal single-qubit set $G_1 = \{\hat{H},\hat{T}\}$. In the appendix~\ref{app:twoqb_gates},  we present a comprehensive analysis of the problem, the main results of which are summarized in the following.

\begin{figure}[!t]
\centering
\includegraphics[width=\linewidth]{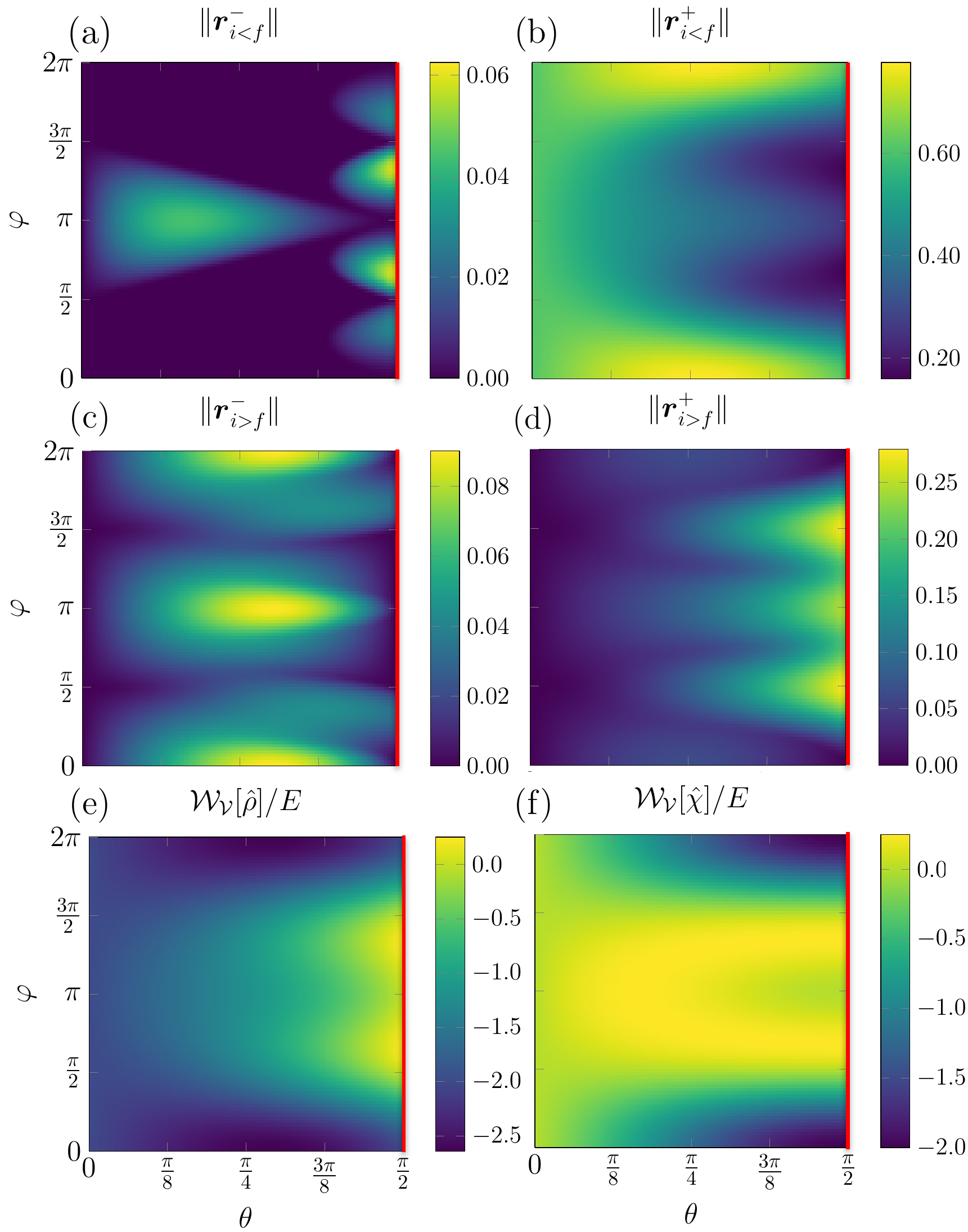}
\caption{Thermodynamic features of the circuit $\hat{\mathcal{V}}=\hat{U}_{\rm CNOT}\hat{H}^{\otimes 2}$ for an input state $\ket{\Psi(\theta,\varphi)}=\ket{\psi(\theta,\varphi)}^{\otimes 2}$, with $\ket{\psi(\theta,\varphi)}=\cos(\theta/2)\ket{\downarrow} + e^{i\varphi}\sin(\theta/2)\ket{\uparrow}$ for different values of $\theta$ and $\varphi$ parameters. Panels (a),(b).(c) and (d) report the norms in Eq.~\eqref{eq:norms} that are defined from the vectors~\eqref{eq:r_i<f} and~\eqref{eq:r_i>f}. Panel (e) reports the total extractable work in the same range of input parameters and panel (f) reports the coherent part of the work. The red line indicates the values of $(\theta,\varphi)$ for which an in-depth analysis is performed in the text.}
\label{fig:CNOTHH_parameters}
\end{figure}
The application of CNOT always produces real and positive (classical) KDQs, for any input state. The application of local operations $\hat{U}\otimes \hat{V}$ on factorized states $\hat{\rho}=\hat{\sigma}\otimes\hat{\tau}$ produces factorized KDQs
\begin{equation}
  q_{(\alpha_i\beta_i)(\alpha_f\beta_f)}^{U\otimes V}(\hat{\sigma}\otimes\hat{\tau})=q_{\alpha_i\alpha_f}^{U}(\hat{\sigma})q_{\beta_i\beta_f}^{V}(\hat{\tau}),
  \label{eq:qp_factor_main}
\end{equation}
where the indices $\alpha_i, \, \beta_i$ and $\alpha_f, \, \beta_f$ are uniquely defined by the indices $i$ and $f$ through the table~\ref{tab:alphabeta} reported in appendix~\ref{app:twoqb_gates}. 

Moreover, any tensor product of local gates can be expressed as a sequence including either a Hadamard or a $\hat{P}_{\phi}$ gates: $\hat{H}\otimes \hat{H}$, $\hat{H}\otimes \hat{P}_{\phi}$, $\hat{P}_{\phi} \otimes \hat{H}$ and $\hat{P}_{\phi} \otimes \hat{P}_{\phi}$, see appendix~\ref{app:twoqb_gates}. Thus, the computation of the KDQs of complex local transformations $\hat{U}\otimes \hat{V}$ possibly passes through the computation of those of their constituent gates to which they are related through Eq.~\eqref{eq:discrepancy_KDQ}.

We observe that the 10 KDQs of interest are identically zero if both the unitaries applied to the two qubits have the $\hat{P}_{\phi}$ structure, for any input state. On the other hand, the application of transformations of the form $\hat{H}\otimes \hat{P}_{\phi}$ and $\hat{P}_{\phi}\otimes \hat{H}$ to a factorized state produces (see appendix~\ref{app:twoqb_gates})
\begin{align}
   q_{if}^{H\otimes P_{\phi}}(\hat{\sigma}\otimes\hat{\tau})&= q_{(\alpha_i\beta_i)(\alpha_f\beta_f)}^{H\otimes P_{\phi}}(\hat{\sigma}\otimes\hat{\tau})=q_{\alpha_i\alpha_f}^{H}(\hat{\sigma})q_{\beta_i\beta_f}^{P_{\phi}}(\hat{\tau})\nonumber\\&=
  \begin{cases}
      0 \,\,\, \textrm{if} \,\,\, \beta_i \neq \beta_f \\
      p \,q_{\alpha_i\alpha_f}^H(\hat{\sigma}) \,\,\, \textrm{if} \,\,\, \beta_i =\beta_f=\uparrow, \\
      (1-p)\,q_{\alpha_i\alpha_f}^H(\hat{\sigma}) \,\,\, \textrm{if} \,\,\, \beta_i =\beta_f=\downarrow,\\
  \end{cases}
  \label{eq:qp_factor_HP}
\end{align}
\begin{align}
   q_{if}^{P_{\phi}\otimes H}(\hat{\sigma}\otimes\hat{\tau})&= q_{(\alpha_i\beta_i)(\alpha_f\beta_f)}^{P_{\phi}\otimes H}(\hat{\sigma}\otimes\hat{\tau})=q_{\alpha_i\alpha_f}^{P_{\phi}}(\hat{\sigma})q_{\beta_i\beta_f}^{H}(\hat{\tau})\nonumber\\&=
  \begin{cases}
      0 \,\,\, \textrm{if} \,\,\, \alpha_i \neq \alpha_f \\
      p \,q_{\beta_i\beta_f}^H(\hat{\tau}) \,\,\, \textrm{if} \,\,\, \alpha_i =\alpha_f=\uparrow, \\
      (1-p)\,q_{\beta_i\beta_f}^H(\hat{\tau}) \,\,\, \textrm{if} \,\,\, \alpha_i =\alpha_f=\downarrow.\\
  \end{cases}
  \label{eq:qp_factor_PH}
\end{align}   
\begin{figure*}[t!]
\centering
\includegraphics[width=\linewidth]{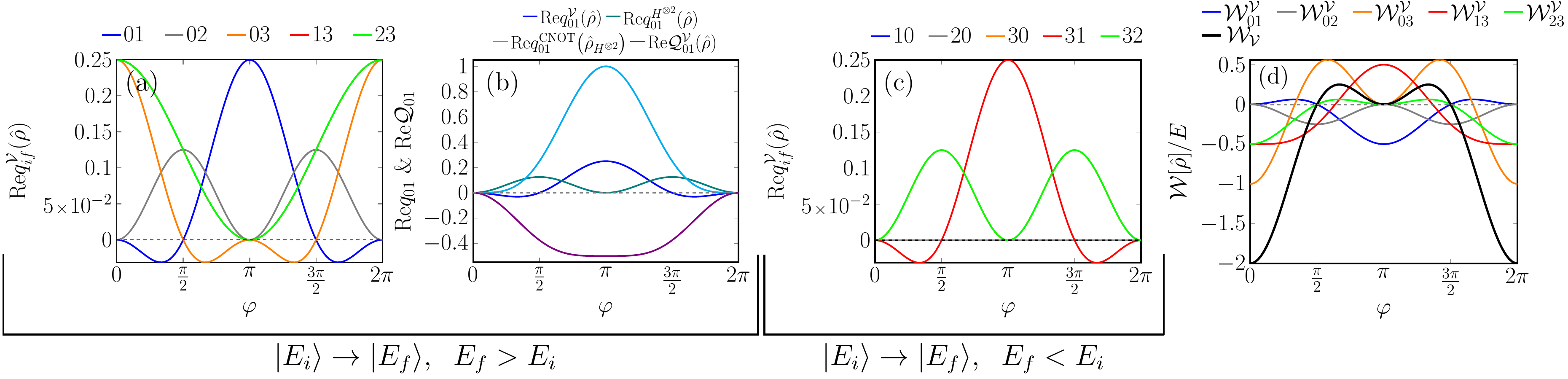}
\caption{Real part of KDQs (or MHQs) and extractable work in a two-qubit circuit made of $\hat{H}^{\otimes 2}$ followed by the CNOT gate. The unitary describing the circuit is $\hat{\mathcal{V}}=\hat{U}_{\rm CNOT}\hat{H}^{\otimes2}$. The input state is $\ket{\Psi}=\ket{\psi}^{\otimes 2}$ with $\ket{\psi}=\frac{1}{\sqrt{2}}(\ket{\uparrow}+e^{i\varphi}\ket{\downarrow})$. Panel (a): MHQs associated to processes $if=01,02,03,13,23$ (different colors, $13$ and $23$ are superposed) under the gate $\hat{H}^{\otimes 2}$ as a function of the phase $\varphi$. Panel (b): MHQs associated to the processes $if=01$ as a function of the phase. Different curves report the MHQs and corrections defined in Eq.~\eqref{eq:_qp_and_correction_} associated to the transformation $\hat{\mathcal{V}}$ as a function of the phase $\varphi$. Panel (c): MHQs as a function of the phase $\varphi$ for the processes $if=10,20,30,31,32$, the MHQ of the processe $if=10$ corresponds to that of the process $if=20$ and the MHQ of $if=30$ corresponds to that of $if=32$. Panel (d): Extractable work and its components as a function of the phase $\varphi$. The black thick curve is the total extractable work, the colored curves are its components defined in Eq.~\eqref{eq:work_components}.} 
\label{fig:qp_CNOTHH}
\end{figure*}
Furthermore, the application of $\hat{H}\otimes\hat{H}$ to a factorized state produces KDQs that can be written as the product of two single-qubit Hadamard KDQs as well; however,
due to the mixing of the imaginary parts of the KDQs associated to the two single-qubit gates with positive MHQs, $\hat{H}\otimes\hat{H}$ can produce more intriguing effects, possibly leading to negative MHQs on the full transformation. Thus, it is possible that the work overall has contributions from anomalous processes, even in the absence of population inversion. 

Conversely, the application of transformations of the form $\hat{H}\otimes \hat{P}_{\phi}$, $\hat{P}_{\phi}\otimes \hat{H}$ and $\hat{H}\otimes\hat{H}$ to an entangled state yields KDQs with a complex structure that cannot be decomposed into a product of single-qubit KDQs. Thus, the presence of entangled input states prevents simple decompositions of the KDQs and requires case-by-case analysis. 

\section{Practical example of a simple two-qubit circuit}\label{sec:example}
In this section, we apply our findings on a simple quantum circuit composed of two gates.
The circuit is defined by the unitary transformation $\hat{\mathcal{V}}$ made of the gate $\hat{H}^{\otimes 2}$ followed by the CNOT gate, thus $\hat{\mathcal{V}}=\hat{U}_{\rm CNOT}\hat{H}^{\otimes 2}$. The extractable work through a generic two-qubit unitary transformation $\hat{\mathcal{U}}$, is given by Eq.~\eqref{eq:work_2qbt}. 

Unlike the single-qubit case, a large number of processes now contribute to work extraction. Therefore, it is useful to introduce quantities that can collectively quantify the overall positive and negative contributions of the MHQs to the extractable work. We introduce the two vectors
\begin{align}
    &\boldsymbol{r}_{i<f}=(\mathrm{Re}\,q_{01}^{\mathcal{V}},\mathrm{Re}\,q_{02}^{\mathcal{V}},2\mathrm{Re}\,q_{03}^{\mathcal{V}},\mathrm{Re}\,q_{13}^{\mathcal{V}},\mathrm{Re}\,q_{23}^{\mathcal{V}}),\label{eq:r_i<f}
    \\&\boldsymbol{r}_{i>f}=(\mathrm{Re}\,q_{10}^{\mathcal{V}},\mathrm{Re}\,q_{20}^{\mathcal{V}},2\mathrm{Re}\,q_{30}^{\mathcal{V}},\mathrm{Re}\,q_{31}^{\mathcal{V}},\mathrm{Re}\,q_{32}^{\mathcal{V}}),\label{eq:r_i>f}
\end{align}
that contain respectively all the MHQs associated to processes with $E_i<E_f$ and $E_i>E_f$. For brevity, we have omitted the dependence on the input state $\hat{\rho}$. We note that processes with $if=03,30$ have double weight because they contribute with a factor $2$ to the work extraction, see Eq.~\eqref{eq:work_2qbt}. Secondly, we introduce the negative parts of the vectors, $\boldsymbol{r}_{i<f}^-$ and $\boldsymbol{r}_{i>f}^-$, which are vectors filled with zeros where the MHQs are positive and with $-\mathrm{Re}\,q_{if}$ where $\mathrm{Re}\,q_{if}<0$. At the same time, we introduce the positive parts of the vectors, $\boldsymbol{r}_{i<f}^+$ and $\boldsymbol{r}_{i>f}^+$, which are vectors filled with zeros where the MHQs are negative and with $\mathrm{Re}\,q_{if}$ where $\mathrm{Re}\,q_{if}>0$. Thus, the euclidean norms
\begin{equation}
    \| \boldsymbol{r}_{i<f}^-\|, \quad
    \| \boldsymbol{r}_{i<f}^+\|,
    \quad
    \| \boldsymbol{r}_{i>f}^-\|, \quad
    \| \boldsymbol{r}_{i>f}^+\|
    \label{eq:norms}
\end{equation}
quantify how much negative and positive MHQs contribute to the work extraction. The presence of non-zero $\| \boldsymbol{r}_{i<f}^-\|$ and $\| \boldsymbol{r}_{i>f}^-\|$ indicates the presence of anomalous processes beneficial and detrimental for work extraction, respectively.

As input state, we consider the most general factorized and symmetric pure state $\ket{\Psi(\theta,\varphi)}=\ket{\psi(\theta,\varphi)}^{\otimes 2}$, with $\ket{\psi(\theta,\varphi)}=\cos(\theta/2)\ket{\downarrow} + e^{i\varphi}\sin(\theta/2)\ket{\uparrow}$. The behavior of the four norms for different values of the input parameters $\theta$ and $\varphi$ is shown in panels (a),(b),(c),(d) of Fig.~\ref{fig:CNOTHH_parameters}, panels (e) and (f), instead, display the total extractable work and its coherent part, respectively. We note that the extractable work is positive in the neighborhood of $\theta = \pi/2$ and for specific values of $\varphi$, which define a region of the parameter space characterized by peaks in $\|\boldsymbol{r}_{i>f}^+\|$ and $\|\boldsymbol{r}_{i<f}^-\|$, which signals the presence of beneficial anomalous processes. Elsewhere, $\|\boldsymbol{r}_{i<f}^+\|$ dominates and the extractable work is basically negative. Conversely, the coherent part of the work is positive in a large region of the parameter space, but not large enough to counterbalance the negative contribution of the population part of the work. As $\theta \to \pi/2$, the population part of the input state approaches the infinite temperature state $\frac{1}{4}\hat{\mathds{1}}$, from which no work can be extracted, as it remains invariant under unitary transformations. Therefore, the extractable work is purely coherent, and within a specific interval of $\varphi$, it becomes positive. For these reasons, we conduct a detailed analysis of the MHQs in the region described above.

More specifically, we fix $\theta = \pi/2$ and vary $\varphi \in [0, 2\pi]$, corresponding to the region marked with a red bar in Fig.~\ref{fig:CNOTHH_parameters}. Thus, the input state is $\ket{\Psi}=\ket{\psi}^{\otimes 2}$, with $\ket{\psi}=\frac{1}{\sqrt{2}}(\ket{\uparrow}+e^{i\varphi}\ket{\downarrow})$. Let us first consider the processes $\ket{E_i}\rightarrow\ket{E_f}$ with $E_i < E_f$, which are described by the KDQs $q_{01}^\mathcal{U}, q_{02}^\mathcal{U}, q_{03}^\mathcal{U}, q_{13}^\mathcal{U}, q_{23}^\mathcal{U}$. We observe that $[\hat{\Pi}_2,\hat{U}_{\rm CNOT}]=[\hat{\Pi}_3,\hat{U}_{\rm CNOT}]=0$ and so, by using Eq.~\eqref{eq:-U}, we obtain
\begin{align}
    &q_{02}^{\mathcal{V}}(\hat{\rho})=q_{02}^{H^{\otimes 2}}(\hat{\rho}), \quad
    q_{03}^{\mathcal{V}}(\hat{\rho})=q_{03}^{H^{\otimes }}(\hat{\rho}), \nonumber\\&
    q_{13}^{\mathcal{V}}(\hat{\rho})=q_{13}^{H^{\otimes 2}}(\hat{\rho}), \quad
    q_{23}^{\mathcal{V}}(\hat{\rho})=q_{23}^{H^{\otimes 2}}(\hat{\rho}),
\end{align}
while, using Eq.~\eqref{eq:_qp_and_correction_}, we have
\begin{equation}
q_{01}^{\mathcal{V}}(\hat{\rho})= \frac{q_{01}^{H^{\otimes 2}}(\hat{\rho})+q_{01}^{\rm CNOT}(\hat{\rho}_{H^{\otimes 2}}) + \mathcal{Q}_{01}^{\mathcal{V}}(\hat{\rho})}{2}, 
\end{equation}
where we have defined $\hat{\rho}_{H^{\otimes 2}}=\hat{H}^{\otimes 2}\hat{\rho}\hat{H}^{\otimes 2}$. Thus, the presence of the CNOT gate simplifies most of the KDQs considered. However, since $[\hat{\Pi}_1, \hat{U}_{\rm CNOT}]\neq 0$ and $[\hat{\Pi}_0, \hat{H}^{\otimes 2}]\neq 0$, the correction $\mathcal{Q}_{01}^{\mathcal{V}}(\hat{\rho})$ plays a central role. In Fig.~\ref{fig:qp_CNOTHH}(a) we report the MHQs associated to the process $if=01$ together with those associated to the processes $if=02,03,13,23$ that can be directly derived from the single qubit Hadamard MHQs. In fact, by using Eq.~\eqref{eq:qp_factor_main}, we obtain
\begin{align}
    \mathrm{Re} q_{if}^{H^{\otimes 2}}(\hat{\rho})=&  q_{(\alpha_i\beta_i)(\alpha_f\beta_f)}^{H^{\otimes 2}}(\ket{\psi}\bra{\psi}\otimes \ket{\psi}\bra{\psi})=\nonumber\\&
\mathrm{Re}q_{\alpha_i \alpha_f}^H(\ket{\psi}\bra{\psi})\:\mathrm{Re}q_{\beta_i \beta_f}^H(\ket{\psi}\bra{\psi})-\nonumber\\& \mathrm{Im}q_{\alpha_i \alpha_f}^H(\ket{\psi}\bra{\psi})\:\mathrm{Im}q_{\beta_i \beta_f}^H(\ket{\psi}\bra{\psi}).
\end{align}
The indces $\alpha_i, \beta_i$ and $\alpha_f, \beta_f$ are uniquely defined by the indeces $i$ and $f$ according to the table~\ref{tab:alphabeta}, reported in appendix~\ref{app:twoqb_gates}. We observe that the processes $if=01,03$ are those that possibly lead to negative values of the MHQs. Specifically, for $\varphi\in(\pi/2,\pi)\cup (\pi,3\pi/2)$, we have $\mathrm{Re}q_{03}^{\mathcal{V}}(\hat{\rho})<0$, and for $\varphi\in(0,\pi/2)\cup (3\pi/2,2\pi)$ we have $\mathrm{Re}q_{01}^{\mathcal{V}}(\hat{\rho})<0$. The other processes are characterized by positive MHQs for any value of the phase. 

In Fig.~\ref{fig:qp_CNOTHH}(b) we focus on the KDQ of the $if=01$ process, that cannot be written as those of the Hdamard gate and so it cannot be factorized. We report the full MHQ of the process, those of its constituent gates and the correction $\mathrm{Re}\mathcal{Q}_{01}^{\mathcal{V}}(\hat{\rho})$.
We observe that the latter is always negative (except for $\varphi = 0$), and since the MHQs of the constituent gates are never negative, it can be identified as the source of the negativity in the overall MHQ of the circuit. This negativity arises for $\varphi \in (0, \pi/2) \cup (3\pi/2, 2\pi)$, where the correction becomes sufficiently negative to counterbalance the positive contributions from the MHQs of the constituent gates.

As a further step, we explore the behavior of the MHQs associated to the processes $\ket{E_i}\rightarrow\ket{E_f}$ with $E_i>E_f$, that increase the extractable work in the case in which the work distribution is classical. Their behavior is reported in Fig.~\ref{fig:qp_CNOTHH}(c). We observe that the processes $if = 10$ and $if = 20$ are entirely suppressed, exhibiting zero MHQs. On the other hand, the processes $if=30$ and $if=32$ have always positive MHQs, and so positively contribute to the work extraction. The only process that is detrimental for the work extraction is $if=31$, whose MHQ is negative for $\varphi\in(0,\pi/2)\cup(3\pi/2,2\pi)$.

Finally, we analyze the extractable work. Since at $p=1/2$ the population part of the input state does not allow for work extraction, work is fully due to quantum coherences, and we are specifically interested in understanding which type of processes contribute to the work extraction. To this end, we split the work (Eq.~\eqref{eq:work_2qbt}) in different components $\mathcal{W}_{if}^\mathcal{U}[\hat{\rho}]$ such that
\begin{equation}
    \mathcal{W}_\mathcal{U}[\hat{\rho}]=\sum_{i<f}\mathcal{W}_{if}^\mathcal{U}[\hat{\rho}],
    \label{eq:work_components}
\end{equation}
where $\mathcal{W}_{if}^\mathcal{U}[\hat{\rho}]\coloneqq 2E(1+\delta_{i,0}\delta_{f,3})(q_{fi}^\mathcal{U}(\hat{\rho})-q^\mathcal{U}_{if}(\hat{\rho}))$. Each $\mathcal{W}_{if}$ takes into account the contribution to the work of all the processes that couple the eigenstates $\ket{E_i}$ and $\ket{E_f}$, the results are reported in Fig.~\ref{fig:qp_CNOTHH}(d). We observe that there are well separated regions in which the extractable work takes different sign. For $\varphi\in(0,\pi/2)\cup(3\pi/2,2\pi)$, the presence of coherences is detrimental, since the extractable work is largely negative and so the transformation incurs a significant energetic cost. On the other hand, $\varphi\in(\pi/2,\pi)\cup(\pi,3\pi/2)$, the extractable work becomes positive and the presence of coherences becomes advantageous. Specifically, the positive extractable work is mainly affected by $\mathcal{W}_{03}$, which is the work contribution coming from energy exchange processes between the lowest and the highest energy states. The latter brings a significant positive contribution that derives from the negativity of the MHQ (see Fig.~\ref{fig:qp_CNOTHH}(a)), and, as a result, a process of the form $\ket{E_i} \rightarrow \ket{E_f}$ (with $E_i < E_f$)occurs, which is anomalously responsible for the emergence of positive extractable work, although not exclusively.

\section{Conclusions and outlook}
\label{secconclu}

We have developed a general framework for analyzing the thermodynamic properties of quantum circuits using Kirkwood–Dirac quasiprobabilities (KDQs). Their real parts, the Margenau-Hill quasiprobabilities (MHQs), capture the first moment of work and reveal the thermodynamic role of quantum coherence. Notably, a negative coherent contribution to the MHQ can cause transitions that would normally be beneficial to become unfavorable and those that would typically be detrimental to become favorable when sufficiently strong. Such negativity acts as a signature of nonclassical, anomalous processes that could facilitate the extraction of work beyond classical thermodynamic bounds.

In this paper, we first explored the thermodynamic features of the KDQs for single-qubit gates, beginning with a general transformation and then focusing on specific examples, such as the Hadamard and $\pi/8$ gates. We introduced a systematic method for characterizing work statistics in quantum circuits by decomposing the KDQs of an entire circuit into those of its elementary gates, as shown in Eq.~\eqref{eq:discrepancy_KDQ}. This gate-level decomposition establishes a direct connection between the thermodynamic behavior of complex circuits and the properties of their basic components. In deep circuits, structural compatibilities often simplify the resulting KDQs. Notably, for single-qubit operations, the $HTH$ gate sequence serves as a minimal example where the KDQs of the circuit cannot be simply reduced to those of its constituent gates, revealing genuinely nonclassical anomalous features even when its constituent parts do not exhibit such behavior individually. We also analyzed the thermodynamic properties of two-qubit gates, such as the CNOT, for which the KDQs are real and positive, signaling thermodynamically classical behavior. More generally, the KDQs of two-qubit gates factorize under local operations with separable inputs, while entangled inputs introduce intrinsically quantum features that require case-specific investigation. As a practical demonstration, we provided an operational example of a simple two-qubit quantum circuit, whose thermodynamic behavior was analyzed using the decomposition of KDQs and the specific properties of two-qubit gates discussed in the previous sections.

The framework we have introduced provides a tool for exploring the quantum thermodynamics of cyclic transformations, modeled as quantum circuits, in the presence of initial coherence, potentially leading to nonclassical work statistics. Our method for computing and decomposing KDQs of arbitrary quantum circuits  establishes a foundation for investigating quantum thermodynamics in the context of quantum computation through the lens of quasiprobability theory. Beyond quantum computation, our framework is also applicable to the study of nonclassical thermodynamic phenomena in out-of-equilibrium many-body quantum systems, such as driven-dissipative phase transitions or ergodicity breaking phenomena. 

\section{Acknowledgements}
We thank A. Palamara for useful discussions. This research was supported by the PNRR MUR project PE0000023-NQSTI through the secondary project ``ThAnQ".

\appendix

\section{Real and imaginary parts of KDQs}\label{app:real_im}
In this appendix, we derive the explicit expressions for the real and imaginary parts of the KDQs, emphasizing their contributions to the first and second moments of the work. The protocol is a cyclic transformation, with initial/final Hamiltonian $\hat{\mathcal{H}} =\sum_k E_k \hat{\Pi}_k$, while $\hat{\mathcal{U}}$ is the unitary evolution operator.  To explicitly obtain real and imaginary parts of the KDQs, we follow Ref.~\cite{gherardini2024quasiprobabilities} and recall the following identities:
\begin{widetext}
\begin{equation}
    \begin{cases}
        \mathrm{Tr}\left[\hat{\mathcal{U}}^{\dagger}\hat{\Pi}_f\hat{\mathcal{U}}\hat{\Pi}_i\hat{\rho}\right]=\mathrm{Re}\Bigl(\mathrm{Tr}\left[\hat{\mathcal{U}}^{\dagger}\hat{\Pi}_f\hat{\mathcal{U}}\hat{\Pi}_i\hat{\rho}\right]\Bigr)+i\,\mathrm{Im}\Bigr(\mathrm{Tr}\left[\hat{\mathcal{U}}^{\dagger}\hat{\Pi}_f\hat{\mathcal{U}}\hat{\Pi}_i\hat{\rho}\right]\Bigr),\\ 
        \mathrm{Tr}\left[\hat{\Pi}_i\hat{\mathcal{U}}^{\dagger}\hat{\Pi}_f\hat{\mathcal{U}}\hat{\rho}\right]=\mathrm{Re}\Bigl(\mathrm{Tr}\left[\hat{\mathcal{U}}^{\dagger}\hat{\Pi}_f\hat{\mathcal{U}}\hat{\Pi}_i\hat{\rho}\right]\Bigr)-i\, \mathrm{Im}\Bigl(\mathrm{Tr}\left[\hat{\mathcal{U}}^{\dagger}\hat{\Pi}_f\hat{\mathcal{U}}\hat{\Pi}_i\hat{\rho}\right]\Bigr),        
    \end{cases}
\end{equation}
\end{widetext}
by adding the two equations, we obtain the real part of the KDQ, or MHQ, that reads
\begin{equation}
    \mathrm{Re} q_{if}^{\mathcal{U}}(\hat{\rho})=\frac{1}{2} \mathrm{Tr}\left[\Bigl\{\hat{\mathcal{U}}^{\dagger}\hat{\Pi}_f\hat{\mathcal{U}},\hat{\Pi}_i\Bigr\}\hat{\rho}\right].
\end{equation}
On the other hand, by performing the difference between the two equations, we obtain the imaginary part of the KDQ
\begin{equation}
\mathrm{Im} q_{if}^{\mathcal{U}}(\hat{\rho})=\frac{1}{2i}\mathrm{Tr}\left[\Bigl[\hat{\mathcal{U}}^{\dagger}\hat{\Pi}_f\hat{\mathcal{U}},\hat{\Pi}_i\Bigr]\hat{\rho}\right].
\label{eq:Im_q_if_rho}
\end{equation}
Recalling that the extractable work reads
\begin{equation}
    \mathcal{W}_{\mathcal{U}}[\hat{\rho}]=\operatorname{Tr}\left[\hat{\mathcal{H}}\hat{\rho}\right] - \operatorname{Tr}\left[\hat{\mathcal{H}}\hat{\mathcal{U}}\hat{\rho}\hat{\mathcal{U}}^{\dagger}\right];
\end{equation}
we now use $\hat{\mathcal{H}}=\sum_i E_i \hat{\Pi}_i$ and write the initial mean energy as $\operatorname{Tr}\left[\hat{\mathcal{H}}\hat{\rho}\right]=\sum_i E_i \operatorname{Tr}\left[\hat{\Pi}_i\hat{\rho}\right]$. Using the identity $\hat{\mathds{1}}=\sum_f \hat{\mathcal{U}}\hat{\Pi}_f \hat{\mathcal{U}}^\dagger$, we obtain $\operatorname{Tr}\left[\hat{\rho}\hat{\mathcal{H}}\right]=\sum_{if} E_i \operatorname{Tr}\left[\hat{\mathcal{U}}^{\dagger}\hat{\Pi}_f \hat{\mathcal{U}} \hat{\Pi}_i \hat{\rho} \right]=\sum_{if}E_i q_{if}^{\mathcal{U}}(\hat{\rho})$; note that at this level the indices $i$ and $f$ do not have any different meaning, they both span on all the Hamiltonian eigenstates. At the same time, we can insert $\hat{\mathcal{H}}=\sum_f E_f \hat{\Pi}_f$ in $\operatorname{Tr}\left[\hat{\mathcal{H}}\hat{\mathcal{U}}\hat{\rho}\hat{\mathcal{U}}^{\dagger}\right]$, to obtain $\operatorname{Tr}\left[\hat{\mathcal{H}}\hat{\mathcal{U}}\hat{\rho}\hat{\mathcal{U}}^{\dagger}\right]=\operatorname{Tr}\left[\hat{\mathcal{U}}^{\dagger}\hat{\mathcal{H}}\hat{\mathcal{U}}\hat{\rho}\right]=\sum_f E_f\operatorname{Tr}\left[\hat{\mathcal{U}}^{\dagger}\hat{\Pi}_f\hat{\mathcal{U}}\hat{\rho}\right]$. With the help of the identity $\hat{\mathds{1}}=\sum_i\hat{\Pi}_i$, the final mean energy becomes $\operatorname{Tr}\left[\hat{\mathcal{H}}\hat{\mathcal{U}}\hat{\rho}\hat{\mathcal{U}}^{\dagger}\right]=\sum_{if} E_f\operatorname{Tr}\left[\hat{\mathcal{U}}^{\dagger}\hat{\Pi}_f\hat{\mathcal{U}}\hat{\Pi}_i\hat{\rho}\right]=\sum_{if}E_f q_{if}^{\mathcal{U}}(\hat{\rho})$. Both initial and final mean energies are expectation values of quantum observables, so they are real. This can be shown explicitly by using Eq.~\eqref{eq:Im_q_if_rho}:
\begin{widetext}
\begin{align}
    \sum_{if}\mathrm{Im} q_{if}^{\mathcal{U}}(\hat{\rho})(E_i - E_f)=\frac{1}{2i}\Biggl(\mathrm{Tr}\left[\Bigl[\hat{\mathds{1}},\hat{\mathcal{H}}\Bigr]\hat{\rho}\right]-\nonumber\mathrm{Tr}\left[\Bigl[\hat{\mathcal{U}}^{\dagger}\hat{\mathcal{H}}\hat{\mathcal{U}},\hat{\mathds{1}}\Bigr]\hat{\rho}\right]    \Biggr)=0,
\end{align}
\end{widetext}
where we have used $\sum_f E_f \hat{\mathcal{U}}^{\dagger}\hat{\Pi}_f\hat{\mathcal{U}}=\hat{\mathcal{U}}^{\dagger}\hat{\mathcal{H}}\hat{\mathcal{U}}$ and $\sum_i E_i \hat{\Pi}_i=\hat{\mathcal{H}}$.

Thus, the extractable work can be written in terms of KDQs (or MHQs) as
\begin{equation} 
    \mathcal{W}_{\mathcal{U}}[\hat{\rho}]=    \sum_{if}q_{if}^{\mathcal{U}}(\hat{\rho})(E_i - E_f)=\sum_{if}\mathrm{Re}q_{if}^{\mathcal{U}}(\hat{\rho})(E_i - E_f),
\end{equation}
and it can be interpreted as the sum over all possible energy transition processes from initial eigenstates with energy $E_i$ to final energy eigenstates with energy $E_f$, each weighted by the associated MHQ. As mentioned in the main text, the MHQs can be negative, and so processes of the type $\ket{E_i}\rightarrow \ket{E_f}$, with $E_i < E_f$, can anomalously positively contribute to the work extraction. 

The imaginary parts of the KDQs affects higher order moments. For instance, the work variance, that is related to the second moment, cannot be described by the MHQs~\cite{gherardini2024quasiprobabilities}. Specifically, the work variance $(\Delta W)^2=\braket{W^2}-\braket{W}^2$ has an imaginary part that is directly affected by the imaginary part of the KDQ:
\begin{equation}
    \mathrm{Im}(\Delta W_{\mathcal{U}})^2=i\mathrm{Tr}\left[\Bigl[\hat{\mathcal{U}}^{\dagger}\hat{\mathcal{H}}\hat{\mathcal{U}},\hat{\mathcal{H}}\Bigr]\hat{\rho}\right]=-2\sum_{if}E_i E_f \mathrm{Im} q_{if}^{\mathcal{U}}(\hat{\rho}).
\end{equation}

\section{Population and coherent parts of KDQs, correction to Jarzynski equality}\label{app:popcoh}
The input quantum state can be written as $\hat{\rho}=\sum_{ik}\lambda_{ik}\ket{E_i}\bra{E_k}=\Delta(\hat{\rho})+\hat{\chi}$, where $\Delta(\hat{\rho})=\sum_i \lambda_{ii}\ket{E_i}\bra{E_i}$ is its diagonal dephased version in the initial/final Hamiltonian eigenbasis and $\hat{\chi}=\sum_{i\neq k}\lambda_{ik}\ket{E_i}\bra{E_k}$ is its coherent part. As a result, the KDQs can be split in population and coherence contributions
\begin{equation}
q_{if}^\mathcal{U}(\hat{\rho})=q_{if}^\mathcal{U}(\Delta(\hat{\rho})) + q_{if}^\mathcal{U}(\hat{\chi}),    
\end{equation}
where
\begin{align}
    &q_{if}^\mathcal{U}(\Delta(\hat{\rho})) =\left(\hat{\mathcal{U}}^{\dagger}\hat{\Pi}_f\hat{\mathcal{U}}\right)_{ii}\lambda_{ii},
    \\&
    q_{if}^\mathcal{U}(\Delta(\hat{\rho})) =\sum_k{}^{'}\left(\hat{\mathcal{U}}^{\dagger}\hat{\Pi}_f\hat{\mathcal{U}}\right)_{ki}\lambda_{ik},
\end{align}
the primed sum $\sum_k{}^{'}$ indicates summation over all the $k\neq i$, and $\left(\hat{\mathcal{O}}\right)_{ki}=\braket{E_k|\hat{\mathcal{O}}|E_i}$. Let us now observe that 
\begin{equation}
    \left(\hat{\mathcal{U}}^{\dagger}\hat{\Pi}_f\hat{\mathcal{U}}\right)_{ki}=\braket{E_k|\hat{\mathcal{U}}^{\dagger}|E_f}\braket{E_f|\hat{\mathcal{U}}|E_i}\coloneqq \left(k_{kf}^{\mathcal{U}}\right)^{*}k_{if}^{\mathcal{U}},
\end{equation}
where we have used $\left(\braket{\psi|\hat{\mathcal{O}}|\phi}\right)^*=\braket{\phi|\hat{\mathcal{O}}^{\dagger}|\psi}$, $\hat{\Pi}_f = \ket{E_f}\bra{E_f}$ and we have introduced the transition amplitude $k_{if}^{\mathcal{U}}=\braket{E_f|\hat{\mathcal{U}}|E_i}$. Thus, we can compactly write the KDQs as
\begin{equation}
        q_{if}^\mathcal{U}(\Delta(\hat{\rho}))=\bigl| k_{if}^{\mathcal{U}}\bigr|^2\lambda_{ii},
    \quad
    q_{if}^\mathcal{U}(\hat{\chi})=\sum_k{}^{'}\left(k_{kf}^{\mathcal{U}}\right)^{*}k_{if}^{\mathcal{U}}\lambda_{ik};
    \label{eq:q_pop_coh}
\end{equation}
the population part of the KDQs is given by the transition probability of the process $\ket{E_i}\rightarrow\ket{E_f}$ weighted by the population of the initial state $\lambda_{ii}$ and always contributes positively to $q_{if}^{\mathcal{U}}(\hat{\rho})$. On the other hand, the coherent part of the KDQs is a combination of the  $\ket{E_i}\rightarrow\ket{E_f}$ and the time-reversed $\ket{E_f}\rightarrow\ket{E_k}$ (with $k\neq i$) transition amplitudes each weighted with the associated coherences $\lambda_{ik}$; it is responsible for possible nonpositivity of the full KDQs. 

Let us now take a step back and take the input state to be thermal in the basis of the initial Hamiltonian, i.e. $\Delta(\hat{\rho})=\hat{\rho}^{\beta}=e^{-\beta\hat{\mathcal{H}}}/Z$, with $Z=\operatorname{Tr}\left[ e^{-\beta\hat{\mathcal{H}}}\right]$, and $\hat{\chi}=0$. Consider the work $W_\mathcal{U}$ performed on the system during the process $\mathcal{U}$. Then, the average of the $e^{-\beta W_\mathcal{U}}$ satisfies the Jarzynski equality (JE)~\cite{jarzynski1997nonequilibrium,jarzynski1997equilibrium}
\begin{equation}
    \braket{e^{-\beta W_\mathcal{U}}}=e^{-\beta \Delta F},
\end{equation}
where $\Delta F$ is the difference between the free energy of the final thermal state and that of the initial thermal state, both at temperature $\beta$. Moreover, the JE implies
\begin{equation}
    \braket{W_\mathcal{U}}\geq \Delta F.
\end{equation}
In a cyclic transformation, initial and final Hamiltonians coincide, and so 
\begin{equation}
    \braket{e^{-\beta W_\mathcal{U}}}=1,
    \quad
    \braket{W_\mathcal{U}}\geq 0,
    \label{eq:Jarz_cyclic}
\end{equation}
which means that it is not possible to extract work from a thermal state through a cyclic transformation. 

The presence of coherences can potentially lead to the violation the JE. To see this, let us suppose that $\hat{\rho}$ has thermal populations, so that $\Delta(\hat{\rho})$ is thermal, but coherences are present in the state ($\hat{\chi} \ne 0$). In this case, given a cyclic transformation, we have~\cite{gherardini2024quasiprobabilities}
\begin{equation}
   \braket{e^{-\beta W_\mathcal{U}}} = \operatorname{Tr}\left[\bigl(\hat{\rho}^{\beta}\bigr)^{-1}\hat{\rho} \hat{\mathcal{U}}^{\dagger}\hat{\rho}^{\beta}\hat{\mathcal{U}}^{\dagger}\right] e^{-\beta \Delta F};
\end{equation}
we now write $\hat{\rho}^{\rm \beta}=\sum_k \lambda_k^{\rm \beta}\hat{\Pi}_k$, with $\lambda_k^{\rm \beta}=\frac{e^{-\beta E_k}}{Z}$, and we obtain
\begin{align}
 \braket{e^{-\beta W_\mathcal{U}}}=&\sum_{if} \frac{\lambda_f^{\beta}}{\lambda_i^{\beta}}\operatorname{Tr}\left[\hat{\Pi}_i\hat{\rho} \hat{\mathcal{U}}^{\dagger}\hat{\Pi}_f\hat{\mathcal{U}}\right]=\nonumber\\&\sum_{if}q_{if}^\mathcal{U}(\hat{\rho})e^{-\beta(E_f - E_i)}. 
\end{align}
Finally, explicitly taking $\hat{\rho}=\hat{\rho}^{\beta}+\hat{\chi}$, we get
\begin{align}
    \braket{e^{-\beta W_\mathcal{U}}}=&\sum_{if}q_{if}^\mathcal{U}(\hat{\rho}^{\beta})e^{-\beta(E_f - E_i)} +\nonumber\\& \sum_{if}q_{if}^{\mathcal{U}}(\hat{\chi})e^{-\beta(E_f - E_i)}=1+\Gamma_\mathcal{U}^{\beta}(\hat{\chi}),
\end{align}
where we have used JE to get $\sum_{if}q_{if}^\mathcal{U}(\hat{\rho}^{\beta})e^{-\beta(E_f - E_i)}=1$, and we have introduced the correction $\Gamma_\mathcal{U}^{\beta}(\hat{\chi})\coloneqq\sum_{if}q_{if}^{\mathcal{U}}(\hat{\chi})e^{-\beta(E_f - E_i)}$ that can possibly be negative or complex. The JE being corrected by a non-thermal component, the bound $\braket{W_\mathcal{U}}\geq 0$ does not hold anymore and cases where $\braket{W_\mathcal{U}}<0$ (i.e. positive work extraction) are admitted.

In a two-level system with Hilbert space spanned by $\{ \ket{\downarrow},\ket{\uparrow}\}$ with initial/final Hamiltonian $\hat{\mathcal{H}}=E_{\uparrow}\hat{\Pi}_{\uparrow} + E_{\downarrow}\hat{\Pi}_{\downarrow}$ ($E_{\uparrow} > E_{\downarrow}$), any incoherent quantum state $\Delta(\hat{\rho})=p\ket{\uparrow}\bra{\uparrow} +(1-p)\ket{\downarrow}\bra{\downarrow}$ admits a thermal description with $p= \frac{e^{-\beta E_{\uparrow}}}{Z}$ and $1-p= \frac{e^{-\beta E_{\downarrow}}}{Z}$,  whose temperature, $\beta = - \frac{1}{E_{\uparrow}-E_{\downarrow}}\ln \left( \frac{p}{1-p}\right)$,  is positive ($\beta \geq 0$) in the absence of population inversion, $p \leq \frac{1}{2}$. Thus, under this condition, it is impossible to get positive extractable work, and the injection of coherence is a necessary condition for $\braket{W_\mathcal{U}} <0$.

\section{KDQs and work extraction in a generic single-qubit transformation}\label{app:single_qb}
We consider a cyclic protocol that evolves the state $\hat{\rho}$ under the most generic unitary $\hat{\mathcal{U}}$ with initial and final Hamiltonian $\hat{\mathcal{H}}=\sum_k E_k \hat{\Pi}_k$, whose KDQs are
\begin{equation}
    q_{if}^{\mathcal{U}}(\hat{\rho})=\operatorname{Tr}\bigl[\hat{\mathcal{U}}^{\dagger}\hat{\Pi}_f \hat{\mathcal{U}} \Pi_i \hat{\rho}\bigr],
\end{equation}
where the indices $i$ and $f$ refer to the same basis. Let us consider the most generic single qubit circuit, described by the unitary transformation 
\begin{equation}
    \hat{\mathcal{U}}(\alpha,\theta,\boldsymbol{n})=e^{i\alpha}\hat{R}(\theta,\boldsymbol{n}),
\end{equation}
where $\hat{R}(\theta,\boldsymbol{n})$ is given by
\begin{equation}
    \hat{R}(\theta,\boldsymbol{n})=e^{-i (\theta/2)\boldsymbol{n}\cdot \hat{\boldsymbol{\sigma}}}=\hat{\mathds{1}}\cos\left(\frac{\theta}{2}\right)-i\left( \boldsymbol{n}\cdot \hat{\boldsymbol{\sigma}}\right)\sin\left(\frac{\theta}{2}\right).
    \label{eq:rotation}
\end{equation}
The form of the KDQs does not depend on the global phase and reads
\begin{equation}
    q_{if}^{R}(\hat{\rho})=\operatorname{Tr}\bigl[\hat{R}(\theta,\boldsymbol{n})^{\dagger}\hat{\Pi}_f\hat{R}(\theta,\boldsymbol{n}) \hat{\Pi}_i \hat{\rho}\bigr].
\end{equation}
In the energy basis, $\{\ket{\uparrow},\ket{\downarrow}\}$, the rotation operator $    \hat{R}(\theta,\boldsymbol{n})$ reads
\begin{equation}
\hat{R}(\theta,\boldsymbol{n}) =
 \begin{bmatrix}
  u_{\uparrow\uparrow} &  u_{\uparrow\downarrow} \\
  u_{\downarrow\uparrow} & u_{\downarrow\downarrow}
 \end{bmatrix},
\end{equation}
with $u_{\uparrow\uparrow}=\cos\left(\frac{\theta}{2}\right)-in_z\sin\left(\frac{\theta}{2}\right)$, $u_{\uparrow\downarrow}=-i(n_x-in_y)\sin\left(\frac{\theta}{2}\right)$, $u_{\downarrow\uparrow} =  -i(n_x+in_y)\sin\left(\frac{\theta}{2}\right)$ and $u_{\downarrow\downarrow}=\cos\left(\frac{\theta}{2}\right)+in_z\sin\left(\frac{\theta}{2}\right)$. 
The projectors and their rotated versions are
\begin{align}
&\hat{\Pi}_{\downarrow}=
\begin{bmatrix}
0 & 0 \\ 0 & 1    
\end{bmatrix},
\quad
\hat{R}^{\dagger}\hat{\Pi}_{\downarrow}\hat{R}=
\begin{bmatrix}
|u_{\downarrow\uparrow}|^2 & u_{\downarrow\uparrow}^*u_{\downarrow\downarrow} \\ u_{\downarrow\uparrow}u_{\downarrow\downarrow}^* & |u_{\downarrow\downarrow}|^2    
\end{bmatrix},
\quad
\\&\hat{\Pi}_{\uparrow}=
\begin{bmatrix}
1 & 0 \\ 0 & 0    
\end{bmatrix},
\quad
\hat{R}^{\dagger}\hat{\Pi}_{\uparrow}\hat{R}=
\begin{bmatrix}
|u_{\uparrow\uparrow}|^2 & u_{\uparrow\uparrow}^*u_{\uparrow\downarrow} \\ u_{\uparrow\uparrow}u_{\uparrow\downarrow}^* & |u_{\uparrow\downarrow}|^2    
\end{bmatrix}.
\end{align}
Finally, using 
\begin{equation}
    \hat{\rho}=
    \begin{bmatrix}
        p & \gamma \\
        \gamma^* & 1-p 
    \end{bmatrix},
\end{equation}
we can compute the KDQs:
\begin{align}
& q^{R}_{\downarrow\downarrow}=i|\gamma|e^{-i\varphi}f_{xy}^*g_z(\theta)\sin(\theta/2)+(1-p)|g_z(\theta)|^2\nonumber\\
& q^{R}_{\downarrow\uparrow}=-i|\gamma|e^{-i\varphi}f_{xy}^*g_z(\theta)\sin(\theta/2)+(1-p)|f_{xy}|^2\sin^2(\theta/2)\nonumber\\
& q^{R}_{\uparrow\downarrow}=-i|\gamma|e^{i\varphi}f_{xy}g_z(\theta)^*\sin(\theta/2)+p|f_{xy}|^2\sin^2(\theta/2)\nonumber\\
& q^{R}_{\uparrow\uparrow}=i|\gamma|e^{i\varphi}f_{xy}g_z(\theta)^*\sin(\theta/2)+p|g_z(\theta)|^2,
\label{eq:qp_2lev}
\end{align}
where we have have used $\gamma=|\gamma|e^{i\varphi}$, omitted the dependence of KDQ on the input state $\hat{\rho}$, and introduced the functions $f_{xy}\coloneqq n_x+in_y$, $g_z(\theta)\coloneqq \cos(\theta/2)+in_z\sin(\theta/2)$.

Using the superscripts $\mathfrak{R}$ and $\mathfrak{I}$ to indicate respectively real and imaginary parts of $f$ and $g$, we obtain
\begin{widetext}
\begin{align}
q^{R}_{\downarrow\downarrow}=&\left[-|\gamma|\cos\varphi f^{\mathfrak{R}}g^{\mathfrak{I}} \sin\left(\frac{\theta}{2}\right) + |\gamma|\cos\varphi f^{\mathfrak{I}} g^{\mathfrak{R}} \sin\left(\frac{\theta}{2}\right) +|\gamma|\sin\varphi f^{\mathfrak{R}}g^{\mathfrak{R}} \sin\left(\frac{\theta}{2}\right)  + |\gamma|\sin\varphi f^{\mathfrak{I}} g^{\mathfrak{I}} \sin\left(\frac{\theta}{2}\right)  + \right. \nonumber \\
& \left. (1-p) |g|^2\right] +\nonumber\\
& i\left[|\gamma|\cos\varphi f^{\mathfrak{R}}g^{\mathfrak{R}} \sin\left(\frac{\theta}{2}\right)  + |\gamma|\cos\varphi f^{\mathfrak{I}} g^{\mathfrak{I}} \sin\left(\frac{\theta}{2}\right) +|\gamma|\sin\varphi f^{\mathfrak{R}}g^{\mathfrak{I}} \sin\left(\frac{\theta}{2}\right)  - |\gamma|\sin\varphi f^{\mathfrak{I}} g^{\mathfrak{R}} \sin\left(\frac{\theta}{2}\right) \right] \label{eq:q_dd} \\
q^{R}_{\uparrow\uparrow}=&\left[|\gamma|\cos\varphi f^{\mathfrak{R}}g^{\mathfrak{I}} \sin\left(\frac{\theta}{2}\right)  - |\gamma|\cos\varphi f^{\mathfrak{I}} g^{\mathfrak{R}} \sin\left(\frac{\theta}{2}\right)  -|\gamma|\sin\varphi f^{\mathfrak{R}}g^{\mathfrak{R}} \sin\left(\frac{\theta}{2}\right)  - |\gamma|\sin\varphi f^{\mathfrak{I}} g^{\mathfrak{I}} \sin\left(\frac{\theta}{2}\right)  + p |g|^2\right] +\nonumber \\
& i\left[|\gamma|\cos\varphi f^{\mathfrak{R}}g^{\mathfrak{R}} \sin\left(\frac{\theta}{2}\right)  + |\gamma|\cos\varphi f^{\mathfrak{I}} g^{\mathfrak{I}} \sin\left(\frac{\theta}{2}\right)  +|\gamma|\sin\varphi f^{\mathfrak{R}}g^{\mathfrak{I}} \sin\left(\frac{\theta}{2}\right)  - |\gamma|\sin\varphi f^{\mathfrak{I}} g^{\mathfrak{R}} \sin\left(\frac{\theta}{2}\right) \right] \label{eq:q_uu} \\
q^{R}_{\downarrow\uparrow}=&\left[|\gamma|\cos\varphi f^{\mathfrak{R}}g^{\mathfrak{I}} \sin(\frac{\theta}{2}) - |\gamma|\cos\varphi f^{\mathfrak{I}} g^{\mathfrak{R}} \sin(\frac{\theta}{2}) -|\gamma|\sin\varphi f^{\mathfrak{R}}g^{\mathfrak{R}} \sin(\frac{\theta}{2}) - |\gamma|\sin\varphi f^{\mathfrak{I}} g^{\mathfrak{I}} \sin(\frac{\theta}{2})\right. +\nonumber \\
&\left. (1-p) |f|^2\sin^2\left(\frac{\theta}{2}\right) \right]+ \nonumber \\
&i\left[-|\gamma|\cos\varphi f^{\mathfrak{R}}g^{\mathfrak{R}} \sin\left(\frac{\theta}{2}\right)  - |\gamma|\cos\varphi f^{\mathfrak{I}} g^{\mathfrak{I}} \sin\left(\frac{\theta}{2}\right) -|\gamma|\sin\varphi f^{\mathfrak{R}}g^{\mathfrak{I}} \sin\left(\frac{\theta}{2}\right)  + |\gamma|\sin\varphi f^{\mathfrak{I}} g^{\mathfrak{R}} \sin\left(\frac{\theta}{2}\right) \right] \label{eq:q_du}\\
q^{R}_{\uparrow\downarrow}=&\left[-|\gamma|\cos\varphi f^{\mathfrak{R}}g^{\mathfrak{I}} \sin\left(\frac{\theta}{2}\right)  + |\gamma|\cos\varphi f^{\mathfrak{I}} g^{\mathfrak{R}} \sin\left(\frac{\theta}{2}\right)  +|\gamma|\sin\varphi f^{\mathfrak{R}}g^{\mathfrak{R}} \sin\left(\frac{\theta}{2}\right)  + |\gamma|\sin\varphi f^{\mathfrak{I}} g^{\mathfrak{I}} \sin(\frac{\theta}{2}) + \right. \nonumber\\
&\left. p|f|^2\sin^2\left(\frac{\theta}{2}\right) \right] +\nonumber \\
&i\left[-|\gamma|\cos\varphi f^{\mathfrak{R}}g^{\mathfrak{R}} \sin\left(\frac{\theta}{2}\right)  - |\gamma|\cos\varphi f^{\mathfrak{I}} g^{\mathfrak{I}} \sin\left(\frac{\theta}{2}\right)  -|\gamma|\sin\varphi f^{\mathfrak{R}}g^{\mathfrak{I}} \sin\left(\frac{\theta}{2}\right)  + |\gamma|\sin\varphi f^{\mathfrak{I}} g^{\mathfrak{R}} \sin\left(\frac{\theta}{2}\right) \right] \label{eq:q_ud} 
\end{align}
\end{widetext}
from which it follows that
\begin{equation}
\mathrm{Im}q^{R}_{\uparrow\uparrow}=\mathrm{Im}q^{R}_{\downarrow\downarrow} ,  
\quad
\mathrm{Re} q^{R}_{\downarrow\downarrow} = -\mathrm{Re}q^{R}_{\uparrow\uparrow}+|g|^2
\end{equation}
and also
\begin{equation}
\mathrm{Im}q^{R}_{\downarrow\uparrow}=\mathrm{Im}q^{R}_{\uparrow\downarrow} ,  
\quad
\mathrm{Re} q^{R}_{\downarrow\uparrow} = -\mathrm{Re}q^{R}_{\uparrow\downarrow}+|f|^2\sin^2(\theta/2).
\label{eq:q_updndnup}
\end{equation}
We further observe that
\begin{eqnarray}
&& \mathrm{Re}q^{R}_{\downarrow\downarrow}-(1-p)|g|^2 = \mathrm{Re}q^{R}_{\uparrow\downarrow} - p|f|^2 \sin^2(\theta/2) \\
&& \mathrm{Re}q^{R}_{\uparrow\uparrow}-p|g|^2=\mathrm{Re}q^{R}_{\downarrow\uparrow}-(1-p)|f|^2\sin^2(\theta/2)
\end{eqnarray}
implying, respectively
\begin{eqnarray}
&& \mathrm{Re}q^{R}_{\downarrow\downarrow} = \mathrm{Re}q^{R}_{\uparrow\downarrow} - p|f|^2 \sin^2(\theta/2)+(1-p)|g|^2, \\
&& \mathrm{Re}q^{R}_{\uparrow\uparrow} = \mathrm{Re}q^{R}_{\downarrow\uparrow}-(1-p)|f|^2\sin^2(\theta/2) + p |g|^2.
\end{eqnarray}
From Eqs.~\eqref{eq:q_dd},\eqref{eq:q_uu},\eqref{eq:q_du},\eqref{eq:q_ud} we can also extract a relation between different coherent parts of the KDQs. More specifically, we first observe that $|\gamma|$ only enters $q_{if}^{R}(\hat{\chi})$, while $p$ or $(1-p)$ only enter $q_{if}^{R}(\Delta(\hat{\rho}))$. Then:
\begin{align}
&\mathrm{Re}q_{\uparrow\uparrow}^{R}(\hat{\chi})=\mathrm{Re}q_{\downarrow\uparrow}^{R}(\hat{\chi});
    \quad
\mathrm{Re}q_{\downarrow\downarrow}^{R}(\hat{\chi})=\mathrm{Re}q_{\uparrow\downarrow}^{R}(\hat{\chi}),   
\label{eq:Reqp_coh_1bt}
\\&\mathrm{Im}q_{\uparrow\uparrow}^{R}(\hat{\chi})=-\mathrm{Im}q_{\downarrow\uparrow}^{R}(\hat{\chi});
    \quad
\mathrm{Im}q_{\downarrow\downarrow}^{R}(\hat{\chi})=-\mathrm{Im}q_{\uparrow\downarrow}^{R}(\hat{\chi}).  
\label{eq:qp_coh_1bt}
\end{align}

With the help of these relations for KDQs, we can write the extractable work as
\begin{align}
\mathcal{W}_R[\hat{\rho}]=&\sum_{if}(E_i - E_f)q^{R}_{if}(\hat{\rho})=\nonumber\\&(E_{\uparrow}-E_{\downarrow})q^{R}_{\uparrow\downarrow}(\hat{\rho})  + (E_{\downarrow}-E_{\uparrow})q^{R}_{\downarrow\uparrow}(\hat{\rho}),   
\end{align}
or, using $E_{\uparrow}=-E_{\downarrow}=E$,
\begin{equation}
\mathcal{W}_R[\hat{\rho}]=2E (q^{R}_{\uparrow\downarrow}(\hat{\rho}) - q^{R}_{\downarrow\uparrow}(\hat{\rho}))=2E(\mathrm{Re}q^{R}_{\uparrow\downarrow}(\hat{\rho})-\mathrm{Re}q^{R}_{\downarrow\uparrow}(\hat{\rho})),
\end{equation}
where we made use the first relation in Eq.~\eqref{eq:q_updndnup}. By using the second relation in Eq.~\eqref{eq:q_updndnup}, we obtain
\begin{equation}
\mathcal{W}_R[\hat{\rho}]=2E\left[(n_x^2+n_y^2)\sin^2(\theta/2)-2\mathrm{Re}q^{R}_{\downarrow\uparrow}(\hat{\rho})\right].   
\label{eq:W_R_1qbt}
\end{equation}
We now focus on the coherent part of the extractable work. To obtain it, we first observe that for a generic unitary $\hat{\mathcal{U}}$, the coherent part of the work is
\begin{widetext}
\begin{align}
    \mathcal{W}_{\mathcal{U}}[\hat{\chi}]&=\sum_{if}(E_i-E_f)q_{if}^{\mathcal{U}}(\hat{\chi})=\sum_{if} E_i \operatorname{Tr}\left[\hat{\mathcal{U}}^\dagger \hat{\Pi}_f \hat{\mathcal{U}}\hat{\Pi}_i \hat{\chi}\right] -\sum_f E_fq_{if}^{\mathcal{U}}(\hat{\chi})=\nonumber\\&=\sum_{i} E_i \operatorname{Tr}\left[\hat{\mathcal{U}}^\dagger \sum_f\hat{\Pi}_f \hat{\mathcal{U}}\hat{\Pi}_i \hat{\chi}\right]-\sum_f E_fq_{if}^{\mathcal{U}}(\hat{\chi})=\sum_{i} E_i \operatorname{Tr}\left[\hat{\Pi}_i \hat{\chi}\right]-\sum_f E_fq_{if}^{\mathcal{U}}(\hat{\chi})=-\sum_f E_fq_{if}^{\mathcal{U}}(\hat{\chi}).
\end{align}
In the specific case of the single-qubit rotation we have
\begin{equation}
\mathcal{W}_R [\hat{\chi}]=-\sum_f E_f \sum_i q_{if}^{R}(\hat{\chi})=-E\left[\left(q_{\downarrow\uparrow}^{R}(\hat{\chi}) + q_{\uparrow\uparrow}^{R}(\hat{\chi})\right)  -
\left(q_{\downarrow\downarrow}^{R}(\hat{\chi}) + q_{\uparrow\downarrow}^{R}(\hat{\chi})\right)\right]=
-2E\left(\mathrm{Re}q_{\downarrow\uparrow}^{R}(\hat{\chi}) - \mathrm{Re}q_{\uparrow\downarrow}^{R}(\hat{\chi})\right),
\end{equation}
\end{widetext}
where we have used the relations between the real and imaginary parts of the coherent parts of KDQs in Eqs.~\eqref{eq:Reqp_coh_1bt},\eqref{eq:qp_coh_1bt}. From Eq.~\eqref{eq:q_du} and Eq.~\eqref{eq:q_ud} we also note that $\mathrm{Re}q_{\uparrow\downarrow}^{R}(\hat{\chi})=-\mathrm{Re}q_{\downarrow\uparrow}^{R}(\hat{\chi})$, which implies
\begin{equation}
    \mathcal{W}_R [\hat{\chi}]=-4E\:\mathrm{Re}q_{\downarrow\uparrow}^{R}(\hat{\chi}).
\end{equation}
Thus, the coherent part of the work can be uniquely defined by the off-diagonal coherent KDQ $q_{\downarrow\uparrow}(\hat{\chi})$.

For completeness, we mention that the population part of the work reads
\begin{equation}
    \mathcal{W}_R [\Delta(\hat{\rho})]=2E\left(\mathrm{Re}q_{\uparrow\downarrow}^{R}(\Delta(\hat{\rho}))-\mathrm{Re}q_{\downarrow\uparrow}^{R}(\Delta(\hat{\rho})) \right);
    \label{eq:pop_work_app}
\end{equation}
by using Eqs.~\eqref{eq:q_du} and~\eqref{eq:q_ud}, we obtain
\begin{align}
& \textrm{Re} q_{\downarrow\uparrow}^R(\Delta(\hat{\rho}))=(1-p)|f|^2 \sin^2 \left(\frac{\theta}{2} \right), \\
& \textrm{Re} q_{\uparrow\downarrow}^R(\Delta(\hat{\rho}))=p|f|^2 \sin^2 \left(\frac{\theta}{2} \right),
\end{align}
which, inserted in Eq.~\eqref{eq:pop_work_app}, lead to 
\begin{align}
    \mathcal{W}_R [\Delta(\hat{\rho})]=2E|f|^2\sin^2(\theta/2) (2p-1);
\end{align}
This derivation confirms the negativity of extractable work from a incoherent state, in the absence of population inversion ($p\leq \frac{1}{2}$).

Since $\mathcal{W}_R [\Delta(\hat{\rho})]$ is always negative, positive work extraction is only possible if the coherent contribution, $\mathcal{W}_R [\hat{\chi}]$, is positive and large enough to outweigh the negative term from $\mathcal{W}_R [\Delta(\hat{\rho})]$. Thus, in the absence of population inversion, $\mathrm{Re}q_{\downarrow\uparrow}^{R}(\hat{\chi})<0$ (corresponding to $\mathcal{W}_R [\hat{\chi}]>0$) is a necessary but not sufficient condition for having positive work extraction overall. The necessary and sufficient condition is $\mathcal{W}_R [\hat{\chi}]>|\mathcal{W}_R [\Delta(\hat{\rho})]|$, which corresponds to
\begin{equation}
|\mathrm{Re}q_{\downarrow\uparrow}^{R}(\hat{\chi})|>\frac{1}{2}|f|^2\sin^2(\theta/2)(1-2p)
\quad
\textrm{with} \; \; \mathrm{Re}q_{\downarrow\uparrow}^{R}(\hat{\chi})<0.
\end{equation}

\section{Explicit derivation of: KDQs and work extraction in the Hadamard-like time evolution}\label{app:explicit_Hadamard-like}
The Hadamard gate $\hat{H}=\frac{1}{\sqrt{2}}\left(\hat{X}+\hat{Z}\right)$ can be implemented through the time evolution of a quantum system governed by an appropriate Hamiltonian $\hat{\mathcal{H}}_h=h\left(\hat{X}+\hat{Z}\right)$ (with $h>0$) until time $t_H =\frac{\pi}{2\sqrt{2}h}$. To understand why this is the case, let us write the Hamiltonian as $\hat{\mathcal{H}}_h=\omega\: \boldsymbol{n}_h\cdot\hat{\boldsymbol{\sigma}}$, with $\omega\coloneqq \sqrt{2}h$, $\boldsymbol{n}_h=\left(\frac{1}{\sqrt{2}},0,\frac{1}{\sqrt{2}}\right)$ and $\hat{\boldsymbol{\sigma}}=\left(\hat{X},\hat{Y},\hat{Z} \right)$. The unitary evolution operator is $\hat{\mathcal{U}}_h(t)=e^{-i(\omega t) \boldsymbol{n}_h\cdot\hat{\boldsymbol{\sigma}}}=\cos(\omega t)\hat{\mathds{1}}-i(\boldsymbol{n}_h\cdot\hat{\boldsymbol{\sigma}})\sin(\omega t)$. At time $t_H$, the latter unitary becomes 
\begin{equation}
\hat{\mathcal{U}}_{h}(t_H)\coloneqq\hat{\mathcal{U}}_H=-i\hat{H}, 
\end{equation}
and so it reproduces the Hadamard gate up to a global phase that does not affect the quasiprobabilities and, in general, the time evolution of the system. Thus, the action of the Hadamard gate corresponds to the action of the unitary
\begin{equation}
    \hat{\mathcal{U}}_H = e^{-i\hat{\mathcal{O}}_H},
    \quad
    \hat{\mathcal{O}}_H = \frac{\pi}{2\sqrt{2}} \left(\hat{X}+\hat{Z}\right).
\end{equation}
We now evaluate KDQs and extractable work for transformations governed by an Hadamard-like time evolution, namely an evolution governed by an Hamiltonian
$\hat{\mathcal{H}}_h=h\left(\hat{X}+\hat{Z}\right)$. Since $\hat{\mathcal{U}}_h(t)=e^{-i(\omega t) \boldsymbol{n}_h\cdot\hat{\boldsymbol{\sigma}}}$ is a rotation~\eqref{eq:rotation} of an angle $\theta=2\omega t$ around the unit vector $\boldsymbol{n}_h$, we can use Eq.~\eqref{eq:qp_2lev} and obtain
\begin{equation}
\begin{aligned}
& q^{\mathcal{U}_h(t)}_{\downarrow\downarrow}(\hat{\rho})=i|\gamma|e^{-i\varphi}\frac{1}{\sqrt{2}}g(\omega t)\sin(\omega t)+(1-p)|g(\omega t)|^2\\
& q^{\mathcal{U}_h(t)}_{\downarrow\uparrow}(\hat{\rho})=-i|\gamma|e^{-i\varphi}\frac{1}{\sqrt{2}}g(\omega t)\sin(\omega t)+(1-p)\frac{1}{2}\sin^2(\omega t)\\
& q^{\mathcal{U}_h(t)}_{\uparrow\downarrow}(\hat{\rho})=-i|\gamma|e^{i\varphi}\frac{1}{\sqrt{2}}g(\omega t)^*\sin(\omega t)+p\frac{1}{2}\sin^2(\omega t)\\
& q^{\mathcal{U}_h(t)}_{\uparrow\uparrow}(\hat{\rho})=i|\gamma|e^{i\varphi}\frac{1}{\sqrt{2}}g(\omega t)^*\sin(\omega t)+p|g(\omega t)|^2,
\end{aligned}
\label{eq:qp_Hdm}
\end{equation}
where we have used $f_{xy}=\frac{1}{\sqrt{2}}$ and $g_z(\theta)\coloneqq g(\omega t)=\cos(\omega t)+i\frac{1}{\sqrt{2}}\sin(\omega t)$. 

We can fix the time to $t_H$ and obtain
\begin{align}
&q^H_{\downarrow\downarrow}(\hat{\rho})=\frac{(1-p)-|\gamma|e^{-i\varphi}}{2},
\quad
q^H_{\downarrow\uparrow}(\hat{\rho})=\frac{(1-p)+|\gamma|e^{-i\varphi}}{2},\\
&q^H_{\uparrow\downarrow}(\hat{\rho})=\dfrac{p-|\gamma|e^{i\varphi}}{2},
\quad
q^H_{\uparrow\uparrow}(\hat{\rho})=\dfrac{p+|\gamma|e^{i\varphi}}{2}.
\quad
\end{align}
The latter result can be also derived by using $q_{if}^H(\hat{\rho})=\operatorname{Tr}\bigl[\hat{H}\hat{\Pi}_f\hat{H}\hat{\Pi}_i \hat{\rho}\bigr]$. The extractable work is
\begin{align}
    \mathcal{W}_H [\hat{\rho}]=&2E\left[(n_x^2+n_y^2)\sin^2(\theta/2)-2\mathrm{Re}q^{H}_{\downarrow\uparrow}(\hat{\rho})\right] =\nonumber\\&\underbrace{E(2p-1)}_{\mathcal{W}_H[\Delta(\hat{\rho})]} + \underbrace{(-2|\gamma|E\cos\varphi)}_{\mathcal{W}_H[\hat{\chi}]}, 
\end{align}
where we have used $n_x=\frac{1}{\sqrt{2}}$, $n_y=0$ and $\theta/2=\omega t_H=\pi/2$. $\mathcal{W}_H[\hat{\chi}]$ can be alternatively found by using $\mathcal{W}_H[\hat{\chi}]=-4E\:\mathrm{Re}q_{\downarrow\uparrow}^{H}(\hat{\chi})$. Positive extractable work is achieved for $\cos\varphi < 0$ and $|\cos\varphi|>\frac{1-2p}{2|\gamma|}$. We also observe that $\max_{\varphi}    \mathcal{W}_H [\hat{\rho}]=E_{\uparrow}(2p-1) +2E_{\uparrow}|\gamma|$ is obtained at $\varphi=\pi$ and it is positive for $2|\gamma|>1-2p$, which means that the coherence needs to be larger than the population imbalance between the ground $\ket{\downarrow}$ and excited state $\ket{\uparrow}$, up to a factor two. 

We now ask ourselves: under which conditions processes of the form $\ket{\downarrow}\rightarrow\ket{\uparrow}$ anomalously positively contribute to the extractable work? We need $\mathrm{Re}q_{\downarrow\uparrow}^H(\hat{\rho})<0$ and so $|\gamma||\cos\varphi|>1-p$ (with $\cos\varphi < 0$), meaning that the amplitude of the coherence corrected by an interference factor $\cos \varphi$ must dominate on the ground state population. However, to preserve the positivity of the state, the coherences must satisfy $|\gamma|\leq \sqrt{p(1-p)}$, which implies that $\mathrm{Re}q_{\downarrow\uparrow}^H(\hat{\rho})$ is negative when $\sqrt{p(1-p)}|\cos\varphi|\geq|\gamma||\cos\varphi|>1-p$ that cannot be satisfied for $p\leq \frac{1}{2}$. Thus, in the absence of population inversion, the aforementioned anomalous processes does not contribute to the work extraction through an Hadamard gate. This means that, for negative $\cos\varphi$, the coherences can suppress $\ket{\downarrow}\rightarrow\ket{\uparrow}$ processes, but not to the extent of assigning them a statistically negative weight.

At $t\neq t_H$, It is possible that the MHQ $\mathrm{Re} q_{\downarrow\uparrow}(\hat{\rho})$ takes negative values, which leads to anomalous positive contributions to the
extractable work. In fact, by using Eq.~\eqref{eq:qp_Hdm}, we obtain
\begin{widetext}
\begin{align}
\mathrm{Re} q^{\mathcal{U}_h(t)}_{\downarrow\uparrow}(\hat{\rho})=\underbrace{\frac{1-p}{2}\sin^2(\omega t)}_{\mathrm{Re} q^{\mathcal{U}_h(t)}_{\downarrow\uparrow}(\Delta(\hat{\rho}))}+\underbrace{\frac{|\gamma|}{\sqrt{2}}\sin(\omega t)\left[\frac{1}{\sqrt{2}}\cos\varphi \sin(\omega t) - \sin\varphi \cos(\omega t) \right]}_{\mathrm{Re} q^{\mathcal{U}_h(t)}_{\downarrow\uparrow}(\hat{\chi})},
\label{eq:Req_du^Ut}
\end{align}
\end{widetext}
which can be negative in the absence of population inversion. For example, the pure state with $p=|\gamma|=\sqrt{p(1-p)}=\frac{1}{2}$ and $\varphi=\frac{\pi}{2}$, at time $\omega t = \frac{\pi}{4}$, has $\mathrm{Re} q^{\mathcal{U}_h(t)}_{\downarrow\uparrow}(\hat{\rho})=\frac{1-\sqrt{2}}{8}<0$.

Given Eq.~\eqref{eq:Req_du^Ut}, we can compute the extractable work by using Eq.~\eqref{eq:W_R_1qbt} with $n_x=\frac{1}{\sqrt{2}}$, $n_y=0$ and $\theta/2=\omega t$:
\begin{widetext}
\begin{align}
\mathcal{W}_t[\hat{\rho}]=&2E\left[\frac{1}{2}\sin^2(\omega t)-2\mathrm{Re}q^{\mathcal{U}_h(t)}_{\downarrow\uparrow}(\hat{\rho})\right]=\underbrace{E(2p - 1)\sin^2(\omega t)}_{\mathcal{W}_t[\Delta(\hat{\rho})]} + \underbrace{\frac{4E |\gamma|}{\sqrt{2}}\left[ \sin\varphi \cos(\omega t)\sin(\omega t) - \frac{1}{\sqrt{2}}\cos\varphi \sin^2(\omega t) \right]}_{\mathcal{W}_t[\hat{\chi}]}.
    \label{eq:W(t)}
\end{align}
\end{widetext}
Finally, for completeness, we also compute the MHQ associated to the processes $\ket{\uparrow}\rightarrow\ket{\downarrow}$ that is detrimental for the work extraction when negative. In order to do that, we use Eq.~\eqref{eq:q_updndnup} with $|f|^2=n_x^2=\frac{1}{2}$ and $\theta/2=\omega t$, and we obtain
\begin{widetext}
\begin{align}
\mathrm{Re} q^{\mathcal{U}_h(t)}_{\uparrow\downarrow}(\hat{\rho})=&\frac{p}{2}\sin^2(\omega t)-\frac{|\gamma|}{\sqrt{2}}\sin(\omega t)\left[\frac{1}{\sqrt{2}}\cos\varphi \sin(\omega t) - \sin\varphi \cos(\omega t) \right].
\end{align}    
\end{widetext}

\section{Decomposition of KDQs of deep quantum circuits: explicit derivation}\label{app:decomposition}
Let us consider a cyclic transformation with initial/final Hamiltonian $\hat{\mathcal{H}}=\sum_k E_k \hat{\Pi}_k$ in which the input quantum state $\hat{\rho}$ undergoes a unitary transformation $\hat{\mathcal{U}}$ made of $N$ consecutive gates, that we identify as constituent gates such that $\hat{\mathcal{U}}=\hat{U}_N\ldots\hat{U}_1$. We want to establish a relation between the full KDQ 
\begin{equation}
    q_{if}^{\mathcal{U}}(\hat{\rho})=q_{if}^{U_N \ldots U_1}(\hat{\rho})=\operatorname{Tr}\bigl[ \bigl( \hat{U}_N\ldots\hat{U}_1\bigr)^{\dagger}\hat{\Pi}_f\bigl(\hat{U}_N\ldots\hat{U}_1\bigr)\hat{\Pi}_i\hat{\rho} \bigr]
    \label{eq:qp_deep}
\end{equation}
and those associated to the gates $\hat{U}_j$ that decompose $\hat{\mathcal{U}}$. For instance, if we are interested in the second gate, the quasiprobability considered will be $q_{if}^{U_2}\bigl(\hat{U}_1\hat{\rho}\hat{U}_1^{\dagger}\bigr)$, that is the KDQ associated to the unitary $\hat{U}_2$, applied to the state that has been already evolved by $\hat{U}_1$. More generally, we want to establish a relation between $q_{if}^{\mathcal{U}}(\hat{\rho})$ and the KDQs associated to the unitary $\hat{U}_{j+1}$ applied to the state that has been already evolved by the application of $\hat{U}_j \ldots \hat{U}_1$, namely $q_{if}^{U_{j+1}}((\hat{U}_j \ldots \hat{U}_1)\hat{\rho}(\hat{U}_j \ldots \hat{U}_1)^{\dagger})\coloneqq q_{if}^{U_{j+1}}(\hat{\rho}_j)$. We consider $j=0,\ldots,N-1$, with $\hat{\rho}_{j=0}=\hat{\rho}$, and identify three main cases:
\begin{enumerate}[label=\roman*)]
    \item $j=1,\ldots,N-2$: the KDQ of the constituent gate is $q_{if}^{U_{j+1}}(\hat{\rho}_j)$; the transformation $\hat{U}_{j+1}$ applied to the state $\hat{\rho}_j=\bigl(\hat{U}_j\ldots\hat{U}_1\bigr)\hat{\rho}\bigl(\hat{U}_j\ldots\hat{U}_1\bigr)^{\dagger} \neq \hat{\rho}$ is followed by at least one gate $\hat{U}_{j+2}$; $\hat{U}_{j+1}$ is not on the border of $\hat{\mathcal{U}}$.
    \item $j=0$: the constituent gate KDQs are of the form $q_{if}^{U_{1}}(\hat{\rho})$, that are those associated with the first transformation $\hat{U}_1$ applied to the state $\hat{\rho}$; $\hat{U}_1$ is followed by at least another gate $\hat{U}_{2}$.
    \item $j=N-1$: the constituent gate KDQs are of the form $q_{if}^{U_{N}}(\hat{\rho}_{N-1})$, that are those associated with the last transformation $\hat{U}_N$ applied to the state $\hat{\rho}_{N-1}$; since $\hat{U}_{N}$ is the last gate, it is not followed by any other transformation.
\end{enumerate}
The three cases are pictorially represented in Fig.~\ref{fig:nonclassical_sketch}(a) of the main text. We first consider the case (i) and write the KDQ associated to the constituent gate as
\begin{align}
    q_{if}^{U_{j+1}}(\hat{\rho}_j)&=\operatorname{Tr}\bigl[  \hat{U}_{j+1}^{\dagger}\hat{\Pi}_f\hat{U}_{j+1}\hat{\Pi}_i\bigl(\hat{U}_j\ldots\hat{U}_1\bigr)\hat{\rho}\bigl(\hat{U}_j\ldots\hat{U}_1\bigr)^{\dagger} \bigr]\nonumber\\&=\operatorname{Tr}\bigl[ \bigl(\hat{U}_{j+1}\ldots\hat{U}_1\bigr)^{\dagger} \hat{\Pi}_f\hat{U}_{j+1}\hat{\Pi}_i\bigl(\hat{U}_j\ldots\hat{U}_1\bigr)\hat{\rho}\bigr];
\end{align}
we insert the identity $\hat{\mathds{1}}=\bigl(\hat{U}_N\ldots\hat{U}_{j+2}\bigr)^{\dagger}\bigl(\hat{U}_N\ldots\hat{U}_{j+2}\bigr)$ and we obtain
\begin{widetext}
\begin{equation}
    q_{if}^{U_{j+1}}(\hat{\rho}_j)=\operatorname{Tr}\bigl[ \bigl(\hat{U}_{N}\ldots\hat{U}_1\bigr)^{\dagger} \bigl(\hat{U}_N\ldots\hat{U}_{j+2}\bigr) \hat{\Pi}_f\hat{U}_{j+1}\hat{\Pi}_i\bigl(\hat{U}_j\ldots\hat{U}_1\bigr)\hat{\rho}\bigr].
\end{equation}
\end{widetext}
We now exchange $\bigl(\hat{U}_N\ldots\hat{U}_{j+2}\bigr)$ with $\hat{\Pi}_f$ and $\hat{\Pi}_i$ with $\bigl(\hat{U}_j\ldots\hat{U}_1\bigr)$ by paying two commutators and, by comparing the result with Eq.~\eqref{eq:qp_deep}, we obtain
\begin{widetext}
\begin{align}
    &q_{if}^{\mathcal{U}}(\hat{\rho}) - q_{if}^{U_{j+1}}(\hat{\rho}_j) = \operatorname{Tr} \bigl[ \hat{\mathcal{M}}_{if}^{U_{j+1}} \hat{\rho} \bigr].\nonumber\\
    &\hat{\mathcal{M}}_{if}^{U_{j+1}}\coloneqq   \bigl(\hat{U}_{N}\ldots\hat{U}_1\bigr)^{\dagger} \bigl(\bigl[ \hat{\Pi}_f,\hat{U}_{N}\ldots\hat{U}_{j+2} \bigr]\hat{U}_{j+1}\hat{\Pi}_i \hat{U}_{j}\ldots\hat{U}_1 - 
    \hat{\Pi}_f\hat{U}_{N}\ldots\hat{U}_{j+1}\bigl[\hat{\Pi}_i, \hat{U}_{j}\ldots\hat{U}_1 \bigr]
    \bigr).
    \label{eq:single_center}
\end{align}
\end{widetext}
Secondly, we consider the case (ii), in which the constituent gate KDQ is the one associated to the the first transformation:
\begin{align}
    q_{if}^{U_1}(\hat{\rho})&=\operatorname{Tr}\bigl[\hat{U}_1^{\dagger}\hat{\Pi}_f \hat{U}_1 \hat{\Pi}_i \hat{\rho} \bigr]\nonumber\\&=\operatorname{Tr}\bigl[\bigl(\hat{U}_N \ldots\hat{U}_1\bigr)^{\dagger} \bigl(\hat{U}_N \ldots\hat{U}_2\bigr)\hat{\Pi}_f \hat{U}_1 \hat{\Pi}_i \hat{\rho} \bigr],
\end{align}

we exchange $\bigl(\hat{U}_N \ldots\hat{U}_2\bigr)$ with $\hat{\Pi}_f$ by paying a commutator and, by comparing the result with Eq.~\eqref{eq:qp_deep}, we get
\begin{align}
    &q_{if}^{\mathcal{U}}(\hat{\rho}) - q_{if}^{U_1}(\hat{\rho})=\operatorname{Tr}\left[\hat{\mathcal{M}}_{if}^{U_1}\hat{\rho} \right].
    \nonumber\\ 
    &\hat{\mathcal{M}}_{if}^{U_1}\coloneqq \bigl(\hat{U}_N \ldots\hat{U}_1\bigr)^{\dagger} \bigl[ \hat{\Pi}_f, \hat{U}_N \ldots\hat{U}_2 \bigr]\hat{U}_1\hat{\Pi}_i 
    \label{eq:single_initial}
\end{align}
Finally, we consider the case (iii), in which the constituent gate KDQ is the one associated to the the last transformation:
\begin{widetext}
    \begin{align}
    q_{if}^{U_N}(\hat{\rho}_{N-1})=\operatorname{Tr}\bigl[\hat{U}_N^{\dagger}\hat{\Pi}_f \hat{U}_N \hat{\Pi}_i \bigl(\hat{U}_{N-1}\ldots\hat{U}_1\bigr)\hat{\rho}\bigl(\hat{U}_{N-1}\ldots\hat{U}_1\bigr)^{\dagger} \bigr] = \operatorname{Tr}\bigl[\bigl(\hat{U}_{N}\ldots\hat{U}_1\bigr)^{\dagger}\hat{\Pi}_f \hat{U}_N \hat{\Pi}_i \bigl(\hat{U}_{N-1}\ldots\hat{U}_1\bigr)\hat{\rho} \bigr];
\end{align}
\end{widetext}

we exchange $\hat{\Pi}_i$ with $\bigl(\hat{U}_{N-1}\ldots\hat{U}_1\bigr)$ by paying a commutator and, by comparing the result with Eq.~\eqref{eq:qp_deep}, we get
\begin{align}
     &q_{if}^{\mathcal{U}}(\hat{\rho}) - q_{if}^{U_N}(\hat{\rho}_{N-1}) =  \operatorname{Tr}\bigl[\hat{\mathcal{M}}_{if}^{U_N}\hat{\rho} \bigr], \nonumber \\
     & \hat{\mathcal{M}}_{if}^{U_N}\coloneqq -\bigl(\hat{U}_{N}\ldots\hat{U}_1\bigr)^{\dagger}\hat{\Pi}_f \hat{U}_N \bigl[ \hat{\Pi}_i, \hat{U}_{N-1}\ldots\hat{U}_1\bigr].
          \label{eq:single_final}
\end{align}
We can compactly write Eqs.~\eqref{eq:single_center},~\eqref{eq:single_initial},~\eqref{eq:single_final} by defining the $\hat{\mathcal{M}}_{if}^{U_{j+1}}$ operator as
\begin{widetext}
    \begin{align}
&q_{if}^{\mathcal{U}}(\hat{\rho}) - q_{if}^{U_{j+1}}(\hat{\rho}_j) = \operatorname{Tr} \bigl[ \hat{\mathcal{M}}_{if}^{U_{j+1}} \hat{\rho} \bigr],
\label{eq:decomposition}\\
&\hat{\mathcal{M}}_{if}^{U_{j+1}}=
    \begin{cases}
        \bigl(\hat{U}_N \ldots\hat{U}_1\bigr)^{\dagger} \bigl[ \hat{\Pi}_f, \hat{U}_N \ldots\hat{U}_2 \bigr]\hat{U}_1\hat{\Pi}_i  \; \; \;\; \textrm{if} \;\; j=0,\\
        \bigl(\hat{U}_{N}\ldots\hat{U}_1\bigr)^{\dagger} \bigl(\bigl[ \hat{\Pi}_f,\hat{U}_{N}\ldots\hat{U}_{j+2} \bigr]\hat{U}_{j+1}\hat{\Pi}_i \hat{U}_{j}\ldots\hat{U}_1 - 
    \hat{\Pi}_f\hat{U}_{N}\ldots\hat{U}_{j+1}\bigl[\hat{\Pi}_i, \hat{U}_{j}\ldots\hat{U}_1 \bigr]
    \bigr) \; \; \;\; \textrm{if} \;\; j=1,\ldots,N-2,\\
    -\bigl(\hat{U}_{N}\ldots\hat{U}_1\bigr)^{\dagger}\hat{\Pi}_f \hat{U}_N \bigl[ \hat{\Pi}_i, \hat{U}_{N-1}\ldots\hat{U}_1\bigr] \; \; \;\; \textrm{if} \;\; j=N-1,
    \end{cases}
    \label{eq:M_operator}
\end{align}
\end{widetext}
The message contained in the latter equation is the following: the difference between the full KDQ and the one associated to a specific constituent gate depends on the incompatibility between the remaining gates and the projectors of the initial/final Hamiltonian. Specifically, it depends on the incompatibility between the gates that precede the constituent gate and the projectors on the initial Hamiltonian together with the incompatibility between the gates that follow the constituent gate and the projectors on the final Hamiltonian. The equation is pictorially represented in Fig.~\ref{fig:nonclassical_sketch}(b) of the main text.

We can render Eqs.~\eqref{eq:single_center},~\eqref{eq:single_initial},~\eqref{eq:single_final} more compact by summing their LHS and RHS; regarding Eq.~\eqref{eq:single_center}, we consider all the possible cases, namely $j=1,\ldots,N-2$. We have
\begin{equation}
    q_{if}^{\mathcal{U}}(\hat{\rho})=\frac{1}{N}\sum_{j=0}^{N-1} q_{if}^{U_{j+1}}(\hat{\rho}_j)+\frac{1}{N}\mathcal{Q}_{if}^{\mathcal{U}}(\hat{\rho}),
    \label{eq:tot_and_correction}
\end{equation}
with $\mathcal{Q}_{if}^{\mathcal{U}}(\hat{\rho})\coloneqq\frac{1}{N}\sum_{j=0}^{N-1} \operatorname{Tr}\Bigl[\hat{\mathcal{M}}_{if}^{U_{j+1}} \hat{\rho} \Bigr]$. Thus, the full KDQ can be written as the weighted sum of the KDQs associated to all the single gates that decompose $\hat{\mathcal{U}}$, corrected by a term, $\mathcal{Q}_{if}^{\mathcal{U}}(\hat{\rho})$, depending on the incompatibility between Hamiltonian projectors and combined constituent gates.

The simplest case is the one in which the full transformation is decomposed in two gates, that we name $\hat{U}$ and $\hat{V}$. The full circuit is $\hat{\mathcal{U}}=\hat{V}\hat{U}$ and the evolved state is $\hat{\rho}_{VU}=\hat{V}\hat{U} \hat{\rho} (\hat{V}\hat{U})^{\dagger}$. By using Eq.~\eqref{eq:single_final}, we obtain
\begin{equation}
    q_{if}^{VU}(\hat{\rho})-q_{if}^V (\hat{\rho}_U) = -\operatorname{Tr}[(\hat{V}\hat{U})^{\dagger}\hat{\Pi}_f \hat{V}[\hat{\Pi}_i ,\hat{U}]\hat{\rho}].
    \label{eq:-V}
\end{equation}   
At the same time, by using Eq.~\eqref{eq:single_initial}, we obtain
\begin{equation}
    q_{if}^{VU}(\hat{\rho})-q_{if}^U (\hat{\rho})=\operatorname{Tr}[ (\hat{V}\hat{U})^{\dagger}[\hat{\Pi}_f,\hat{V}] \hat{U}\hat{\Pi}_i \hat{\rho}].
    \label{eq:-U}
\end{equation}
The compatibility between initial projectors and gate $\hat{U}$ reduces the full KDQ to the one associated to gate $\hat{V}$, and, vice versa, the compatibility between final projectors and the gate $\hat{V}$ reduces the full KDQ to the one associated to gate $\hat{U}$. If gates and projectors are not compatible, the relation between the full KDQ and those of the constituent gates is not trivial and reads
\begin{equation}
    q_{if}^{VU}(\hat{\rho})=\frac{q_{if}^U(\hat{\rho})+q_{if}^V(\hat{\rho}_U)}{2}+\frac{1}{2}\mathcal{Q}_{if}^{VU}(\hat{\rho}),
    \label{eq:_qp_and_correction_}
\end{equation} 
where the correction factor is $\mathcal{Q}^{VU}_{if}(\hat{\rho})\coloneqq \operatorname{Tr}\left[(\hat{V}\hat{U})^{\dagger}\left([\hat{\Pi}_f,\hat{V}]\hat{U}\hat{\Pi}_i -\hat{\Pi}_f\hat{V}[\hat{\Pi}_i,\hat{U}] \right)\hat{\rho}\right]$.

Another interesting case is the one in which the full circuit is decomposed into three gates. We first use Eq.~\eqref{eq:single_center} and get
    \begin{align}
q_{if}^{MVU}(\hat{\rho})-q_{if}^V (\hat{\rho}_U) = & \operatorname{Tr}[(\hat{M
       }\hat{V}\hat{U})^{\dagger}[\hat{\Pi}_f,\hat{M}] \hat{V}\hat{\Pi}_i \hat{U}\hat{\rho}]-\nonumber \\&\operatorname{Tr}[(\hat{M
       }\hat{V}\hat{U})^{\dagger}\hat{\Pi}_f\hat{M}\hat{V}[\hat{\Pi}_i, \hat{U}]\hat{\rho}],
       \label{eq:MVU-V}
\end{align} 

while using Eq.~\eqref{eq:single_initial} and Eq.~\eqref{eq:single_final}, we obtain, respectively,
\begin{equation}
    q_{if}^{MVU}(\hat{\rho})-q_{if}^U (\hat{\rho})=\operatorname{Tr}[(\hat{M
       }\hat{V}\hat{U})^{\dagger}[\hat{\Pi}_f,\hat{M
       }\hat{V}] \hat{U}\hat{\Pi}_i \hat{\rho}]
              \label{eq:MVU-U}
\end{equation}
and
\begin{align}
        q_{if}^{MVU}(\hat{\rho})-q_{if}^M (\hat{\rho}_{UV})=-\operatorname{Tr}\left[(\hat{M}\hat{V}\hat{U})^{\dagger} \hat{\Pi}_f \hat{M }[\hat{\Pi}_i, \hat{V}\hat{U}]\hat{\rho}\right].
        \label{eq:MVU-M}
\end{align}    
Thus, by summing Eqs.~\eqref{eq:MVU-U},~\eqref{eq:MVU-V}~\eqref{eq:MVU-M}, we obtain the relation between the total KDQ and that of the constituent gates:
\begin{equation}
q_{if}^{MVU}(\hat{\rho})=\frac{q_{if}^{U}(\hat{\rho})+q_{if}^{V}(\hat{\rho}_U)+q_{if}^{M}(\hat{\rho}_{UV})}{3}+\frac{1}{3}\mathcal{Q}_{if}^{MVU}(\hat{\rho}),  
\end{equation}
where the correction is now given by
\begin{widetext}
    \begin{align}
\mathcal{Q}_{if}^{MVU}(\hat{\rho})\coloneqq\operatorname{Tr}[(\hat{M
       }\hat{V}\hat{U})^{\dagger}[\hat{\Pi}_f,\hat{M
       }\hat{V}] \hat{U}\hat{\Pi}_i \hat{\rho}] + \nonumber\operatorname{Tr}[(\hat{M
       }\hat{V}\hat{U})^{\dagger}([\hat{\Pi}_f,\hat{M}] \hat{V}\hat{\Pi}_i \hat{U}-\nonumber\hat{\Pi}_f\hat{M}\hat{V}[\hat{\Pi}_i, \hat{U}])\hat{\rho}] -\nonumber \operatorname{Tr}[ (\hat{M}\hat{V}\hat{U})^{\dagger}\hat{\Pi}_f\hat{M}[\hat{\Pi}_i,\hat{V}\hat{U}] \hat{\rho}].
\end{align}
\end{widetext}

\section{KDQs of two qubit-gates: detailed analysis}
\label{app:twoqb_gates}
In this section, we provide a in-depth analysis of the features of the KDQs for two-qubit gates. In a two-qubit system the initial/final Hamiltonian is $\hat{\mathcal{H}}=E\bigl(\hat{Z}\otimes\hat{\mathds{1}} + \hat{\mathds{1}}\otimes \hat{Z} \bigr)$. Labelling the eigenstates as $\{\ket{E_0},\ket{E_1}, \ket{E_2}, \ket{E_3} \}$, with associated eigenergies $\{E_0 = -2E, E_1 = 0, E_2 = 0, E_3=2E\}$, (see main text), a generic two-qubit input state can be written as
\begin{align}
    \hat{\rho}&=\sum_{\alpha ', \alpha '', \beta ', \beta ''=\uparrow\downarrow}\lambda_{(\alpha ' \beta ')(\alpha '' \beta '')} \ket{\alpha '}\bra{\alpha ''}\otimes\ket{\beta '}\bra{\beta ''}\nonumber\\
    &=\sum_{m,l=1}^4 \lambda_{ml}\ket{E_m}\bra{E_l}.
    \label{eq:rho2qb_state_app}
\end{align}
The extractable work through a generic two-qubit unitary cyclic transformation $\hat{\mathcal{U}}$ is, then,
\begin{align}
    \mathcal{W}_\mathcal{U}[\hat{\rho}]=&\sum_{i\neq f}q^\mathcal{U}_{if} (E_i - E_f)=\nonumber\\&2E\bigl[(q_{10}^\mathcal{U}-q_{01}^\mathcal{U}) + (q_{20}^\mathcal{U}-q_{02}^\mathcal{U}) + 2(q_{30}^\mathcal{U}-q_{03}^\mathcal{U}) +\nonumber\\& (q_{31}^\mathcal{U}-q_{13}^\mathcal{U}) + (q_{32}^\mathcal{U}-q_{23}^\mathcal{U})\bigr].
\end{align}
It is convenient to group the 10 (out of 16) KDQs contributing to the work extraction in the vector 
\begin{equation}
\boldsymbol{q}^\mathcal{U}=(q_{01}^\mathcal{U}, q_{02}^\mathcal{U}, q_{03}^\mathcal{U},q_{13}^\mathcal{U}, q_{23}^\mathcal{U}, q_{10}^\mathcal{U}, q_{20}^\mathcal{U},q_{30}^\mathcal{U},q_{31}^\mathcal{U},q_{32}^\mathcal{U}).    
\end{equation}

Let us now recall that any unitary operation can be approximated to arbitrary accuracy by the set of gates $\bigl\{\hat{H},\hat{T},\hat{U}_{\rm CNOT}\}$, where $\hat{U}_{\rm CNOT}=\ket{E_3}\bra{E_3}+\ket{E_2}\bra{E_2}+\ket{E_1}\bra{E_0}+\ket{E_0}\bra{E_1}$ is the controlled NOT gate that flips (keeps unchanged) the second spin if the first one is in the state $\ket{\downarrow}$ ($\ket{\uparrow}$). Among the gates in the aforementioned set, the CNOT gate is the only one that acts on two qubits and can generate entanglement.
The action of the CNOT gate on the projectors $\hat{\Pi}_{3}$ and $\hat{\Pi}_{2}$ keeps them unchanged and exchanges the other two: $\hat{U}_{\rm CNOT}\hat{\Pi}_{1}\hat{U}_{\rm CNOT}=\hat{\Pi}_{0}$ and $\hat{U}_{\rm CNOT}\hat{\Pi}_{0}\hat{U}_{\rm CNOT}=\hat{\Pi}_{1}$ ($\hat{U}_{\rm CNOT}^{\dagger}=\hat{U}_{\rm CNOT}$). Thus, the only nonzero KDQs are
\begin{align}
&q_{33}^{\rm CNOT}(\hat{\rho})=\operatorname{Tr}\bigl[\hat{\Pi}_{3}\hat{\rho} \bigr]=\lambda_{33}\in \mathbb{R}^+, 
 \nonumber\\
&q_{22}^{\rm CNOT}(\hat{\rho})=\operatorname{Tr}\bigl[\hat{\Pi}_{2}\hat{\rho} \bigr]=\lambda_{22}\in \mathbb{R}^+, \nonumber\\
&q_{01}^{\rm CNOT}(\hat{\rho})=\operatorname{Tr}\bigl[\hat{\Pi}_{0}\hat{\rho} \bigr]=\lambda_{00}\in \mathbb{R}^+,\nonumber
\\
&q_{10}^{\rm CNOT}(\hat{\rho})=\operatorname{Tr}\bigl[\hat{\Pi}_{1}\hat{\rho} \bigr]=\lambda_{11}\in \mathbb{R}^+,
\end{align}
which are real and not affected by possible coherences of the initial state. Among these non-zero KDQs, only $q_{01}^{\rm CNOT}$ and $q_{10}^{\rm CNOT}$ contribute to the extractable work. Albeit generating quantum correlations between different quantum states, from the thermodynamic point of view, the CNOT gate has classical features given by a real and positive distribution of the work.

The CNOT gate is the only entangling two-qubit gate of the set $\{\hat{H},\hat{T},\hat{U}_{\rm CNOT}\}$, thus, the remaining gates that decompose a generic two-qubit transformation are (bi-)local and have the form $\hat{U}\otimes \hat{V}$. Let us first consider the case in which the input state is factorized $\hat{\rho}=\hat{\sigma}\otimes \hat{\tau}$, $\hat{\sigma}$ and $\hat{\tau}$ being generic one-qubit states. The KDQ reads
\begin{align}
    q_{if}^{U\otimes V}(\hat{\sigma}\otimes\hat{\tau})=&q_{(\alpha_i\beta_i)(\alpha_f\beta_f)}^{U\otimes V}(\hat{\sigma}\otimes\hat{\tau})=\nonumber\\&\mathrm{Tr}\bigl[\bigl(\hat{U}\otimes \hat{V}\bigr)^{\dagger}\hat{\Pi}_{\alpha_f\beta_f} \hat{U}\otimes \hat{V} \hat{\Pi}_{\alpha_i\beta_i} (\hat{\sigma}\otimes\hat{\tau})  \bigr],
\end{align}
where $\alpha_k, \beta_k$ are uniquely defined by the value of $k$ as in the the table~\ref{tab:alphabeta}.
\begin{table}[ht]
\centering
\begin{tabular}{ | c | c| c | } 
  \hline
  $k$& $\alpha_k$ & $\beta_k$ \\ 
  \hline
  0 & $\downarrow$ & $\downarrow$ \\ 
  \hline
  1 & $\downarrow$ & $\uparrow$ \\ 
  \hline
  2 & $\uparrow$ & $\downarrow$ \\ 
  \hline
   3 & $\uparrow$ & $\uparrow$ \\ 
  \hline
\end{tabular}
\caption{Association of the indices $\alpha_k$ and $\beta_k$ to any index $k$.}
\label{tab:alphabeta}
\end{table}

We now observe that $\hat{\Pi}_{\alpha_k\beta_k}=\hat{\Pi}_{\alpha_k}\otimes\hat{\Pi}_{\beta_k}$ and we obtain
\begin{equation}
  q_{(\alpha_i\beta_i)(\alpha_f\beta_f)}^{U\otimes V}(\hat{\sigma}\otimes\hat{\tau})=q_{\alpha_i\alpha_f}^{U}(\hat{\sigma})q_{\beta_i\beta_f}^{V}(\hat{\tau});
  \label{eq:qp_factor}
\end{equation}
namely, that the two-qubit KDQ is the product of the KDQs of the single transformations. Thus, from Eq.~\eqref{eq:qp_factor}, it follows that
\begin{align}
    & q_{01}^{U\otimes V}(\hat{\sigma}\otimes\hat{\tau})=q_{\downarrow\downarrow}^U(\hat{\sigma})q_{\downarrow\uparrow}^V(\hat{\tau}),\nonumber \\&
    q_{10}^{U\otimes V}(\hat{\sigma}\otimes\hat{\tau})=q_{\downarrow\downarrow}^U(\hat{\sigma})q_{\uparrow\downarrow}^V(\hat{\tau}), \nonumber\\
    & q_{02}^{U\otimes V}(\hat{\sigma}\otimes\hat{\tau})=q_{\downarrow\uparrow}^U(\hat{\sigma})q_{\downarrow\downarrow}^V(\hat{\tau}), \nonumber \\ &
    q_{20}^{U\otimes V}(\hat{\sigma}\otimes\hat{\tau})=q_{\uparrow\downarrow}^U(\hat{\sigma})q_{\downarrow\downarrow}^V(\hat{\tau}),\nonumber \\
    &q_{03}^{U\otimes V}(\hat{\sigma}\otimes\hat{\tau})=q_{\downarrow\uparrow}^U(\hat{\sigma})q_{\downarrow\uparrow}^V(\hat{\tau}),\nonumber \\&
    q_{30}^{U\otimes V}(\hat{\sigma}\otimes\hat{\tau})=q_{\uparrow\downarrow}^U(\hat{\sigma})q_{\uparrow\downarrow}^V(\hat{\tau}),\nonumber \\
    &q_{13}^{U\otimes V}(\hat{\sigma}\otimes\hat{\tau})=q_{\downarrow\uparrow}^U(\hat{\sigma})q_{\uparrow\uparrow}^V(\hat{\tau}),\nonumber \\& 
    q_{31}^{U\otimes V}(\hat{\sigma}\otimes\hat{\tau})=q_{\uparrow\downarrow}^U(\hat{\sigma})q_{\uparrow\uparrow}^V(\hat{\tau}), \nonumber\\
    &q_{23}^{U\otimes V}(\hat{\sigma}\otimes\hat{\tau})=q_{\uparrow\uparrow}^U(\hat{\sigma})q_{\downarrow\uparrow}^V(\hat{\tau}),\nonumber \\&
    q_{32}^{U\otimes V}(\hat{\sigma}\otimes\hat{\tau})=q_{\uparrow\uparrow}^U(\hat{\sigma})q_{\uparrow\downarrow}^V(\hat{\tau}). 
    \label{eq:factorization_app}
\end{align}
Since sequences of Hadamard gates satisfy $\hat{H}^{2n} = \hat{\mathds{1}}$ and $\hat{H}^{2n+1} = \hat{H}$ for integer $n$, and powers of the $\hat{T}$ gate generate phase gates of the form $\hat{P}_{\phi} = \ket{\uparrow}\bra{\uparrow} + e^{i\phi}\ket{\downarrow}\bra{\downarrow}$ (with the identity $\hat{\mathds{1}}$ corresponding to $\hat{P}_0$), any factorized two-qubit gate $\hat{U}\otimes\hat{V}$ can be expressed as a sequence of gates in each of which either a Hadamard or $\hat{P}_{\phi}$ gates act on both qubits: $\hat{H}\otimes \hat{H}$, $\hat{H}\otimes \hat{P}_{\phi}$, $\hat{P}_{\phi} \otimes \hat{H}$ and $\hat{P}_{\phi} \otimes \hat{P}_{\phi}$.

Thus, the KDQs of non-entangling gates can be decomposed in those of their constituent gates whose possible structure is described by the $2 \times 20$ matrix
\begin{equation}
    \boldsymbol{Q}(\hat{\rho})=
    \begin{bmatrix}
    \boldsymbol{q}^{P_{\phi}\otimes P_{\phi}}(\hat{\rho}) & \boldsymbol{q}^{P_{\phi}\otimes H}(\hat{\rho}) \\
        \boldsymbol{q}^{H\otimes P_{\phi}}(\hat{\rho}) & \boldsymbol{q}^{H\otimes H}(\hat{\rho}) \\
    \end{bmatrix}.
    \label{eq:qp_matrix}
\end{equation}
We now consider a factorized input state $\hat{\rho}=\hat{\sigma}\otimes \hat{\tau}$ and observe that 
$\hat{P}_{\phi}$ commutes with both $\hat{\Pi}_{\uparrow}$ and $\hat{\Pi}_{\downarrow}$ and so we get the trivial KDQs
\begin{equation}
q_{\downarrow\uparrow}^{P_{\phi}}(\hat{\sigma})=q_{\uparrow\downarrow}^{P_{\phi}}(\hat{\sigma})=0, 
    \quad
    q_{\downarrow\downarrow}^{P_{\phi}}(\hat{\sigma})=1-p, \quad 
    q_{\uparrow\uparrow}^{P_{\phi}}(\hat{\sigma})=p,
    \label{eq:qp_Pgate}
\end{equation}
for any single-qubit state $\hat{\sigma}$. Since the KDQs $q_{if}^{U\otimes V}(\hat{\sigma}\otimes\hat{\tau})$ depend on at least one of the two off-diagonal single qubit KDQs ($q_{\uparrow\downarrow}$ and/or $q_{\downarrow\uparrow}$), see Eq.~\eqref{eq:factorization_app}, for transformations in which gates with $\hat{P}_{\phi}$ structure act on both qubits, all the $10$ KDQs of interest are identically zero. This result simplifies the matrix~\eqref{eq:qp_matrix}, which, for transformations applied to factorized states, turns out to be
\begin{equation}
    \boldsymbol{Q}(\hat{\sigma}\otimes \hat{\tau})=
    \begin{bmatrix}
    0 & \boldsymbol{q}^{P_{\phi}\otimes H}(\hat{\sigma}\otimes \hat{\tau}) \\
        \boldsymbol{q}^{H\otimes P_{\phi}}(\hat{\sigma}\otimes \hat{\tau}) & \boldsymbol{q}^{H\otimes H}(\hat{\sigma}\otimes \hat{\tau}) \\
    \end{bmatrix}.
    \label{eq:Matrix_simplified}
\end{equation}
For transformations of the type $\hat{H}\otimes \hat{P}_{\phi}$, the full KDQs can be written as 
\begin{widetext}
\begin{equation}
  q_{(\alpha_i\beta_i)(\alpha_f\beta_f)}^{H\otimes P_{\phi}}(\hat{\sigma}\otimes\hat{\tau})=q_{\alpha_i\alpha_f}^{H}(\hat{\sigma})q_{\beta_i\beta_f}^{P_{\phi}}(\hat{\tau})=
  \begin{cases}
      0 \,\,\, \textrm{if} \,\,\, \beta_i \neq \beta_f \\
      p \,q_{\alpha_i\alpha_f}^H(\hat{\sigma}) \,\,\, \textrm{if} \,\,\, \beta_i =\beta_f=\uparrow, \\
      (1-p)\,q_{\alpha_i\alpha_f}^H(\hat{\sigma}) \,\,\, \textrm{if} \,\,\, \beta_i =\beta_f=\downarrow,\\
  \end{cases}
  \label{eq:qp_factor_HP}
\end{equation}
where we have used Eqs.~\eqref{eq:qp_factor},~\eqref{eq:qp_Pgate} and used the table~\ref{tab:alphabeta} as reference. At the same time, we can use the same procedure for transformations of the type $\hat{P}_{\phi}\otimes \hat{H}$, and we obtain
\begin{equation}
  q_{(\alpha_i\beta_i)(\alpha_f\beta_f)}^{P_{\phi}\otimes H}(\hat{\sigma}\otimes\hat{\tau})=q_{\alpha_i\alpha_f}^{P_{\phi}}(\hat{\sigma})q_{\beta_i\beta_f}^{H}(\hat{\tau})=
  \begin{cases}
      0 \,\,\, \textrm{if} \,\,\, \alpha_i \neq \alpha_f \\
      p \,q_{\beta_i\beta_f}^H(\hat{\tau}) \,\,\, \textrm{if} \,\,\, \alpha_i =\alpha_f=\uparrow, \\
      (1-p)\,q_{\beta_i\beta_f}^H(\hat{\tau}) \,\,\, \textrm{if} \,\,\, \alpha_i =\alpha_f=\downarrow,\\
  \end{cases}
  \label{eq:qp_factor_PH}
\end{equation}
\end{widetext}
On the other hand, the transformation $\hat{H}\otimes \hat{H}$ can lead to less trivial effects. For instance, it is possible that anomalous processes with negative MHQs emerge even if separately the single-qubit MHQs do not admit them. As an example, we can consider the initial state $\ket{\psi}^{\otimes 2}$ with $\ket{\psi}=\frac{1}{\sqrt{2}}(\ket{\uparrow} + e^{i\pi/4}\ket{\downarrow})$. The single-qubit KDQs associated to the Hadamard gate are
\begin{align}
&q_{\downarrow\downarrow}^H(\ket{\psi}\bra{\psi})=\frac{2+\sqrt{2}}{8} + i \frac{\sqrt{2}}{8}, \nonumber\\
&  
q_{\downarrow\uparrow}^H(\ket{\psi}\bra{\psi})=\frac{2-\sqrt{2}}{8} - i \frac{\sqrt{2}}{8}, \nonumber\\
& q_{\uparrow\downarrow}^H(\ket{\psi}\bra{\psi})=\frac{2+\sqrt{2}}{8} - i \frac{\sqrt{2}}{8}, 
\nonumber\\
& 
q_{\uparrow\uparrow}^H(\ket{\psi}\bra{\psi})=\frac{2-\sqrt{2}}{8} + i \frac{\sqrt{2}}{8},
\end{align}
and thus all the MHQs are positive, implying the absence of anomalous processes. However, the full KDQ $q_{03}^{H\otimes H}(\ket{\psi}\bra{\psi}\otimes \ket{\psi}\bra{\psi})=\bigl(q_{\downarrow\uparrow}^H(\ket{\psi}\bra{\psi})\bigr)^2$ has $\mathrm{Re}q_{03}^{H\otimes H}(\ket{\psi}\bra{\psi}\otimes \ket{\psi}\bra{\psi})\approx -0.0258$, which means that the process $\ket{\downarrow\downarrow}\rightarrow\ket{\uparrow\uparrow}$ anomalously positively contributes to the extractable work.

Finally, we are interested in how quasiprobabilities behave for generic input states that might be entangled. If the input state is not factorized, the two-qubit KDQs do not simply reduce to the product of two single-qubit KDQs; their structure is more complex. However, we can observe that for a generic two-qubit state $\hat{\rho}$ defined in Eq.~\eqref{eq:rho2qb_state_app}, the full KDQs associated to local transformations $\hat{U}\otimes \hat{V}$
\begin{widetext}
\begin{equation}
  q_{if}^{U\otimes V}(\hat{\rho})=q_{(\alpha_i\beta_i)(\alpha_f\beta_f)}^{U\otimes V}(\hat{\rho})=\sum_{\alpha ', \alpha '', \beta ', \beta ''=\uparrow\downarrow}\lambda_{(\alpha ' \beta ')(\alpha '' \beta '')} q_{\alpha_i\alpha_f}^{U}(\ket{\alpha '}\bra{\alpha ''})q_{\beta_i\beta_f}^{V}(\ket{\beta '}\bra{\beta ''});
  \label{eq:q_nonfactor}
\end{equation}
\end{widetext}
$q_{\alpha_i\alpha_f}^{U}(\ket{\alpha '}\bra{\alpha ''})$ and $q_{\beta_i\beta_f}^{V}(\ket{\beta '}\bra{\beta ''})$ are not quasiprobabilities, unless $\alpha ' = \alpha ''$ and $\beta ' = \beta ''$. However, if the single qubit transformations to which they are associated have the $\hat{P}_{\phi}$ structure, and one of the couples $\alpha_i \alpha_f$, $\beta_i \beta_f$ is $\uparrow\downarrow$ or $\downarrow\uparrow$, they are indentically zero. The reason is the same as above: both the Hamiltonian projectors commute with $\hat{P}_{\phi}$ and so $q_{\downarrow\uparrow}^{P_{\phi}}(\ket{\alpha '}\bra{\alpha ''})$, $q_{\uparrow\downarrow}^{P_{\phi}}(\ket{\alpha '}\bra{\alpha ''})$, $q_{\downarrow\uparrow}^{P_{\phi}}(\ket{\beta '}\bra{\beta ''})$, $q_{\uparrow\downarrow}^{P_{\phi}}(\ket{\beta '}\bra{\beta ''})$ are identically zero since they contain products between orthogonal projectors. In any of the 10 KDQs~\eqref{eq:q_nonfactor} composing the vector $\boldsymbol{q}$, at least one of the couples $\alpha_i \alpha_f$, $\beta_i \beta_f$ is $\uparrow\downarrow$ or $\downarrow\uparrow$ and so, if both $\hat{U}$ and $\hat{V}$ have the $\hat{P}_{\phi}$ structure, all the KDQs of interest are identically zero. Thus, the structure of the matrix in Eq.~\eqref{eq:Matrix_simplified} is preserved also for non-factorized states. The remaining nonzero KDQs cannot be factorized, thus those associated to transformations $\hat{H}\otimes \hat{P}_{\phi}$, $\hat{P}_{\phi}\otimes\hat{H}$ and $\hat{H}\otimes \hat{H}$ are not simply product of single-qubit KDQs and require a case-by-case analysis.


\begin{thebibliography}{60}%
\makeatletter
\providecommand \@ifxundefined [1]{%
 \@ifx{#1\undefined}
}%
\providecommand \@ifnum [1]{%
 \ifnum #1\expandafter \@firstoftwo
 \else \expandafter \@secondoftwo
 \fi
}%
\providecommand \@ifx [1]{%
 \ifx #1\expandafter \@firstoftwo
 \else \expandafter \@secondoftwo
 \fi
}%
\providecommand \natexlab [1]{#1}%
\providecommand \enquote  [1]{``#1''}%
\providecommand \bibnamefont  [1]{#1}%
\providecommand \bibfnamefont [1]{#1}%
\providecommand \citenamefont [1]{#1}%
\providecommand \href@noop [0]{\@secondoftwo}%
\providecommand \href [0]{\begingroup \@sanitize@url \@href}%
\providecommand \@href[1]{\@@startlink{#1}\@@href}%
\providecommand \@@href[1]{\endgroup#1\@@endlink}%
\providecommand \@sanitize@url [0]{\catcode `\\12\catcode `\$12\catcode
  `\&12\catcode `\#12\catcode `\^12\catcode `\_12\catcode `\%12\relax}%
\providecommand \@@startlink[1]{}%
\providecommand \@@endlink[0]{}%
\providecommand \url  [0]{\begingroup\@sanitize@url \@url }%
\providecommand \@url [1]{\endgroup\@href {#1}{\urlprefix }}%
\providecommand \urlprefix  [0]{URL }%
\providecommand \Eprint [0]{\href }%
\providecommand \doibase [0]{https://doi.org/}%
\providecommand \selectlanguage [0]{\@gobble}%
\providecommand \bibinfo  [0]{\@secondoftwo}%
\providecommand \bibfield  [0]{\@secondoftwo}%
\providecommand \translation [1]{[#1]}%
\providecommand \BibitemOpen [0]{}%
\providecommand \bibitemStop [0]{}%
\providecommand \bibitemNoStop [0]{.\EOS\space}%
\providecommand \EOS [0]{\spacefactor3000\relax}%
\providecommand \BibitemShut  [1]{\csname bibitem#1\endcsname}%
\let\auto@bib@innerbib\@empty
\bibitem [{\citenamefont {Streltsov}\ \emph {et~al.}(2017)\citenamefont
  {Streltsov}, \citenamefont {Adesso},\ and\ \citenamefont
  {Plenio}}]{streltsov2017colloquium}%
  \BibitemOpen
  \bibfield  {author} {\bibinfo {author} {\bibfnamefont {A.}~\bibnamefont
  {Streltsov}}, \bibinfo {author} {\bibfnamefont {G.}~\bibnamefont {Adesso}},\
  and\ \bibinfo {author} {\bibfnamefont {M.~B.}\ \bibnamefont {Plenio}},\
  }\bibfield  {title} {\bibinfo {title} {Colloquium: Quantum coherence as a
  resource},\ }\href {https://doi.org/10.1103/RevModPhys.89.041003} {\bibfield
  {journal} {\bibinfo  {journal} {Rev. Mod. Phys.}\ }\textbf {\bibinfo {volume}
  {89}},\ \bibinfo {pages} {041003} (\bibinfo {year} {2017})}\BibitemShut
  {NoStop}%
\bibitem [{\citenamefont {Xi}\ \emph {et~al.}(2015)\citenamefont {Xi},
  \citenamefont {Li},\ and\ \citenamefont {Fan}}]{xi2015quantum}%
  \BibitemOpen
  \bibfield  {author} {\bibinfo {author} {\bibfnamefont {Z.}~\bibnamefont
  {Xi}}, \bibinfo {author} {\bibfnamefont {Y.}~\bibnamefont {Li}},\ and\
  \bibinfo {author} {\bibfnamefont {H.}~\bibnamefont {Fan}},\ }\bibfield
  {title} {\bibinfo {title} {Quantum coherence and correlations in quantum
  system},\ }\href {https://doi.org/10.1038/srep10922} {\bibfield  {journal}
  {\bibinfo  {journal} {Scientific Reports}\ }\textbf {\bibinfo {volume} {5}},\
  \bibinfo {pages} {10922} (\bibinfo {year} {2015})}\BibitemShut {NoStop}%
\bibitem [{\citenamefont {Streltsov}\ \emph {et~al.}(2015)\citenamefont
  {Streltsov}, \citenamefont {Singh}, \citenamefont {Dhar}, \citenamefont
  {Bera},\ and\ \citenamefont {Adesso}}]{streltsov2015measuring}%
  \BibitemOpen
  \bibfield  {author} {\bibinfo {author} {\bibfnamefont {A.}~\bibnamefont
  {Streltsov}}, \bibinfo {author} {\bibfnamefont {U.}~\bibnamefont {Singh}},
  \bibinfo {author} {\bibfnamefont {H.~S.}\ \bibnamefont {Dhar}}, \bibinfo
  {author} {\bibfnamefont {M.~N.}\ \bibnamefont {Bera}},\ and\ \bibinfo
  {author} {\bibfnamefont {G.}~\bibnamefont {Adesso}},\ }\bibfield  {title}
  {\bibinfo {title} {Measuring quantum coherence with entanglement},\ }\href
  {https://doi.org/10.1103/PhysRevLett.115.020403} {\bibfield  {journal}
  {\bibinfo  {journal} {Phys. Rev. Lett.}\ }\textbf {\bibinfo {volume} {115}},\
  \bibinfo {pages} {020403} (\bibinfo {year} {2015})}\BibitemShut {NoStop}%
\bibitem [{\citenamefont {Li}\ and\ \citenamefont {Lin}(2016)}]{li2016quantum}%
  \BibitemOpen
  \bibfield  {author} {\bibinfo {author} {\bibfnamefont {Y.-C.}\ \bibnamefont
  {Li}}\ and\ \bibinfo {author} {\bibfnamefont {H.-Q.}\ \bibnamefont {Lin}},\
  }\bibfield  {title} {\bibinfo {title} {Quantum coherence and quantum phase
  transitions},\ }\href {https://doi.org/10.1038/srep26365} {\bibfield
  {journal} {\bibinfo  {journal} {Scientific Reports}\ }\textbf {\bibinfo
  {volume} {6}},\ \bibinfo {pages} {26365} (\bibinfo {year}
  {2016})}\BibitemShut {NoStop}%
\bibitem [{\citenamefont {Wu}\ \emph {et~al.}(2021)\citenamefont {Wu},
  \citenamefont {Streltsov}, \citenamefont {Regula}, \citenamefont {Xiang},
  \citenamefont {Li},\ and\ \citenamefont {Guo}}]{wu2021experimental}%
  \BibitemOpen
  \bibfield  {author} {\bibinfo {author} {\bibfnamefont {K.}~\bibnamefont
  {Wu}}, \bibinfo {author} {\bibfnamefont {A.}~\bibnamefont {Streltsov}},
  \bibinfo {author} {\bibfnamefont {B.}~\bibnamefont {Regula}}, \bibinfo
  {author} {\bibfnamefont {G.}~\bibnamefont {Xiang}}, \bibinfo {author}
  {\bibfnamefont {C.}~\bibnamefont {Li}},\ and\ \bibinfo {author}
  {\bibfnamefont {G.}~\bibnamefont {Guo}},\ }\bibfield  {title} {\bibinfo
  {title} {Experimental progress on quantum coherence: Detection,
  quantification, and manipulation},\ }\href
  {https://doi.org/10.1002/qute.202100040} {\bibfield  {journal} {\bibinfo
  {journal} {Advanced Quantum Technologies}\ }\textbf {\bibinfo {volume} {4}},\
  \bibinfo {pages} {2100040} (\bibinfo {year} {2021})}\BibitemShut {NoStop}%
\bibitem [{\citenamefont {Kammerlander}\ and\ \citenamefont
  {Anders}(2016)}]{anders2016scirep}%
  \BibitemOpen
  \bibfield  {author} {\bibinfo {author} {\bibfnamefont {P.}~\bibnamefont
  {Kammerlander}}\ and\ \bibinfo {author} {\bibfnamefont {J.}~\bibnamefont
  {Anders}},\ }\bibfield  {title} {\bibinfo {title} {Coherence and measurement
  in quantum thermodynamics},\ }\href {https://doi.org/10.1038/srep22174}
  {\bibfield  {journal} {\bibinfo  {journal} {Sci. Rep.}\ }\textbf {\bibinfo
  {volume} {6}},\ \bibinfo {pages} {22174} (\bibinfo {year}
  {2016})}\BibitemShut {NoStop}%
\bibitem [{\citenamefont {G.~Francica}\ and\ \citenamefont
  {Plastina}(2019)}]{francica2019pre}%
  \BibitemOpen
  \bibfield  {author} {\bibinfo {author} {\bibfnamefont {J.~G.}\ \bibnamefont
  {G.~Francica}}\ and\ \bibinfo {author} {\bibfnamefont {F.}~\bibnamefont
  {Plastina}},\ }\bibfield  {title} {\bibinfo {title} {The role of coherence in
  the non-equilibrium thermodynamics of quantum systems},\ }\href
  {https://doi.org/10.1103/PhysRevE.99.042105} {\bibfield  {journal} {\bibinfo
  {journal} {Phys. Rev. E}\ }\textbf {\bibinfo {volume} {99}},\ \bibinfo
  {pages} {042105} (\bibinfo {year} {2019})}\BibitemShut {NoStop}%
\bibitem [{\citenamefont {Jader P.~Santos}\ and\ \citenamefont
  {Paternostro}(2019)}]{santos2019njp}%
  \BibitemOpen
  \bibfield  {author} {\bibinfo {author} {\bibfnamefont {G.~T.~L.}\
  \bibnamefont {Jader P.~Santos}, \bibfnamefont {Lucas C.~Céleri}}\ and\
  \bibinfo {author} {\bibfnamefont {M.}~\bibnamefont {Paternostro}},\
  }\bibfield  {title} {\bibinfo {title} {The role of quantum coherence in
  non-equilibrium entropy production},\ }\href
  {https://doi.org/10.1038/s41534-019-0138-y} {\bibfield  {journal} {\bibinfo
  {journal} {npj Quantum Inf.}\ }\textbf {\bibinfo {volume} {5}},\ \bibinfo
  {pages} {23} (\bibinfo {year} {2019})}\BibitemShut {NoStop}%
\bibitem [{\citenamefont {Francica}\ \emph {et~al.}(2020)\citenamefont
  {Francica}, \citenamefont {Binder}, \citenamefont {Guarnieri}, \citenamefont
  {Mitchison}, \citenamefont {Goold},\ and\ \citenamefont
  {Plastina}}]{francica2020quantum}%
  \BibitemOpen
  \bibfield  {author} {\bibinfo {author} {\bibfnamefont {G.}~\bibnamefont
  {Francica}}, \bibinfo {author} {\bibfnamefont {F.~C.}\ \bibnamefont
  {Binder}}, \bibinfo {author} {\bibfnamefont {G.}~\bibnamefont {Guarnieri}},
  \bibinfo {author} {\bibfnamefont {M.~T.}\ \bibnamefont {Mitchison}}, \bibinfo
  {author} {\bibfnamefont {J.}~\bibnamefont {Goold}},\ and\ \bibinfo {author}
  {\bibfnamefont {F.}~\bibnamefont {Plastina}},\ }\bibfield  {title} {\bibinfo
  {title} {Quantum coherence and ergotropy},\ }\href
  {https://doi.org/10.1103/PhysRevLett.125.180603} {\bibfield  {journal}
  {\bibinfo  {journal} {Phys. Rev. Lett.}\ }\textbf {\bibinfo {volume} {125}},\
  \bibinfo {pages} {180603} (\bibinfo {year} {2020})}\BibitemShut {NoStop}%
\bibitem [{\citenamefont {Shi}\ \emph {et~al.}(2022)\citenamefont {Shi},
  \citenamefont {Ding}, \citenamefont {Wan}, \citenamefont {Wang},\ and\
  \citenamefont {Yang}}]{shi2022entanglement}%
  \BibitemOpen
  \bibfield  {author} {\bibinfo {author} {\bibfnamefont {H.-L.}\ \bibnamefont
  {Shi}}, \bibinfo {author} {\bibfnamefont {S.}~\bibnamefont {Ding}}, \bibinfo
  {author} {\bibfnamefont {Q.-K.}\ \bibnamefont {Wan}}, \bibinfo {author}
  {\bibfnamefont {X.-H.}\ \bibnamefont {Wang}},\ and\ \bibinfo {author}
  {\bibfnamefont {W.-L.}\ \bibnamefont {Yang}},\ }\bibfield  {title} {\bibinfo
  {title} {Entanglement, coherence, and extractable work in quantum
  batteries},\ }\href {https://doi.org/10.1103/PhysRevLett.129.130602}
  {\bibfield  {journal} {\bibinfo  {journal} {Phys. Rev. Lett.}\ }\textbf
  {\bibinfo {volume} {129}},\ \bibinfo {pages} {130602} (\bibinfo {year}
  {2022})}\BibitemShut {NoStop}%
\bibitem [{\citenamefont {Shi}\ \emph {et~al.}(2020)\citenamefont {Shi},
  \citenamefont {Shi}, \citenamefont {Wang}, \citenamefont {Hu}, \citenamefont
  {Liu}, \citenamefont {Yang},\ and\ \citenamefont {Fan}}]{shi2020quantum}%
  \BibitemOpen
  \bibfield  {author} {\bibinfo {author} {\bibfnamefont {Y.-H.}\ \bibnamefont
  {Shi}}, \bibinfo {author} {\bibfnamefont {H.-L.}\ \bibnamefont {Shi}},
  \bibinfo {author} {\bibfnamefont {X.-H.}\ \bibnamefont {Wang}}, \bibinfo
  {author} {\bibfnamefont {M.-L.}\ \bibnamefont {Hu}}, \bibinfo {author}
  {\bibfnamefont {S.-Y.}\ \bibnamefont {Liu}}, \bibinfo {author} {\bibfnamefont
  {W.-L.}\ \bibnamefont {Yang}},\ and\ \bibinfo {author} {\bibfnamefont
  {H.}~\bibnamefont {Fan}},\ }\bibfield  {title} {\bibinfo {title} {Quantum
  coherence in a quantum heat engine},\ }\href
  {https://doi.org/10.1088/1751-8121/ab6a6b} {\bibfield  {journal} {\bibinfo
  {journal} {Journal of Physics A: Mathematical and Theoretical}\ }\textbf
  {\bibinfo {volume} {53}},\ \bibinfo {pages} {085301} (\bibinfo {year}
  {2020})}\BibitemShut {NoStop}%
\bibitem [{\citenamefont {Gour}(2022)}]{gour2022PRXQ}%
  \BibitemOpen
  \bibfield  {author} {\bibinfo {author} {\bibfnamefont {G.}~\bibnamefont
  {Gour}},\ }\bibfield  {title} {\bibinfo {title} {Role of quantum coherence in
  thermodynamics},\ }\href {https://doi.org/10.1103/PRXQuantum.3.040323}
  {\bibfield  {journal} {\bibinfo  {journal} {PRX Quantum}\ }\textbf {\bibinfo
  {volume} {3}},\ \bibinfo {pages} {040323} (\bibinfo {year}
  {2022})}\BibitemShut {NoStop}%
\bibitem [{\citenamefont {Francica}\ and\ \citenamefont
  {Dell'Anna}(2024)}]{francica2024work}%
  \BibitemOpen
  \bibfield  {author} {\bibinfo {author} {\bibfnamefont {G.}~\bibnamefont
  {Francica}}\ and\ \bibinfo {author} {\bibfnamefont {L.}~\bibnamefont
  {Dell'Anna}},\ }\bibfield  {title} {\bibinfo {title} {Work fluctuation
  theorems with initial quantum coherence},\ }\href
  {https://doi.org/10.1103/PhysRevE.109.064138} {\bibfield  {journal} {\bibinfo
   {journal} {Phys. Rev. E}\ }\textbf {\bibinfo {volume} {109}},\ \bibinfo
  {pages} {064138} (\bibinfo {year} {2024})}\BibitemShut {NoStop}%
\bibitem [{\citenamefont {Rodrigues}\ and\ \citenamefont
  {Lutz}(2024)}]{rodrigues2024}%
  \BibitemOpen
  \bibfield  {author} {\bibinfo {author} {\bibfnamefont {F.~L.~S.}\
  \bibnamefont {Rodrigues}}\ and\ \bibinfo {author} {\bibfnamefont
  {E.}~\bibnamefont {Lutz}},\ }\bibfield  {title} {\bibinfo {title}
  {Nonequilibrium thermodynamics of quantum coherence beyond linear response},\
  }\href {https://doi.org/10.1038/s42005-024-01548-2} {\bibfield  {journal}
  {\bibinfo  {journal} {Commun. Phys.}\ }\textbf {\bibinfo {volume} {7}},\
  \bibinfo {pages} {61} (\bibinfo {year} {2024})}\BibitemShut {NoStop}%
\bibitem [{\citenamefont {Onishchenko}\ \emph {et~al.}(2024)\citenamefont
  {Onishchenko}, \citenamefont {Guarnieri}, \citenamefont {Rosillo-Rodes},
  \citenamefont {Pijn}, \citenamefont {Hilder}, \citenamefont {Poschinger},
  \citenamefont {Perarnau-Llobet}, \citenamefont {Eisert},\ and\ \citenamefont
  {Schmidt-Kaler}}]{onishchenko2024natcom}%
  \BibitemOpen
  \bibfield  {author} {\bibinfo {author} {\bibfnamefont {O.}~\bibnamefont
  {Onishchenko}}, \bibinfo {author} {\bibfnamefont {G.}~\bibnamefont
  {Guarnieri}}, \bibinfo {author} {\bibfnamefont {P.}~\bibnamefont
  {Rosillo-Rodes}}, \bibinfo {author} {\bibfnamefont {D.}~\bibnamefont {Pijn}},
  \bibinfo {author} {\bibfnamefont {J.}~\bibnamefont {Hilder}}, \bibinfo
  {author} {\bibfnamefont {U.~G.}\ \bibnamefont {Poschinger}}, \bibinfo
  {author} {\bibfnamefont {M.}~\bibnamefont {Perarnau-Llobet}}, \bibinfo
  {author} {\bibfnamefont {J.}~\bibnamefont {Eisert}},\ and\ \bibinfo {author}
  {\bibfnamefont {F.}~\bibnamefont {Schmidt-Kaler}},\ }\bibfield  {title}
  {\bibinfo {title} {Probing coherent quantum thermodynamics using a trapped
  ion},\ }\bibfield  {journal} {\bibinfo  {journal} {Nature Communications}\
  }\textbf {\bibinfo {volume} {15}},\ \href
  {https://doi.org/10.1038/s41467-024-51263-3} {10.1038/s41467-024-51263-3}
  (\bibinfo {year} {2024})\BibitemShut {NoStop}%
\bibitem [{\citenamefont {Vinjanampathy}\ and\ \citenamefont
  {Anders}(2016)}]{vinjanampathy2016quantum}%
  \BibitemOpen
  \bibfield  {author} {\bibinfo {author} {\bibfnamefont {S.}~\bibnamefont
  {Vinjanampathy}}\ and\ \bibinfo {author} {\bibfnamefont {J.}~\bibnamefont
  {Anders}},\ }\bibfield  {title} {\bibinfo {title} {Quantum thermodynamics},\
  }\href {https://doi.org/10.1080/00107514.2016.1201896} {\bibfield  {journal}
  {\bibinfo  {journal} {Contemporary Physics}\ }\textbf {\bibinfo {volume}
  {57}},\ \bibinfo {pages} {545–579} (\bibinfo {year} {2016})}\BibitemShut
  {NoStop}%
\bibitem [{\citenamefont {Goold}\ \emph {et~al.}(2016)\citenamefont {Goold},
  \citenamefont {Huber}, \citenamefont {Riera}, \citenamefont {Rio},\ and\
  \citenamefont {Skrzypczyk}}]{goold2016role}%
  \BibitemOpen
  \bibfield  {author} {\bibinfo {author} {\bibfnamefont {J.}~\bibnamefont
  {Goold}}, \bibinfo {author} {\bibfnamefont {M.}~\bibnamefont {Huber}},
  \bibinfo {author} {\bibfnamefont {A.}~\bibnamefont {Riera}}, \bibinfo
  {author} {\bibfnamefont {L.~d.}\ \bibnamefont {Rio}},\ and\ \bibinfo {author}
  {\bibfnamefont {P.}~\bibnamefont {Skrzypczyk}},\ }\bibfield  {title}
  {\bibinfo {title} {The role of quantum information in thermodynamics—a
  topical review},\ }\href {https://doi.org/10.1088/1751-8113/49/14/143001}
  {\bibfield  {journal} {\bibinfo  {journal} {Journal of Physics A:
  Mathematical and Theoretical}\ }\textbf {\bibinfo {volume} {49}},\ \bibinfo
  {pages} {143001} (\bibinfo {year} {2016})}\BibitemShut {NoStop}%
\bibitem [{\citenamefont {Alicki}\ and\ \citenamefont
  {Kosloff}(2018)}]{alicki2018introduction}%
  \BibitemOpen
  \bibfield  {author} {\bibinfo {author} {\bibfnamefont {R.}~\bibnamefont
  {Alicki}}\ and\ \bibinfo {author} {\bibfnamefont {R.}~\bibnamefont
  {Kosloff}},\ }\bibinfo {title} {Introduction to quantum thermodynamics:
  History and prospects},\ in\ \href
  {https://doi.org/10.1007/978-3-319-99046-0_1} {\emph {\bibinfo {booktitle}
  {Thermodynamics in the Quantum Regime}}}\ (\bibinfo  {publisher} {Springer
  International Publishing},\ \bibinfo {year} {2018})\ p.\ \bibinfo {pages}
  {1–33}\BibitemShut {NoStop}%
\bibitem [{\citenamefont {Campbell}\ \emph {et~al.}(2025)\citenamefont
  {Campbell}, \citenamefont {D'Amico}, \citenamefont {Ciampini}, \citenamefont
  {Anders}, \citenamefont {Ares}, \citenamefont {Artini}, \citenamefont
  {Auffèves}, \citenamefont {Oftelie}, \citenamefont {Bettmann}, \citenamefont
  {Bonança}, \citenamefont {Busch}, \citenamefont {Campisi}, \citenamefont
  {Cavalcante}, \citenamefont {Correa}, \citenamefont {Cuestas}, \citenamefont
  {Dag}, \citenamefont {Dago}, \citenamefont {Deffner}, \citenamefont {Campo},
  \citenamefont {Deutschmann-Olek}, \citenamefont {Donadi}, \citenamefont
  {Doucet}, \citenamefont {Elouard}, \citenamefont {Ensslin}, \citenamefont
  {Erker}, \citenamefont {Fabbri}, \citenamefont {Fedele}, \citenamefont
  {Fiusa}, \citenamefont {Fogarty}, \citenamefont {Folk}, \citenamefont
  {Guarnieri}, \citenamefont {Hegde}, \citenamefont {Hernández-Gómez},
  \citenamefont {Hu}, \citenamefont {Iemini}, \citenamefont {Karimi},
  \citenamefont {Kiesel}, \citenamefont {Landi}, \citenamefont {Lasek},
  \citenamefont {Lemziakov}, \citenamefont {Monaco}, \citenamefont {Lutz},
  \citenamefont {Lvov}, \citenamefont {Maillet}, \citenamefont {Mehboudi},
  \citenamefont {Mendonça}, \citenamefont {Miller}, \citenamefont {Mitchell},
  \citenamefont {Mitchison}, \citenamefont {Mukherjee}, \citenamefont
  {Paternostro}, \citenamefont {Pekola}, \citenamefont {Perarnau-Llobet},
  \citenamefont {Poschinger}, \citenamefont {Rolandi}, \citenamefont {Rosa},
  \citenamefont {Sánchez}, \citenamefont {Santos}, \citenamefont {Sarthour},
  \citenamefont {Sela}, \citenamefont {Solfanelli}, \citenamefont {Souza},
  \citenamefont {Splettstoesser}, \citenamefont {Tan}, \citenamefont {Tesser},
  \citenamefont {Vu}, \citenamefont {Widera}, \citenamefont {Halpern},\ and\
  \citenamefont {Zawadzki}}]{campbell2025roadmap}%
  \BibitemOpen
  \bibfield  {author} {\bibinfo {author} {\bibfnamefont {S.}~\bibnamefont
  {Campbell}}, \bibinfo {author} {\bibfnamefont {I.}~\bibnamefont {D'Amico}},
  \bibinfo {author} {\bibfnamefont {M.~A.}\ \bibnamefont {Ciampini}}, \bibinfo
  {author} {\bibfnamefont {J.}~\bibnamefont {Anders}}, \bibinfo {author}
  {\bibfnamefont {N.}~\bibnamefont {Ares}}, \bibinfo {author} {\bibfnamefont
  {S.}~\bibnamefont {Artini}}, \bibinfo {author} {\bibfnamefont
  {A.}~\bibnamefont {Auffèves}}, \bibinfo {author} {\bibfnamefont {L.~B.}\
  \bibnamefont {Oftelie}}, \bibinfo {author} {\bibfnamefont {L.~P.}\
  \bibnamefont {Bettmann}}, \bibinfo {author} {\bibfnamefont {M.~V.~S.}\
  \bibnamefont {Bonança}}, \bibinfo {author} {\bibfnamefont {T.}~\bibnamefont
  {Busch}}, \bibinfo {author} {\bibfnamefont {M.}~\bibnamefont {Campisi}},
  \bibinfo {author} {\bibfnamefont {M.~F.}\ \bibnamefont {Cavalcante}},
  \bibinfo {author} {\bibfnamefont {L.~A.}\ \bibnamefont {Correa}}, \bibinfo
  {author} {\bibfnamefont {E.}~\bibnamefont {Cuestas}}, \bibinfo {author}
  {\bibfnamefont {C.~B.}\ \bibnamefont {Dag}}, \bibinfo {author} {\bibfnamefont
  {S.}~\bibnamefont {Dago}}, \bibinfo {author} {\bibfnamefont {S.}~\bibnamefont
  {Deffner}}, \bibinfo {author} {\bibfnamefont {A.~D.}\ \bibnamefont {Campo}},
  \bibinfo {author} {\bibfnamefont {A.}~\bibnamefont {Deutschmann-Olek}},
  \bibinfo {author} {\bibfnamefont {S.}~\bibnamefont {Donadi}}, \bibinfo
  {author} {\bibfnamefont {E.}~\bibnamefont {Doucet}}, \bibinfo {author}
  {\bibfnamefont {C.}~\bibnamefont {Elouard}}, \bibinfo {author} {\bibfnamefont
  {K.}~\bibnamefont {Ensslin}}, \bibinfo {author} {\bibfnamefont
  {P.}~\bibnamefont {Erker}}, \bibinfo {author} {\bibfnamefont
  {N.}~\bibnamefont {Fabbri}}, \bibinfo {author} {\bibfnamefont
  {F.}~\bibnamefont {Fedele}}, \bibinfo {author} {\bibfnamefont
  {G.}~\bibnamefont {Fiusa}}, \bibinfo {author} {\bibfnamefont
  {T.}~\bibnamefont {Fogarty}}, \bibinfo {author} {\bibfnamefont
  {J.}~\bibnamefont {Folk}}, \bibinfo {author} {\bibfnamefont {G.}~\bibnamefont
  {Guarnieri}}, \bibinfo {author} {\bibfnamefont {A.~S.}\ \bibnamefont
  {Hegde}}, \bibinfo {author} {\bibfnamefont {S.}~\bibnamefont
  {Hernández-Gómez}}, \bibinfo {author} {\bibfnamefont {C.-K.}\ \bibnamefont
  {Hu}}, \bibinfo {author} {\bibfnamefont {F.}~\bibnamefont {Iemini}}, \bibinfo
  {author} {\bibfnamefont {B.}~\bibnamefont {Karimi}}, \bibinfo {author}
  {\bibfnamefont {N.}~\bibnamefont {Kiesel}}, \bibinfo {author} {\bibfnamefont
  {G.~T.}\ \bibnamefont {Landi}}, \bibinfo {author} {\bibfnamefont
  {A.}~\bibnamefont {Lasek}}, \bibinfo {author} {\bibfnamefont
  {S.}~\bibnamefont {Lemziakov}}, \bibinfo {author} {\bibfnamefont {G.~L.}\
  \bibnamefont {Monaco}}, \bibinfo {author} {\bibfnamefont {E.}~\bibnamefont
  {Lutz}}, \bibinfo {author} {\bibfnamefont {D.}~\bibnamefont {Lvov}}, \bibinfo
  {author} {\bibfnamefont {O.}~\bibnamefont {Maillet}}, \bibinfo {author}
  {\bibfnamefont {M.}~\bibnamefont {Mehboudi}}, \bibinfo {author}
  {\bibfnamefont {T.~M.}\ \bibnamefont {Mendonça}}, \bibinfo {author}
  {\bibfnamefont {H.~J.~D.}\ \bibnamefont {Miller}}, \bibinfo {author}
  {\bibfnamefont {A.~K.}\ \bibnamefont {Mitchell}}, \bibinfo {author}
  {\bibfnamefont {M.~T.}\ \bibnamefont {Mitchison}}, \bibinfo {author}
  {\bibfnamefont {V.}~\bibnamefont {Mukherjee}}, \bibinfo {author}
  {\bibfnamefont {M.}~\bibnamefont {Paternostro}}, \bibinfo {author}
  {\bibfnamefont {J.}~\bibnamefont {Pekola}}, \bibinfo {author} {\bibfnamefont
  {M.}~\bibnamefont {Perarnau-Llobet}}, \bibinfo {author} {\bibfnamefont
  {U.}~\bibnamefont {Poschinger}}, \bibinfo {author} {\bibfnamefont
  {A.}~\bibnamefont {Rolandi}}, \bibinfo {author} {\bibfnamefont
  {D.}~\bibnamefont {Rosa}}, \bibinfo {author} {\bibfnamefont {R.}~\bibnamefont
  {Sánchez}}, \bibinfo {author} {\bibfnamefont {A.~C.}\ \bibnamefont
  {Santos}}, \bibinfo {author} {\bibfnamefont {R.~S.}\ \bibnamefont
  {Sarthour}}, \bibinfo {author} {\bibfnamefont {E.}~\bibnamefont {Sela}},
  \bibinfo {author} {\bibfnamefont {A.}~\bibnamefont {Solfanelli}}, \bibinfo
  {author} {\bibfnamefont {A.~M.}\ \bibnamefont {Souza}}, \bibinfo {author}
  {\bibfnamefont {J.}~\bibnamefont {Splettstoesser}}, \bibinfo {author}
  {\bibfnamefont {D.}~\bibnamefont {Tan}}, \bibinfo {author} {\bibfnamefont
  {L.}~\bibnamefont {Tesser}}, \bibinfo {author} {\bibfnamefont {T.~V.}\
  \bibnamefont {Vu}}, \bibinfo {author} {\bibfnamefont {A.}~\bibnamefont
  {Widera}}, \bibinfo {author} {\bibfnamefont {N.~Y.}\ \bibnamefont
  {Halpern}},\ and\ \bibinfo {author} {\bibfnamefont {K.}~\bibnamefont
  {Zawadzki}},\ }\href {https://arxiv.org/abs/2504.20145} {\bibinfo {title}
  {Roadmap on quantum thermodynamics}} (\bibinfo {year} {2025}),\ \Eprint
  {https://arxiv.org/abs/2504.20145} {arXiv:2504.20145 [quant-ph]} \BibitemShut
  {NoStop}%
\bibitem [{\citenamefont {Korzekwa}\ \emph {et~al.}(2016)\citenamefont
  {Korzekwa}, \citenamefont {Lostaglio}, \citenamefont {Oppenheim},\ and\
  \citenamefont {Jennings}}]{korzekwa2016extraction}%
  \BibitemOpen
  \bibfield  {author} {\bibinfo {author} {\bibfnamefont {K.}~\bibnamefont
  {Korzekwa}}, \bibinfo {author} {\bibfnamefont {M.}~\bibnamefont {Lostaglio}},
  \bibinfo {author} {\bibfnamefont {J.}~\bibnamefont {Oppenheim}},\ and\
  \bibinfo {author} {\bibfnamefont {D.}~\bibnamefont {Jennings}},\ }\bibfield
  {title} {\bibinfo {title} {The extraction of work from quantum coherence},\
  }\href {https://doi.org/10.1088/1367-2630/18/2/023045} {\bibfield  {journal}
  {\bibinfo  {journal} {New Journal of Physics}\ }\textbf {\bibinfo {volume}
  {18}},\ \bibinfo {pages} {023045} (\bibinfo {year} {2016})}\BibitemShut
  {NoStop}%
\bibitem [{\citenamefont {Talkner}\ and\ \citenamefont
  {H\"anggi}(2016)}]{talkner2016aspects}%
  \BibitemOpen
  \bibfield  {author} {\bibinfo {author} {\bibfnamefont {P.}~\bibnamefont
  {Talkner}}\ and\ \bibinfo {author} {\bibfnamefont {P.}~\bibnamefont
  {H\"anggi}},\ }\bibfield  {title} {\bibinfo {title} {Aspects of quantum
  work},\ }\href {https://doi.org/10.1103/PhysRevE.93.022131} {\bibfield
  {journal} {\bibinfo  {journal} {Phys. Rev. E}\ }\textbf {\bibinfo {volume}
  {93}},\ \bibinfo {pages} {022131} (\bibinfo {year} {2016})}\BibitemShut
  {NoStop}%
\bibitem [{\citenamefont
  {Allahverdyan}(2014)}]{allahverdyan2014nonequilibrium}%
  \BibitemOpen
  \bibfield  {author} {\bibinfo {author} {\bibfnamefont {A.~E.}\ \bibnamefont
  {Allahverdyan}},\ }\bibfield  {title} {\bibinfo {title} {Nonequilibrium
  quantum fluctuations of work},\ }\href
  {https://doi.org/10.1103/PhysRevE.90.032137} {\bibfield  {journal} {\bibinfo
  {journal} {Phys. Rev. E}\ }\textbf {\bibinfo {volume} {90}},\ \bibinfo
  {pages} {032137} (\bibinfo {year} {2014})}\BibitemShut {NoStop}%
\bibitem [{\citenamefont {Campisi}\ \emph {et~al.}(2011)\citenamefont
  {Campisi}, \citenamefont {H\"anggi},\ and\ \citenamefont
  {Talkner}}]{campisi2011colloquium}%
  \BibitemOpen
  \bibfield  {author} {\bibinfo {author} {\bibfnamefont {M.}~\bibnamefont
  {Campisi}}, \bibinfo {author} {\bibfnamefont {P.}~\bibnamefont {H\"anggi}},\
  and\ \bibinfo {author} {\bibfnamefont {P.}~\bibnamefont {Talkner}},\
  }\bibfield  {title} {\bibinfo {title} {Colloquium: Quantum fluctuation
  relations: Foundations and applications},\ }\href
  {https://doi.org/10.1103/RevModPhys.83.771} {\bibfield  {journal} {\bibinfo
  {journal} {Rev. Mod. Phys.}\ }\textbf {\bibinfo {volume} {83}},\ \bibinfo
  {pages} {771} (\bibinfo {year} {2011})}\BibitemShut {NoStop}%
\bibitem [{\citenamefont {Campaioli}\ \emph {et~al.}(2024)\citenamefont
  {Campaioli}, \citenamefont {Gherardini}, \citenamefont {Quach}, \citenamefont
  {Polini},\ and\ \citenamefont {Andolina}}]{campaioli2024colloquium}%
  \BibitemOpen
  \bibfield  {author} {\bibinfo {author} {\bibfnamefont {F.}~\bibnamefont
  {Campaioli}}, \bibinfo {author} {\bibfnamefont {S.}~\bibnamefont
  {Gherardini}}, \bibinfo {author} {\bibfnamefont {J.~Q.}\ \bibnamefont
  {Quach}}, \bibinfo {author} {\bibfnamefont {M.}~\bibnamefont {Polini}},\ and\
  \bibinfo {author} {\bibfnamefont {G.~M.}\ \bibnamefont {Andolina}},\
  }\bibfield  {title} {\bibinfo {title} {Colloquium: Quantum batteries},\
  }\href {https://doi.org/10.1103/RevModPhys.96.031001} {\bibfield  {journal}
  {\bibinfo  {journal} {Rev. Mod. Phys.}\ }\textbf {\bibinfo {volume} {96}},\
  \bibinfo {pages} {031001} (\bibinfo {year} {2024})}\BibitemShut {NoStop}%
\bibitem [{\citenamefont {Tasaki}(2000)}]{tasaki2000jarzynski}%
  \BibitemOpen
  \bibfield  {author} {\bibinfo {author} {\bibfnamefont {H.}~\bibnamefont
  {Tasaki}},\ }\href {https://arxiv.org/abs/cond-mat/0009244} {\bibinfo {title}
  {Jarzynski relations for quantum systems and some applications}} (\bibinfo
  {year} {2000}),\ \Eprint {https://arxiv.org/abs/cond-mat/0009244}
  {arXiv:cond-mat/0009244 [cond-mat.stat-mech]} \BibitemShut {NoStop}%
\bibitem [{\citenamefont {Plastina}\ \emph {et~al.}(2014)\citenamefont
  {Plastina}, \citenamefont {Alecce}, \citenamefont {Apollaro}, \citenamefont
  {Falcone}, \citenamefont {Francica}, \citenamefont {Galve}, \citenamefont
  {Lo~Gullo},\ and\ \citenamefont {Zambrini}}]{plastina2014irreversible}%
  \BibitemOpen
  \bibfield  {author} {\bibinfo {author} {\bibfnamefont {F.}~\bibnamefont
  {Plastina}}, \bibinfo {author} {\bibfnamefont {A.}~\bibnamefont {Alecce}},
  \bibinfo {author} {\bibfnamefont {T.~J.~G.}\ \bibnamefont {Apollaro}},
  \bibinfo {author} {\bibfnamefont {G.}~\bibnamefont {Falcone}}, \bibinfo
  {author} {\bibfnamefont {G.}~\bibnamefont {Francica}}, \bibinfo {author}
  {\bibfnamefont {F.}~\bibnamefont {Galve}}, \bibinfo {author} {\bibfnamefont
  {N.}~\bibnamefont {Lo~Gullo}},\ and\ \bibinfo {author} {\bibfnamefont
  {R.}~\bibnamefont {Zambrini}},\ }\bibfield  {title} {\bibinfo {title}
  {Irreversible work and inner friction in quantum thermodynamic processes},\
  }\href {https://doi.org/10.1103/PhysRevLett.113.260601} {\bibfield  {journal}
  {\bibinfo  {journal} {Phys. Rev. Lett.}\ }\textbf {\bibinfo {volume} {113}},\
  \bibinfo {pages} {260601} (\bibinfo {year} {2014})}\BibitemShut {NoStop}%
\bibitem [{\citenamefont {Silva}(2008)}]{silva2008statistics}%
  \BibitemOpen
  \bibfield  {author} {\bibinfo {author} {\bibfnamefont {A.}~\bibnamefont
  {Silva}},\ }\bibfield  {title} {\bibinfo {title} {Statistics of the work done
  on a quantum critical system by quenching a control parameter},\ }\href
  {https://doi.org/10.1103/PhysRevLett.101.120603} {\bibfield  {journal}
  {\bibinfo  {journal} {Phys. Rev. Lett.}\ }\textbf {\bibinfo {volume} {101}},\
  \bibinfo {pages} {120603} (\bibinfo {year} {2008})}\BibitemShut {NoStop}%
\bibitem [{\citenamefont {Gherardini}\ and\ \citenamefont
  {De~Chiara}(2024)}]{gherardini2024quasiprobabilities}%
  \BibitemOpen
  \bibfield  {author} {\bibinfo {author} {\bibfnamefont {S.}~\bibnamefont
  {Gherardini}}\ and\ \bibinfo {author} {\bibfnamefont {G.}~\bibnamefont
  {De~Chiara}},\ }\bibfield  {title} {\bibinfo {title} {Quasiprobabilities in
  quantum thermodynamics and many-body systems},\ }\href
  {https://doi.org/10.1103/PRXQuantum.5.030201} {\bibfield  {journal} {\bibinfo
   {journal} {PRX Quantum}\ }\textbf {\bibinfo {volume} {5}},\ \bibinfo {pages}
  {030201} (\bibinfo {year} {2024})}\BibitemShut {NoStop}%
\bibitem [{\citenamefont {Díaz}\ \emph {et~al.}(2020)\citenamefont {Díaz},
  \citenamefont {Guarnieri},\ and\ \citenamefont
  {Paternostro}}]{diaz2020quantum}%
  \BibitemOpen
  \bibfield  {author} {\bibinfo {author} {\bibfnamefont {M.}~\bibnamefont
  {Díaz}}, \bibinfo {author} {\bibfnamefont {G.}~\bibnamefont {Guarnieri}},\
  and\ \bibinfo {author} {\bibfnamefont {M.}~\bibnamefont {Paternostro}},\
  }\bibfield  {title} {\bibinfo {title} {Quantum work statistics with initial
  coherence},\ }\href {https://doi.org/10.3390/e22111223} {\bibfield  {journal}
  {\bibinfo  {journal} {Entropy}\ }\textbf {\bibinfo {volume} {22}},\ \bibinfo
  {pages} {1223} (\bibinfo {year} {2020})}\BibitemShut {NoStop}%
\bibitem [{\citenamefont {Batalh\~ao}\ \emph {et~al.}(2014)\citenamefont
  {Batalh\~ao}, \citenamefont {Souza}, \citenamefont {Mazzola}, \citenamefont
  {Auccaise}, \citenamefont {Sarthour}, \citenamefont {Oliveira}, \citenamefont
  {Goold}, \citenamefont {De~Chiara}, \citenamefont {Paternostro},\ and\
  \citenamefont {Serra}}]{batalhao2014experimental}%
  \BibitemOpen
  \bibfield  {author} {\bibinfo {author} {\bibfnamefont {T.~B.}\ \bibnamefont
  {Batalh\~ao}}, \bibinfo {author} {\bibfnamefont {A.~M.}\ \bibnamefont
  {Souza}}, \bibinfo {author} {\bibfnamefont {L.}~\bibnamefont {Mazzola}},
  \bibinfo {author} {\bibfnamefont {R.}~\bibnamefont {Auccaise}}, \bibinfo
  {author} {\bibfnamefont {R.~S.}\ \bibnamefont {Sarthour}}, \bibinfo {author}
  {\bibfnamefont {I.~S.}\ \bibnamefont {Oliveira}}, \bibinfo {author}
  {\bibfnamefont {J.}~\bibnamefont {Goold}}, \bibinfo {author} {\bibfnamefont
  {G.}~\bibnamefont {De~Chiara}}, \bibinfo {author} {\bibfnamefont
  {M.}~\bibnamefont {Paternostro}},\ and\ \bibinfo {author} {\bibfnamefont
  {R.~M.}\ \bibnamefont {Serra}},\ }\bibfield  {title} {\bibinfo {title}
  {Experimental reconstruction of work distribution and study of fluctuation
  relations in a closed quantum system},\ }\href
  {https://doi.org/10.1103/PhysRevLett.113.140601} {\bibfield  {journal}
  {\bibinfo  {journal} {Phys. Rev. Lett.}\ }\textbf {\bibinfo {volume} {113}},\
  \bibinfo {pages} {140601} (\bibinfo {year} {2014})}\BibitemShut {NoStop}%
\bibitem [{\citenamefont {Mazzola}\ \emph {et~al.}(2013)\citenamefont
  {Mazzola}, \citenamefont {De~Chiara},\ and\ \citenamefont
  {Paternostro}}]{mazzola2013meausring}%
  \BibitemOpen
  \bibfield  {author} {\bibinfo {author} {\bibfnamefont {L.}~\bibnamefont
  {Mazzola}}, \bibinfo {author} {\bibfnamefont {G.}~\bibnamefont {De~Chiara}},\
  and\ \bibinfo {author} {\bibfnamefont {M.}~\bibnamefont {Paternostro}},\
  }\bibfield  {title} {\bibinfo {title} {Measuring the characteristic function
  of the work distribution},\ }\href
  {https://doi.org/10.1103/PhysRevLett.110.230602} {\bibfield  {journal}
  {\bibinfo  {journal} {Phys. Rev. Lett.}\ }\textbf {\bibinfo {volume} {110}},\
  \bibinfo {pages} {230602} (\bibinfo {year} {2013})}\BibitemShut {NoStop}%
\bibitem [{\citenamefont {Dorner}\ \emph {et~al.}(2013)\citenamefont {Dorner},
  \citenamefont {Clark}, \citenamefont {Heaney}, \citenamefont {Fazio},
  \citenamefont {Goold},\ and\ \citenamefont {Vedral}}]{dorner2013extracting}%
  \BibitemOpen
  \bibfield  {author} {\bibinfo {author} {\bibfnamefont {R.}~\bibnamefont
  {Dorner}}, \bibinfo {author} {\bibfnamefont {S.~R.}\ \bibnamefont {Clark}},
  \bibinfo {author} {\bibfnamefont {L.}~\bibnamefont {Heaney}}, \bibinfo
  {author} {\bibfnamefont {R.}~\bibnamefont {Fazio}}, \bibinfo {author}
  {\bibfnamefont {J.}~\bibnamefont {Goold}},\ and\ \bibinfo {author}
  {\bibfnamefont {V.}~\bibnamefont {Vedral}},\ }\bibfield  {title} {\bibinfo
  {title} {Extracting quantum work statistics and fluctuation theorems by
  single-qubit interferometry},\ }\href
  {https://doi.org/10.1103/PhysRevLett.110.230601} {\bibfield  {journal}
  {\bibinfo  {journal} {Phys. Rev. Lett.}\ }\textbf {\bibinfo {volume} {110}},\
  \bibinfo {pages} {230601} (\bibinfo {year} {2013})}\BibitemShut {NoStop}%
\bibitem [{\citenamefont {Oftelie}\ and\ \citenamefont
  {Campisi}(2025)}]{oftelie2025measurement}%
  \BibitemOpen
  \bibfield  {author} {\bibinfo {author} {\bibfnamefont {L.~B.}\ \bibnamefont
  {Oftelie}}\ and\ \bibinfo {author} {\bibfnamefont {M.}~\bibnamefont
  {Campisi}},\ }\bibfield  {title} {\bibinfo {title} {Measurement of the work
  statistics of an open quantum system using a quantum computer},\ }\href
  {https://doi.org/10.1088/2058-9565/adbd6c} {\bibfield  {journal} {\bibinfo
  {journal} {Quantum Science and Technology}\ }\textbf {\bibinfo {volume}
  {10}},\ \bibinfo {pages} {025045} (\bibinfo {year} {2025})}\BibitemShut
  {NoStop}%
\bibitem [{\citenamefont {Solfanelli}\ and\ \citenamefont
  {Defenu}(2025)}]{solfanelli2025universal}%
  \BibitemOpen
  \bibfield  {author} {\bibinfo {author} {\bibfnamefont {A.}~\bibnamefont
  {Solfanelli}}\ and\ \bibinfo {author} {\bibfnamefont {N.}~\bibnamefont
  {Defenu}},\ }\bibfield  {title} {\bibinfo {title} {Universal work statistics
  in long-range interacting quantum systems},\ }\href
  {https://doi.org/10.1103/PhysRevLett.134.030402} {\bibfield  {journal}
  {\bibinfo  {journal} {Phys. Rev. Lett.}\ }\textbf {\bibinfo {volume} {134}},\
  \bibinfo {pages} {030402} (\bibinfo {year} {2025})}\BibitemShut {NoStop}%
\bibitem [{\citenamefont {Hovhannisyan1}\ and\ \citenamefont
  {Imparato}(2024)}]{imparato2024Q}%
  \BibitemOpen
  \bibfield  {author} {\bibinfo {author} {\bibfnamefont {K.~V.}\ \bibnamefont
  {Hovhannisyan1}}\ and\ \bibinfo {author} {\bibfnamefont {A.}~\bibnamefont
  {Imparato}},\ }\bibfield  {title} {\bibinfo {title} {Energy conservation and
  fluctuation theorem are incompatible for quantum work},\ }\href
  {https://doi.org/10.22331/q-2024-05-06-1336} {\bibfield  {journal} {\bibinfo
  {journal} {Quantum}\ }\textbf {\bibinfo {volume} {8}},\ \bibinfo {pages}
  {1336} (\bibinfo {year} {2024})}\BibitemShut {NoStop}%
\bibitem [{\citenamefont {Martí Perarnau-Llobet}\ and\ \citenamefont
  {Acin}(2017)}]{acin2017nogo}%
  \BibitemOpen
  \bibfield  {author} {\bibinfo {author} {\bibfnamefont {K.~V. H. M.~H.}\
  \bibnamefont {Martí Perarnau-Llobet}, \bibfnamefont {Elisa~Bäumer}}\ and\
  \bibinfo {author} {\bibfnamefont {A.}~\bibnamefont {Acin}},\ }\bibfield
  {title} {\bibinfo {title} {No-go theorem for the characterization of work
  fluctuations in coherent quantum systems},\ }\href
  {https://doi.org/10.1103/PhysRevLett.118.07060} {\bibfield  {journal}
  {\bibinfo  {journal} {Phys. Rev. Lett.}\ }\textbf {\bibinfo {volume} {118}},\
  \bibinfo {pages} {070601} (\bibinfo {year} {2017})}\BibitemShut {NoStop}%
\bibitem [{\citenamefont {Leggett}\ and\ \citenamefont
  {Garg}(1985)}]{leggett1985quantum}%
  \BibitemOpen
  \bibfield  {author} {\bibinfo {author} {\bibfnamefont {A.~J.}\ \bibnamefont
  {Leggett}}\ and\ \bibinfo {author} {\bibfnamefont {A.}~\bibnamefont {Garg}},\
  }\bibfield  {title} {\bibinfo {title} {Quantum mechanics versus macroscopic
  realism: Is the flux there when nobody looks?},\ }\href
  {https://doi.org/10.1103/PhysRevLett.54.857} {\bibfield  {journal} {\bibinfo
  {journal} {Phys. Rev. Lett.}\ }\textbf {\bibinfo {volume} {54}},\ \bibinfo
  {pages} {857} (\bibinfo {year} {1985})}\BibitemShut {NoStop}%
\bibitem [{\citenamefont {Kirkwood}(1933)}]{kirkwood1933quantum}%
  \BibitemOpen
  \bibfield  {author} {\bibinfo {author} {\bibfnamefont {J.~G.}\ \bibnamefont
  {Kirkwood}},\ }\bibfield  {title} {\bibinfo {title} {Quantum statistics of
  almost classical assemblies},\ }\href {https://doi.org/10.1103/PhysRev.44.31}
  {\bibfield  {journal} {\bibinfo  {journal} {Phys. Rev.}\ }\textbf {\bibinfo
  {volume} {44}},\ \bibinfo {pages} {31} (\bibinfo {year} {1933})}\BibitemShut
  {NoStop}%
\bibitem [{\citenamefont {Dirac}(1945)}]{dirac1945on}%
  \BibitemOpen
  \bibfield  {author} {\bibinfo {author} {\bibfnamefont {P.~A.~M.}\
  \bibnamefont {Dirac}},\ }\bibfield  {title} {\bibinfo {title} {On the analogy
  between classical and quantum mechanics},\ }\href
  {https://doi.org/10.1103/RevModPhys.17.195} {\bibfield  {journal} {\bibinfo
  {journal} {Rev. Mod. Phys.}\ }\textbf {\bibinfo {volume} {17}},\ \bibinfo
  {pages} {195} (\bibinfo {year} {1945})}\BibitemShut {NoStop}%
\bibitem [{\citenamefont {Arvidsson-Shukur}\ \emph {et~al.}(2024)\citenamefont
  {Arvidsson-Shukur}, \citenamefont {Braasch~Jr}, \citenamefont {De~Bièvre},
  \citenamefont {Dressel}, \citenamefont {Jordan}, \citenamefont {Langrenez},
  \citenamefont {Lostaglio}, \citenamefont {Lundeen},\ and\ \citenamefont
  {Halpern}}]{arvidsson2024properties}%
  \BibitemOpen
  \bibfield  {author} {\bibinfo {author} {\bibfnamefont {D.~R.~M.}\
  \bibnamefont {Arvidsson-Shukur}}, \bibinfo {author} {\bibfnamefont {W.~F.}\
  \bibnamefont {Braasch~Jr}}, \bibinfo {author} {\bibfnamefont
  {S.}~\bibnamefont {De~Bièvre}}, \bibinfo {author} {\bibfnamefont
  {J.}~\bibnamefont {Dressel}}, \bibinfo {author} {\bibfnamefont {A.~N.}\
  \bibnamefont {Jordan}}, \bibinfo {author} {\bibfnamefont {C.}~\bibnamefont
  {Langrenez}}, \bibinfo {author} {\bibfnamefont {M.}~\bibnamefont
  {Lostaglio}}, \bibinfo {author} {\bibfnamefont {J.~S.}\ \bibnamefont
  {Lundeen}},\ and\ \bibinfo {author} {\bibfnamefont {N.~Y.}\ \bibnamefont
  {Halpern}},\ }\bibfield  {title} {\bibinfo {title} {Properties and
  applications of the kirkwood–dirac distribution},\ }\href
  {https://doi.org/10.1088/1367-2630/ada05d} {\bibfield  {journal} {\bibinfo
  {journal} {New Journal of Physics}\ }\textbf {\bibinfo {volume} {26}},\
  \bibinfo {pages} {121201} (\bibinfo {year} {2024})}\BibitemShut {NoStop}%
\bibitem [{\citenamefont {Lostaglio}\ \emph {et~al.}(2023)\citenamefont
  {Lostaglio}, \citenamefont {Belenchia}, \citenamefont {Levy}, \citenamefont
  {Hernández-Gómez}, \citenamefont {Fabbri},\ and\ \citenamefont
  {Gherardini}}]{lostaglio2023kirkwood}%
  \BibitemOpen
  \bibfield  {author} {\bibinfo {author} {\bibfnamefont {M.}~\bibnamefont
  {Lostaglio}}, \bibinfo {author} {\bibfnamefont {A.}~\bibnamefont
  {Belenchia}}, \bibinfo {author} {\bibfnamefont {A.}~\bibnamefont {Levy}},
  \bibinfo {author} {\bibfnamefont {S.}~\bibnamefont {Hernández-Gómez}},
  \bibinfo {author} {\bibfnamefont {N.}~\bibnamefont {Fabbri}},\ and\ \bibinfo
  {author} {\bibfnamefont {S.}~\bibnamefont {Gherardini}},\ }\bibfield  {title}
  {\bibinfo {title} {Kirkwood-dirac quasiprobability approach to the statistics
  of incompatible observables},\ }\href
  {https://doi.org/10.22331/q-2023-10-09-1128} {\bibfield  {journal} {\bibinfo
  {journal} {Quantum}\ }\textbf {\bibinfo {volume} {7}},\ \bibinfo {pages}
  {1128} (\bibinfo {year} {2023})}\BibitemShut {NoStop}%
\bibitem [{\citenamefont {Pezzutto}\ \emph {et~al.}(2025)\citenamefont
  {Pezzutto}, \citenamefont {Chiara},\ and\ \citenamefont
  {Gherardini}}]{pezzutto2025nonpositive}%
  \BibitemOpen
  \bibfield  {author} {\bibinfo {author} {\bibfnamefont {M.}~\bibnamefont
  {Pezzutto}}, \bibinfo {author} {\bibfnamefont {G.~D.}\ \bibnamefont
  {Chiara}},\ and\ \bibinfo {author} {\bibfnamefont {S.}~\bibnamefont
  {Gherardini}},\ }\href {https://arxiv.org/abs/2503.07759} {\bibinfo {title}
  {Non-positive energy quasidistributions in coherent collision models}}
  (\bibinfo {year} {2025}),\ \Eprint {https://arxiv.org/abs/2503.07759}
  {arXiv:2503.07759 [quant-ph]} \BibitemShut {NoStop}%
\bibitem [{\citenamefont {Santini}\ \emph {et~al.}(2023)\citenamefont
  {Santini}, \citenamefont {Solfanelli}, \citenamefont {Gherardini},\ and\
  \citenamefont {Collura}}]{santini2023work}%
  \BibitemOpen
  \bibfield  {author} {\bibinfo {author} {\bibfnamefont {A.}~\bibnamefont
  {Santini}}, \bibinfo {author} {\bibfnamefont {A.}~\bibnamefont {Solfanelli}},
  \bibinfo {author} {\bibfnamefont {S.}~\bibnamefont {Gherardini}},\ and\
  \bibinfo {author} {\bibfnamefont {M.}~\bibnamefont {Collura}},\ }\href
  {https://doi.org/10.1103/PhysRevB.108.104308} {\bibfield  {journal} {\bibinfo
   {journal} {Phys. Rev. B}\ }\textbf {\bibinfo {volume} {108}},\ \bibinfo
  {pages} {104308} (\bibinfo {year} {2023})}\BibitemShut {NoStop}%
\bibitem [{\citenamefont {Pei}\ \emph {et~al.}(2023)\citenamefont {Pei},
  \citenamefont {Chen},\ and\ \citenamefont {Quan}}]{pei2023exploring}%
  \BibitemOpen
  \bibfield  {author} {\bibinfo {author} {\bibfnamefont {J.-H.}\ \bibnamefont
  {Pei}}, \bibinfo {author} {\bibfnamefont {J.-F.}\ \bibnamefont {Chen}},\ and\
  \bibinfo {author} {\bibfnamefont {H.~T.}\ \bibnamefont {Quan}},\ }\bibfield
  {title} {\bibinfo {title} {Exploring quasiprobability approaches to quantum
  work in the presence of initial coherence: Advantages of the margenau-hill
  distribution},\ }\href {https://doi.org/10.1103/PhysRevE.108.054109}
  {\bibfield  {journal} {\bibinfo  {journal} {Phys. Rev. E}\ }\textbf {\bibinfo
  {volume} {108}},\ \bibinfo {pages} {054109} (\bibinfo {year}
  {2023})}\BibitemShut {NoStop}%
\bibitem [{\citenamefont {Chakrabarty}\ \emph {et~al.}(2025)\citenamefont
  {Chakrabarty}, \citenamefont {Mallick}, \citenamefont {Mukherjee},\ and\
  \citenamefont {Maity}}]{chakrabarty2025probing}%
  \BibitemOpen
  \bibfield  {author} {\bibinfo {author} {\bibfnamefont {S.}~\bibnamefont
  {Chakrabarty}}, \bibinfo {author} {\bibfnamefont {B.}~\bibnamefont
  {Mallick}}, \bibinfo {author} {\bibfnamefont {S.}~\bibnamefont {Mukherjee}},\
  and\ \bibinfo {author} {\bibfnamefont {A.~G.}\ \bibnamefont {Maity}},\ }\href
  {https://arxiv.org/abs/2506.08107} {\bibinfo {title} {Probing kirkwood-dirac
  nonpositivity and its operational implications via moments}} (\bibinfo {year}
  {2025}),\ \Eprint {https://arxiv.org/abs/2506.08107} {arXiv:2506.08107
  [quant-ph]} \BibitemShut {NoStop}%
\bibitem [{\citenamefont {Donati}\ \emph {et~al.}(2024)\citenamefont {Donati},
  \citenamefont {Cataliotti},\ and\ \citenamefont
  {Gherardini}}]{donati2024energetics}%
  \BibitemOpen
  \bibfield  {author} {\bibinfo {author} {\bibfnamefont {L.}~\bibnamefont
  {Donati}}, \bibinfo {author} {\bibfnamefont {F.~S.}\ \bibnamefont
  {Cataliotti}},\ and\ \bibinfo {author} {\bibfnamefont {S.}~\bibnamefont
  {Gherardini}},\ }\bibfield  {title} {\bibinfo {title} {Energetics and
  quantumness of fano coherence generation},\ }\href
  {https://doi.org/10.1038/s41598-024-67037-2} {\bibfield  {journal} {\bibinfo
  {journal} {Scientific Reports}\ }\textbf {\bibinfo {volume} {14}},\ \bibinfo
  {pages} {20145} (\bibinfo {year} {2024})}\BibitemShut {NoStop}%
\bibitem [{\citenamefont {Thio}\ \emph {et~al.}(2025)\citenamefont {Thio},
  \citenamefont {Yang}, \citenamefont {Bièvre}, \citenamefont {Barnes},\ and\
  \citenamefont {Arvidsson-Shukur}}]{thio2025kirkwood}%
  \BibitemOpen
  \bibfield  {author} {\bibinfo {author} {\bibfnamefont {J.~J.}\ \bibnamefont
  {Thio}}, \bibinfo {author} {\bibfnamefont {S.}~\bibnamefont {Yang}}, \bibinfo
  {author} {\bibfnamefont {S.~D.}\ \bibnamefont {Bièvre}}, \bibinfo {author}
  {\bibfnamefont {C.~H.~W.}\ \bibnamefont {Barnes}},\ and\ \bibinfo {author}
  {\bibfnamefont {D.~R.~M.}\ \bibnamefont {Arvidsson-Shukur}},\ }\href
  {https://arxiv.org/abs/2506.08092} {\bibinfo {title} {Kirkwood-dirac
  nonpositivity is a necessary resource for quantum computing}} (\bibinfo
  {year} {2025}),\ \Eprint {https://arxiv.org/abs/2506.08092} {arXiv:2506.08092
  [quant-ph]} \BibitemShut {NoStop}%
\bibitem [{\citenamefont {Donelli}\ \emph {et~al.}(2025)\citenamefont
  {Donelli}, \citenamefont {Chiara}, \citenamefont {Scazza},\ and\
  \citenamefont {Gherardini}}]{donelli2025impact}%
  \BibitemOpen
  \bibfield  {author} {\bibinfo {author} {\bibfnamefont {B.}~\bibnamefont
  {Donelli}}, \bibinfo {author} {\bibfnamefont {G.~D.}\ \bibnamefont {Chiara}},
  \bibinfo {author} {\bibfnamefont {F.}~\bibnamefont {Scazza}},\ and\ \bibinfo
  {author} {\bibfnamefont {S.}~\bibnamefont {Gherardini}},\ }\href
  {https://arxiv.org/abs/2508.01444} {\bibinfo {title} {Impact of quantum
  coherence on the dynamics and thermodynamics of quenched free fermions
  coupled to a localized defect}} (\bibinfo {year} {2025}),\ \Eprint
  {https://arxiv.org/abs/2508.01444} {arXiv:2508.01444 [cond-mat.quant-gas]}
  \BibitemShut {NoStop}%
\bibitem [{\citenamefont {Li}\ \emph {et~al.}(2025)\citenamefont {Li},
  \citenamefont {Xie}, \citenamefont {Kwon}, \citenamefont {Zhao},
  \citenamefont {Kim},\ and\ \citenamefont {Zhang}}]{li2025experimental}%
  \BibitemOpen
  \bibfield  {author} {\bibinfo {author} {\bibfnamefont {H.}~\bibnamefont
  {Li}}, \bibinfo {author} {\bibfnamefont {J.}~\bibnamefont {Xie}}, \bibinfo
  {author} {\bibfnamefont {H.}~\bibnamefont {Kwon}}, \bibinfo {author}
  {\bibfnamefont {Y.}~\bibnamefont {Zhao}}, \bibinfo {author} {\bibfnamefont
  {M.~S.}\ \bibnamefont {Kim}},\ and\ \bibinfo {author} {\bibfnamefont
  {L.}~\bibnamefont {Zhang}},\ }\bibfield  {title} {\bibinfo {title}
  {Experimental demonstration of generalized quantum fluctuation theorems in
  the presence of coherence},\ }\href {https://doi.org/10.1126/sciadv.adq6014}
  {\bibfield  {journal} {\bibinfo  {journal} {Science Advances}\ }\textbf
  {\bibinfo {volume} {11}},\ \bibinfo {pages} {eadq6014} (\bibinfo {year}
  {2025})}\BibitemShut {NoStop}%
\bibitem [{\citenamefont {Hern\'andez-G\'omez}\ \emph
  {et~al.}(2024)\citenamefont {Hern\'andez-G\'omez}, \citenamefont
  {Gherardini}, \citenamefont {Belenchia}, \citenamefont {Lostaglio},
  \citenamefont {Levy},\ and\ \citenamefont
  {Fabbri}}]{hernandez2024projective}%
  \BibitemOpen
  \bibfield  {author} {\bibinfo {author} {\bibfnamefont {S.}~\bibnamefont
  {Hern\'andez-G\'omez}}, \bibinfo {author} {\bibfnamefont {S.}~\bibnamefont
  {Gherardini}}, \bibinfo {author} {\bibfnamefont {A.}~\bibnamefont
  {Belenchia}}, \bibinfo {author} {\bibfnamefont {M.}~\bibnamefont
  {Lostaglio}}, \bibinfo {author} {\bibfnamefont {A.}~\bibnamefont {Levy}},\
  and\ \bibinfo {author} {\bibfnamefont {N.}~\bibnamefont {Fabbri}},\
  }\bibfield  {title} {\bibinfo {title} {Projective measurements can probe
  nonclassical work extraction and time correlations},\ }\href
  {https://doi.org/10.1103/PhysRevResearch.6.023280} {\bibfield  {journal}
  {\bibinfo  {journal} {Phys. Rev. Res.}\ }\textbf {\bibinfo {volume} {6}},\
  \bibinfo {pages} {023280} (\bibinfo {year} {2024})}\BibitemShut {NoStop}%
\bibitem [{\citenamefont {Hernández-Gómez}\ \emph {et~al.}(2024)\citenamefont
  {Hernández-Gómez}, \citenamefont {Isogawa}, \citenamefont {Belenchia},
  \citenamefont {Levy}, \citenamefont {Fabbri}, \citenamefont {Gherardini},\
  and\ \citenamefont {Cappellaro}}]{Hernandez2024interferometry}%
  \BibitemOpen
  \bibfield  {author} {\bibinfo {author} {\bibfnamefont {S.}~\bibnamefont
  {Hernández-Gómez}}, \bibinfo {author} {\bibfnamefont {T.}~\bibnamefont
  {Isogawa}}, \bibinfo {author} {\bibfnamefont {A.}~\bibnamefont {Belenchia}},
  \bibinfo {author} {\bibfnamefont {A.}~\bibnamefont {Levy}}, \bibinfo {author}
  {\bibfnamefont {N.}~\bibnamefont {Fabbri}}, \bibinfo {author} {\bibfnamefont
  {S.}~\bibnamefont {Gherardini}},\ and\ \bibinfo {author} {\bibfnamefont
  {P.}~\bibnamefont {Cappellaro}},\ }\bibfield  {title} {\bibinfo {title}
  {Interferometry of quantum correlation functions to access quasiprobability
  distribution of work},\ }\bibfield  {journal} {\bibinfo  {journal} {npj
  Quantum Information}\ }\textbf {\bibinfo {volume} {10}},\ \href
  {https://doi.org/10.1038/s41534-024-00913-x} {10.1038/s41534-024-00913-x}
  (\bibinfo {year} {2024})\BibitemShut {NoStop}%
\bibitem [{\citenamefont {Perarnau-Llobet}\ \emph {et~al.}(2017)\citenamefont
  {Perarnau-Llobet}, \citenamefont {B\"aumer}, \citenamefont {Hovhannisyan},
  \citenamefont {Huber},\ and\ \citenamefont {Acin}}]{perarnau2017nogo}%
  \BibitemOpen
  \bibfield  {author} {\bibinfo {author} {\bibfnamefont {M.}~\bibnamefont
  {Perarnau-Llobet}}, \bibinfo {author} {\bibfnamefont {E.}~\bibnamefont
  {B\"aumer}}, \bibinfo {author} {\bibfnamefont {K.~V.}\ \bibnamefont
  {Hovhannisyan}}, \bibinfo {author} {\bibfnamefont {M.}~\bibnamefont
  {Huber}},\ and\ \bibinfo {author} {\bibfnamefont {A.}~\bibnamefont {Acin}},\
  }\bibfield  {title} {\bibinfo {title} {No-go theorem for the characterization
  of work fluctuations in coherent quantum systems},\ }\href
  {https://doi.org/10.1103/PhysRevLett.118.070601} {\bibfield  {journal}
  {\bibinfo  {journal} {Phys. Rev. Lett.}\ }\textbf {\bibinfo {volume} {118}},\
  \bibinfo {pages} {070601} (\bibinfo {year} {2017})}\BibitemShut {NoStop}%
\bibitem [{\citenamefont {Margenau}\ and\ \citenamefont
  {Hill}(1961)}]{margenau1961correlation}%
  \BibitemOpen
  \bibfield  {author} {\bibinfo {author} {\bibfnamefont {H.}~\bibnamefont
  {Margenau}}\ and\ \bibinfo {author} {\bibfnamefont {R.~N.}\ \bibnamefont
  {Hill}},\ }\bibfield  {title} {\bibinfo {title} {Correlation between
  measurements in quantum theory},\ }\href {https://doi.org/10.1143/ptp.26.722}
  {\bibfield  {journal} {\bibinfo  {journal} {Progress of Theoretical Physics}\
  }\textbf {\bibinfo {volume} {26}},\ \bibinfo {pages} {722–738} (\bibinfo
  {year} {1961})}\BibitemShut {NoStop}%
\bibitem [{\citenamefont {Yi}\ and\ \citenamefont
  {Kim}(2025)}]{yi2025observing}%
  \BibitemOpen
  \bibfield  {author} {\bibinfo {author} {\bibfnamefont {J.}~\bibnamefont
  {Yi}}\ and\ \bibinfo {author} {\bibfnamefont {Y.~W.}\ \bibnamefont {Kim}},\
  }\bibfield  {title} {\bibinfo {title} {Observing quantum work by
  margenau-hill quasiprobability},\ }\href
  {https://doi.org/10.1209/0295-5075/adcc4c} {\bibfield  {journal} {\bibinfo
  {journal} {Europhysics Letters}\ }\textbf {\bibinfo {volume} {150}},\
  \bibinfo {pages} {41001} (\bibinfo {year} {2025})}\BibitemShut {NoStop}%
\bibitem [{\citenamefont {Yoshimura}\ and\ \citenamefont
  {Hamazaki}(2025)}]{yoshimura2025quasiprobability}%
  \BibitemOpen
  \bibfield  {author} {\bibinfo {author} {\bibfnamefont {K.}~\bibnamefont
  {Yoshimura}}\ and\ \bibinfo {author} {\bibfnamefont {R.}~\bibnamefont
  {Hamazaki}},\ }\href {https://arxiv.org/abs/2508.14354} {\bibinfo {title}
  {Quasiprobability thermodynamic uncertainty relation}} (\bibinfo {year}
  {2025}),\ \Eprint {https://arxiv.org/abs/2508.14354} {arXiv:2508.14354
  [quant-ph]} \BibitemShut {NoStop}%
\bibitem [{\citenamefont {Bizzarri}\ \emph {et~al.}(2025)\citenamefont
  {Bizzarri}, \citenamefont {Gherardini}, \citenamefont {Manrique},
  \citenamefont {Bruni}, \citenamefont {Gianani},\ and\ \citenamefont
  {Barbieri}}]{bizzarri2025quasiprobability}%
  \BibitemOpen
  \bibfield  {author} {\bibinfo {author} {\bibfnamefont {G.}~\bibnamefont
  {Bizzarri}}, \bibinfo {author} {\bibfnamefont {S.}~\bibnamefont
  {Gherardini}}, \bibinfo {author} {\bibfnamefont {M.}~\bibnamefont
  {Manrique}}, \bibinfo {author} {\bibfnamefont {F.}~\bibnamefont {Bruni}},
  \bibinfo {author} {\bibfnamefont {I.}~\bibnamefont {Gianani}},\ and\ \bibinfo
  {author} {\bibfnamefont {M.}~\bibnamefont {Barbieri}},\ }\bibfield  {title}
  {\bibinfo {title} {Quasiprobability distributions with weak measurements},\
  }\href {https://doi.org/10.1088/2058-9565/adf573} {\bibfield  {journal}
  {\bibinfo  {journal} {Quantum Science and Technology}\ }\textbf {\bibinfo
  {volume} {10}},\ \bibinfo {pages} {045008} (\bibinfo {year}
  {2025})}\BibitemShut {NoStop}%
\bibitem [{\citenamefont
  {Jarzynski}(1997{\natexlab{a}})}]{jarzynski1997nonequilibrium}%
  \BibitemOpen
  \bibfield  {author} {\bibinfo {author} {\bibfnamefont {C.}~\bibnamefont
  {Jarzynski}},\ }\bibfield  {title} {\bibinfo {title} {Nonequilibrium equality
  for free energy differences},\ }\href
  {https://doi.org/10.1103/PhysRevLett.78.2690} {\bibfield  {journal} {\bibinfo
   {journal} {Phys. Rev. Lett.}\ }\textbf {\bibinfo {volume} {78}},\ \bibinfo
  {pages} {2690} (\bibinfo {year} {1997}{\natexlab{a}})}\BibitemShut {NoStop}%
\bibitem [{\citenamefont
  {Jarzynski}(1997{\natexlab{b}})}]{jarzynski1997equilibrium}%
  \BibitemOpen
  \bibfield  {author} {\bibinfo {author} {\bibfnamefont {C.}~\bibnamefont
  {Jarzynski}},\ }\bibfield  {title} {\bibinfo {title} {Equilibrium free-energy
  differences from nonequilibrium measurements: A master-equation approach},\
  }\href {https://doi.org/10.1103/PhysRevE.56.5018} {\bibfield  {journal}
  {\bibinfo  {journal} {Phys. Rev. E}\ }\textbf {\bibinfo {volume} {56}},\
  \bibinfo {pages} {5018} (\bibinfo {year} {1997}{\natexlab{b}})}\BibitemShut
  {NoStop}%
\bibitem [{\citenamefont {Nielsen}\ and\ \citenamefont
  {Chuang}(2012)}]{nielsen2012quantum}%
  \BibitemOpen
  \bibfield  {author} {\bibinfo {author} {\bibfnamefont {M.~A.}\ \bibnamefont
  {Nielsen}}\ and\ \bibinfo {author} {\bibfnamefont {I.~L.}\ \bibnamefont
  {Chuang}},\ }\href {https://doi.org/10.1017/cbo9780511976667} {\emph
  {\bibinfo {title} {Quantum Computation and Quantum Information: 10th
  Anniversary Edition}}}\ (\bibinfo  {publisher} {Cambridge University Press},\
  \bibinfo {year} {2012})\BibitemShut {NoStop}%
\bibitem [{\citenamefont {Dawson}\ and\ \citenamefont
  {Nielsen}(2005)}]{dawson2005solovay}%
  \BibitemOpen
  \bibfield  {author} {\bibinfo {author} {\bibfnamefont {C.~M.}\ \bibnamefont
  {Dawson}}\ and\ \bibinfo {author} {\bibfnamefont {M.~A.}\ \bibnamefont
  {Nielsen}},\ }\href {https://arxiv.org/abs/quant-ph/0505030} {\bibinfo
  {title} {The solovay-kitaev algorithm}} (\bibinfo {year} {2005}),\ \Eprint
  {https://arxiv.org/abs/quant-ph/0505030} {arXiv:quant-ph/0505030 [quant-ph]}
  \BibitemShut {NoStop}%
\end{thebibliography}
\end{document}